\documentclass[11pt,preprintnumbers]{article}
\usepackage{jheppub}
\usepackage{bbm}
\usepackage{epsfig,bm}
\usepackage{graphics,graphicx,comment,subfigure}
\usepackage{amsmath}
\usepackage{mdframed}

\usepackage{color}

\usepackage{mathtools}
\usepackage{hyperref}

\hypersetup{
    colorlinks=true,       
    linkcolor=red,          
    citecolor=blue,        
    filecolor=magenta,      
    urlcolor=blue           
}

\newcommand{\pslash}{\not{\hbox{\kern-2.3pt $p$}}}
\newcommand{\qslash}{\not{\hbox{\kern-2.3pt $q$}}}
\newcommand{\kslash}{\not{\hbox{\kern-2.3pt $k$}}}
\newcommand{\partialslash}{\not{\hbox{\kern-2.3pt $\partial$}}}

\def \be { \begin{equation} }
\def \ee { \end{equation} }

\def\B{{\cal B}}

\def\N{{\cal N}}
\def\T{{\cal T}}

\def\F{{\mathcal{F}}}

\def\I{{\cal I}}

\def\R{{\mathbb R}}
  
\def\tr{\,{\rm Tr}\,}
\def\Re{{\rm Re}\,}
\def\Im{{\rm Im}\,}

\def\Z{{\mathbb Z}}
\def\Dslash{{\rlap{\raise 1pt \hbox{$\>/$}}D}}

\def\Im{\mathrm {Im}}
\def \( {\left(}
\def \) {\right)}
\def\psit{\tilde{\psi}}
\def\Ut{\tilde{U}}
\def\s{\sigma}
\def\th{\theta}

\def\slashchar#1{\ensuremath{                               %
   \setbox0=\hbox{${}#1{}$}       
   \dimen0=\wd0                                 
   \setbox1=\hbox{/} \dimen1=\wd1               
   \ifdim\dimen0>\dimen1                        
      \rlap{\hbox to \dimen0{\hfil/\hfil}}      
      {}#1{}                                    
   \else                                        
      \rlap{\hbox to \dimen1{\hfil${}#1{}$\hfil}}   
   \fi}}

\DeclareMathOperator{\arccot}{arcCot}

\preprint{{\flushright DAMTP-2014-17\\UMN-TH-2239/14\\FTPI-MINN-14/8\\}}

\title{ 
Decoding  perturbation theory using resurgence:
Stokes phenomena, new saddle points and Lefschetz thimbles
}

\author[1]{Aleksey Cherman,}
\emailAdd{acherman@umn.edu}
\affiliation[1]{%
Fine Theoretical Physics Institute, Department of Physics, 
 University of Minnesota, USA
}
\author[2]{Daniele Dorigoni}
\emailAdd{d.dorigoni@damtp.cam.ac.uk}
\affiliation[2]{%
DAMTP, University  of Cambridge, 
Wilberforce Road, Cambridge CB3 0WA, UK
}
\author[3]{and Mithat \"Unsal}
\emailAdd{unsal.mithat@gmail.com}
\affiliation[3]{%
Department  of Physics,    
North Carolina State University, Raleigh, NC, 27695 USA}

\abstract{%
Resurgence theory implies that  the non-perturbative (NP)     and perturbative (P) data in a QFT are quantitatively related,  and that detailed information about non-perturbative saddle point field configurations of path integrals
can be extracted from 
perturbation theory.  Traditionally, only stable NP saddle points are considered in QFT, and homotopy group considerations are used to classify them.  However, in many QFTs the relevant homotopy groups are trivial, and even when they are non-trivial they leave many NP saddle points undetected.  
Resurgence provides  a refined  classification of NP-saddles, going beyond conventional topological considerations. To demonstrate some of these ideas, we study the  $SU(N)$ principal chiral model (PCM), a two dimensional asymptotically free matrix field theory which has no instantons, because the relevant homotopy group is trivial.   Adiabatic continuity is used to reach 
 a weakly coupled regime where NP effects are calculable.    We then use resurgence theory to uncover the existence and role of novel `fracton' saddle points, which turn out to be the fractionalized constituents of previously observed unstable `uniton' saddle points.   The fractons play a crucial role in the physics of the PCM, and are responsible for the dynamically generated mass gap of the theory.  Moreover, we show that the fracton-anti-fracton events are the weak coupling realization of 't Hooft's renormalons, and argue that the renormalon ambiguities are systematically cancelled in the semi-classical expansion.  Our results motivate the conjecture that the semi-classical expansion of the path integral can be geometrized as a sum over Lefschetz thimbles.
 }


\begin{document}

\maketitle
\section{Introduction}
\label{sec:Introduction}

\subsection{Problems with perturbation theory and the semi-classical expansion}
Consider an observable $\mathcal{O}$ in a quantum  theory with a dimensionless coupling $\lambda$, defined by an appropriately regularized Euclidean path integral:
\begin{align}
\langle \mathcal{O}[\lambda] \rangle = Z[\lambda]^{-1} \int d U e^{-S[U; \lambda]} \mathcal{O},
\end{align}
where $Z[\lambda]$ is the partition function.  If $\lambda$ can be kept small,\footnote{In 
asymptotically free theories, $\lambda$ grows at low energies.  However, its growth can be cut off 
and $\lambda$ can be kept naturally small at low energies by e.g. adjoint Higgsing either by an elementary Higgs scalar or Wilson line expectation value in gauge theories, or by turning on appropriate background fields. }
 one expects to be able to evaluate the path integral using the saddle-point method, so that (schematically)
\begin{align}
\langle \mathcal{O}[\lambda] \rangle = \sum_{n=0}^{\infty} p_{0,n}  \lambda^n + \sum_{c} e^{-S_c/\lambda} \sum^{\infty}_{n=0} p_{c, n} \lambda^{n} .
\label{eq:NaiveTransSeries}
\end{align}
Above $p_{0,n}$ are the perturbative contributions to $\mathcal{O}$, and encode an expansion in fluctuations around the trivial ``perturbative'' saddle-point $U_0$  of the path integral, which has zero action.  There are also contributions  
from non-perturbative saddle point field configurations $U_c$, which have finite actions $S_c$ measured in units of $\lambda$, and contributions from the perturbative fluctuations $p_{c,n}$  around $U_c$. 

Eq.~\eqref{eq:NaiveTransSeries}  is traditionally viewed as the semiclassical \emph{approximation} to the original path integral.
The reason is that in almost all interesting QFTs, and even in simple  quantum mechanics or even simpler ordinary integrals, the perturbative series expansions around both the perturbative saddle $U_0$ as well as $U_c$ are actually divergent asymptotic expansions, with  $p_{0,n},\,p_{c,n} \sim n!$ \cite{Dyson:1952tj,Lipatov:1976ny,Beneke:1998ui}.   
 The standard way to give a meaning to such perturbative series is via Borel transform and resummation. 
After computing the Borel transform of an asymptotic series, and its analytic continuation, one obtains a function with singularities in the `Borel plane'.  The Borel sum of the perturbative series is defined as a Laplace transform of the analytic continuation of the Borel transform.  The issue is that if $p_{0,n},\,p_{c,n} \sim n!$ then there will be singularities on the integration contour in the Borel sum, and the integral --- and hence the sum --- is not well-defined.  Different choices of contour deformations to avoid the singularities give different results for the same physical observable.  This is a reflection of the fact that $\lambda \in \mathbb{R}^+$ is a Stokes line.
As a result,  the Borel sums of all of the perturbative series appearing in Eq.~\eqref{eq:NaiveTransSeries}  are usually not well-defined.  

Another (much less widely appreciated) fact about the semiclassical expansion is that the amplitudes associated with  certain 
  saddle points, for example, correlated  instanton-anti-instanton $[\I \bar\I]$ events,  are  not well-defined  either \cite{Bogomolny:1980ur,ZinnJustin:1981dx} along the $\lambda \in \mathbb{R}^+$ Stokes line.  That is, in addition to the ambiguities in the sum of the perturbative series, the sum over non-perturbative saddle points also suffers from ambiguities. 
    But if every perturbative series  and most of the  non-perturbative factors appearing
    in  our expansion are not well-defined,
     then in what sense, and to what extent, does the semiclassical expansion capture the physics encoded in the original path integral?  How do we give a meaning to a saddle point expansion?   The standard perspective is that the semiclassical approximation has an inherent `fuzziness' defined by the size of the resummation ambiguities, and Eq.~\eqref{eq:NaiveTransSeries} only approximates the value of the original integral up to semiclassically-incalculable corrections of the order of the ambiguities.
     
     Although the inclusion of the contributions of the NP-saddles seems to make the problems in the semiclassical approximation  even worse, we will argue that including the NP saddles is in fact  {\it the solution}  to  defining our saddle point expansion for $\langle \mathcal{O}[\lambda] \rangle$ in an ambiguity-free, meaningful way.   
               

\subsection{Resurgence theory } 
To see how the program of assigning unambiguous meaning to the semiclassical expansion for $\langle \mathcal{O}[\lambda] \rangle$ might work, we note that it has been known for some time that there are special cases when an unambiguous meaning can be assigned to Eq.~\eqref{eq:NaiveTransSeries} by carefully including the contributions of the NP saddles.  For example, in  a double-well  or periodic potential problem in quantum mechanics,  it is known that the leading ambiguity in perturbation theory is cured by 
      the ambiguity in the  $[\I \bar\I]$  amplitude (and fluctuations around it),  
      and the ambiguity in the perturbation theory around  an instanton is cured by the ambiguity in 
      in the  $[\I \I \bar\I]$  amplitude (and fluctuations around it), {\it etc}. 
 \cite{Bogomolny:1980ur,ZinnJustin:1981dx,delabaere1997exact},  and see also 
  \cite{Dunne:2013ada,Dunne:2014bca}. 

Such cancellations of ambiguities may seem magical, but in fact underlying the cancellations there is a systematic mathematical framework called resurgence theory, a term coined in a different context by J. Ecalle in early 80s \cite{Ecalle:1981}. 
Applied to QFT, resurgence theory is a generalization of the venerable idea of Borel resummation of the perturbative expansion around the perturbative saddle  which systematically incorporates Stokes phenomena \cite
{Dingle_asymptotics,berry1990hyperasymptotics,berry1991hyperasymptotics,2007arXiv0706.0137S,costin2008asymptotics}. 
 As described above,  in most interesting quantum mechanical systems and  QFTs, Borel resummation does not work (i.e., gives ambiguities)    due to singularities in the Borel plane. 
If the Borel transform of all perturbative series are  {\it endlessly continuable} (i.e.   the set of singularities in all  Riemann sheets are discrete and there are no natural boundaries),  then trans-series of the form of Eq.~\eqref{eq:NaiveTransSeries} can be viewed as expansions of resurgent functions.     
 Ecalle's work \cite{Ecalle:1981} implies that for such trans-series, all would-be ambiguities of the semiclassical representation cancel, see also \cite{Aniceto:2013fka}.   The key to these cancellations is that in the trans-series representation Eq.~\eqref{eq:NaiveTransSeries}, there are also ambiguities associated with the non-perturbative factors $e^{-S_c/\lambda}$, which exactly cancel  
  the leading ambiguities in the perturbation theory, with further (more intricate) relations amongst the various terms in the trans-series leading to the cancellation of ambiguities at higher orders, in such a way that the trans-series representation is ambiguity-free to all orders\footnote{See Appendix~\ref{sec:ResurgenceAppendix} for a brief overview of Borel summation, trans-series, and resurgence.}.
If we conjecture that observables in QFT are resurgent functions, then resurgence theory implies that the expansions around any given saddle-point must contain exact information concerning the expansions around all other saddle-points of the theory.   In particular, resurgence implies that encoded within the large order terms of \emph{perturbative} series there is exact information about \emph{non-perturbative} saddles. As a suggestive equation, one may call this idea ``P-data = NP-data".

We should emphasize that resurgence suggests a major philosophical shift on the meaning of the semiclassical approximation. If the right hand side of Eq.~\eqref{eq:NaiveTransSeries} can indeed be systematically interpreted in an unambiguous way, then the semiclassical expansion should not be thought of as an approximation.  Instead, when viewed as a resurgent trans-series, the saddle point expansion should be viewed as an exact coded \emph{representation} of the observable $\langle \mathcal{O}(\lambda)\rangle$ in the regime of the QFT which is smoothly connected to the small $\lambda$ semiclassical limit.

In this work, we take resurgence as our guiding principle, and use it to find new saddles in certain QFTs.   We are able to systematically test the predictions of resurgence theory by using the recently developing ideas of adiabatic continuity and weak coupling NP-calculability.

\subsection{Beyond the topological classification of NP saddles} 
In the context of QFT,  it has recently been proposed to use resurgence theory to provide  evidence for a non-perturbative continuum definition in the semi-classical domain ~\cite{Argyres:2012ka,Argyres:2012vv,Dunne:2012ae,Dunne:2012zk}  by invoking the idea of adiabatic continuity  \cite{Shifman:2008ja, Unsal:2008ch}. This program provides  a new insight into 't Hooft's mysterious  renormalon problem \cite{tHooft:1977am, Beneke:1998ui}.
In this context,  resurgence theory has been applied to non-Abelian gauge theories on $\mathbb{R}^3 \times S^1$~
and the $\mathbb {CP}^{N-1}$ non-linear sigma model on $\mathbb{R} \times S^1$. 
  In both cases, the theories involved have a non-trivial homotopy group classifying the stable NP saddle points, and consequently, they also have
instantons, fractionalized instantons\cite{Shifman:1994ce,Lee:1997vp,Kraan:1998pm,Bruckmann:2007zh,Brendel:2009mp,Harland:2009mf}, and composite configurations made from some combination of correlated instanton and fractionalized instanton events.   Using resurgence theory, it has recently been proposed that the ambiguities due to the most severe ``infrared renormalon'' sources of divergences in these asymptotically-free theories cancel against the contributions of the appropriate  neutral bion (fractional instanton-anti-instanton) events with action $\frac{2}{N}$ in units where BPST instanton action is normalized to unity,  in a semi-classical regime of the theory\cite{Argyres:2012ka,Argyres:2012vv,Dunne:2012ae,Dunne:2012zk}.
 If it turns out that the cancellations of ambiguities persist to all orders, resurgence theory would yield a systematic non-perturbative semi-classical definition of asymptotically-free theories.  

It is important  to note that resurgence provides a classification of NP saddles which is more refined than the traditional topological classification of saddle points, based on $\pi_{3}[G]$ in 4D gauge theories with gauge group $G$, and on $\pi_2[T]$ in 2D non-linear sigma models with target space $T$.
If two saddles are in the same conjugacy class in these homotopy groups then they carry the same topological charge. So topology cannot be used to distinguish them. 
On the other hand, if these two topologically identical saddle points have different actions, then the non-analyticities in the coupling $\lambda$ of their contributions to the path integral are different, and hence they are distinguishable using resurgence theory. 
For example, the perturbative saddle and the instanton saddle by definition constitute two different conjugacy classes according to homotopy, call them ${\cal C}_0$ and ${\cal C}_{+1}$. The elements of these conjugacy classes are
 \begin{align} 
{\cal C}_0:& \qquad  \Big \{ \;  [0], [\I \bar {\I}], [\I^2 \bar {\I}^2], [\I^3 \bar {\I}^3], \ldots,  [\I^n \bar {\I}^n], \ldots  \Big   \}  \cr
{\cal C}_{+1}:& \qquad  \Big \{ \;  [\I], [\I^2 \bar {\I}], [\I^3 \bar {\I}^2], [\I^4 \bar {\I}^3], \ldots, [\I^{n+1}  \bar {\I}^n], \ldots  \Big   \}\; , \qquad 
\label{NP-events}
\end{align} 
The elements within each class are not distinguished by topological considerations.  However, the elements of these conjugacy classes can be distinguished according to resurgence theory.
 This is the motivation of the ``resurgence triangle" classification of saddle points discussed in \cite{Dunne:2012ae,Dunne:2012zk}.  

We should emphasize that all of the NP saddle points appearing in the homotopy conjugacy classes above, except for the perturbative saddle $[0]$ and the instanton saddle $[\mathcal{I}]$, are actually \emph{quasi}-saddle-points.  To see what is meant by this, recall that in the semiclassical limit path integrals become dominated by field configurations which come as close as possible to satisfying the equations of motion of the classical action and have finite action.   So field configurations that are exact solutions of the equations of motion are of course important in the semiclassical limit, and often they are only field configurations considered.  However, while it is much less widely appreciated, in the semiclassical limit $\lambda \ll 1$ there are generally also quasi-solutions of the equations of motion, which come parametrically close to satisfying the equations of motion and have finite action.  Some prominent examples of such configurations are e.g. \emph{correlated} multi-instanton events in quantum mechanics and QFT, and magnetic bions\cite{Unsal:2007jx} in gauge theory.  We refer to such finite-action quasi-solution field configurations as quasi-saddle-points.  While quasi-saddle points are typically not distinguished from exact saddle points by homotopy theory considerations, they are distinguishable using resurgence theory, and are categorized in the resurgence triangle classification of NP saddles.  We find that quasi-saddle-points make critical contributions to QFTs in the semiclassical limit.

In this work, we give a more dramatic realization of the idea of the resurgence triangle classification of saddle points. 
Some interesting QFTs have a trivial homotopy group. Relatedly, they do not possess
any known topologically-stable finite-action field configurations like instantons, and hence cannot have fractionalized instantons either!  So one might naively think that a semi-classical calculation of observables in such theories would include contributions \emph{only} from the trivial perturbative saddle point.  

However,  high-order factorial divergences of perturbation theory are ubiquitous and are known to occur even in theories without instantons.   If the resurgence formalism is the right way to think about the semiclassical representation of path integrals, it implies that there must \emph{always} be finite-action configurations that contribute to the path integral  whenever the sum of the perturbative series is ambiguous.  This must be the case even when homotopy considerations leave no room for contributions from stable instantons or their constituents.    This is an even sharper illustration of the point that resurgence theory provides a much more refined classification of the finite-action field configurations that can contribute to path integrals than conventional homotopy-theoretic methods.  Understanding how this works in detail is a major focus of this paper.  Previous works related to this question include  \cite{Dunne:2012ae,Dunne:2012zk,Dabrowski:2013kba} in the context of the $\mathbb{CP}^{N-1}$ model and \cite{Argyres:2012ka,Argyres:2012vv} in QCD(adj) and deformed YM, which are  theories with a non-trivial homotopy group,  and \cite{Cherman:2013yfa} in a theory  with trivial homotopy group.  Indeed, the present paper is a detailed exposition of the results briefly announced in our joint work with G.~V.~Dunne \cite{Cherman:2013yfa}.  

We also note a related work \cite{Pasquetti:2009jg} in the context of matrix models and topological string theory.   Indeed, there is an important body of work applying resurgence theory to matrix models and string theory\cite{Marino:2007te,Marino:2008ya,Marino:2008vx,Garoufalidis:2010ya,Aniceto:2011nu,Schiappa:2013opa}, where resurgent analysis is used to find new non-perturbative sectors which must also be taken into account in order to construct full non-perturbative solutions. 
  For a recent review emphasizing resurgence in quantum mechanics and matrix models see \cite{Marino:2012zq}.  Ideas from resurgence theory have also recently appeared in the context of 4D $\mathcal{N}=2$ supersymmetric gauge theories in the body of work triggered by \cite{Gaiotto:2008cd,Gaiotto:2009hg}, which gave a physical derivation of the Kontsevich-Soibelman (KS) wall-crossing formula\cite{Kontsevich:2008fj}.

\subsection{Application: Principal Chiral Model}
A convenient toy model which we will use to examine the above issues is the two dimensional $SU(N)$ principal chiral model (PCM).  Some basic facts and expectations about the dynamics of PCM are summarized in 
Section~\ref{basic}, and Section 2.1 of Polyakov's book \cite{Polyakov:1987ez}.
This theory shares many of the most important features of Yang-Mills theory, with the following ``similarity list'':
\begin{itemize}
\item[\bf  S1)] Asymptotic freedom;
\item[\bf S2)] Matrix-like large-$N$ limit (as opposed to vector-like), planar dominance;
\item[\bf S3)] Dynamically generated mass gap;
\item[\bf S4)]  Confinement:  free energy $\mathcal{F}/N^2$ goes to zero in the {\it low} temperature regime;
\item[\bf S5)] Deconfinement: $O(N^2)$ free energy in the {\it  high} temperature regime;
\item[\bf S6)] Large-order structure of perturbation theory,  presence of renormalon singularities.
\end{itemize}
    These features {\it might} have made the PCM an especially useful toy model for Yang-Mills. 
    In fact, the above similarities are much more pronounced between Yang-Mills theory and the PCM compared to the similarities between  YM and the $\mathbb {CP}^{N-1}$ model.  However, historically, the $\mathbb{CP}^{N-1}$ model has been studied much more intensively.  A major reason behind this 
    is the fact that the PCM has some properties which are in stark contrast with Yang-Mills theory. So, we also make a  ``contrast list":
  \begin{itemize}
\item[\bf C1)]  $\pi_2[SU(N)] = 0$, so the homotopy group which would classify instantons is trivial;
\item[\bf C2)] Absence of topological charge and  topologically stable instantons;
 \item[\bf C3)] Presence of uniton saddle (with no proposed quantum interpretation until \cite{Cherman:2013yfa}) 
\end{itemize}
Furthermore,  to actually be useful, a toy model has to be more   friendly  than the original theory, QCD, at least in some limit. But unlike other (vector-like) sigma models, such as the $\mathbb{CP}^{N-1}$ model, for which explicit first-principles analytic solutions are known at large-$N$,  there is no known first-principles large-$N$ analytic solution based on expanding around a large-$N$ saddle 
point despite heroic efforts by Polyakov \cite{Polyakov:1987ez},  Section 8.2.
\footnote{ Despite not being a first-principles solvable model at large-$N$, in the sense that one can go directly from the Lagrangian to the physical spectrum and correlation functions without making assumptions about the dynamics, the PCM turns out to be quantum mechanically integrable for any $N$, in contrast to the $\mathbb{CP}^{N-1}$ model, which is not integrable at the quantum level for $N>2$.  
The integrability of the model, together with a few plausible assumptions, most crucially the existence of a mass gap, leads to a solution for the minimal factorized $S$-matrix for the PCM\cite{Polyakov:1983tt,Wiegmann:1984ec} which was discovered a long time ago. For some recent work see e.~g.~\cite{Cubero:2012xi,Cubero:2013iga,Cubero:2014hla}. The spectrum as given by integrability contains  $N-1$ massive particles. 
While integrability techniques lead to a solution for the PCM $S$-matrix, 
the relation of the resulting solution for the S-matrix to the microscopic physics is quite opaque.  Moreover, while the assumptions necessary for the derivation of the S-matrix using integrability --- such as the presence of the mass gap --- are certainly reasonable, they have yet to be demonstrated from first principles.  Finally, the integrability-based approaches do not yield any information about the interpretation of the divergences of renormalized perturbation theory, which are a major focus of our work.}

The consequences of the application of resurgence theory to the PCM are rather striking. Resurgence tells us that if the PCM model 
exist as a quantum theory, item {\bf S6)} in the similarity list implies that 
all the elements in the contrast list  must be consequence of {\it superficial} reasoning. At a deeper level, the similarity of the large-order growth of perturbation theory in PCM and YM theory makes it {\it impossible} that the  principal chiral model only has a trivial perturbative saddle point. In fact, it implies that it must possess a plethora
of NP-saddles which is just as rich as in Yang-Mills theory. 
In this work, we confirm this resurgence theory expectation by explicit calculations.

\subsection{Outline}
The organization of this somewhat lengthy paper is as follows.  In Section~\ref{sec:examples},  we provide a zero dimensional toy example, related to the 2d theory via dimensional reduction, which exhibits 
Borel non-summability,  Stokes phenomena and the cancellation of ambiguities upon Borel-Ecalle (BE)resummation. In combination with Appendix~\ref{sec:ResurgenceAppendix}, we hope that this provides a gentle introduction to some of the methods of resurgence theory. In Section~\ref{sec:examples}, we also point out the relation between semiclassical expansions and Lefschetz thimbles, giving a geometric perspective on resurgence. Section~\ref{basic} summarizes some basic facts and expectations about the dynamics of the PCM  on $\R^2$.
 In Section~\ref{sec:SmallLLimit} we explain the construction of the unique weakly-coupled small-$L$ limit of the PCM  on $ \R \times S^1_L$ which is continuously connected to the theory on $\mathbb{R}^2$.  Our analysis is inspired by the one in \cite{Dunne:2012ae,Dunne:2012zk} for the $\mathbb {CP}^{N-1}$ model.
  The weak-coupling parameter turns out to be $\frac{N L \Lambda}{2\pi} \ll 1$,  similarly to deformed YM and  QCD(adj) \cite{Unsal:2008ch} and 3d YM with adjoint matter or twisted boundary conditions  \cite{Armoni:2011dw,Perez:2013dra}. 
  At leading order in $N L \Lambda$, low-energy observables in the 2D PCM can be described by a simple one-dimensional effective field theory, which is just quantum mechanics.  In Section~\ref{sec:pert-th} we study large-order perturbative behavior of the  weakly-coupled QM limit of the $SU(N)$ PCM. By using resurgence theory techniques, we identify the non-perturbative ambiguity in the Borel resummation and interpret it as pointing to the presence of new NP-saddles in the problem. 
    In Section~\ref{sec:FractonsUnitons}, we show that the model indeed has the predicted non-perturbative saddle points, the unitons and fractons, and describe their properties.   In Section~\ref{sec:Resurgence} we show that the amplitudes of correlated fracton-anti-fracton events (which we often refer to as neutral bions) have ambiguous parts on the Stokes line, and these ambiguous parts cancel the renormalon ambiguities of perturbation theory.    
   Thus, we interpret the neutral bion as the semi-classical realization of the infrared renormalon.  
     In Section~\ref{sec:Borelflow} we show that the fractons are responsible for the generation of the mass gap in the bosonic PCM at small-$L$. We describe a plausible flow of the mass gap as the radius is dialed from small to large-$L$.  Relatedly, we point out that the Borel plane singularities on $\R^2$ are {\it twice}   as dense as compared to the location of singularities on 
  $\R \times S^1$, and argue that there should exist a smooth flow  in the location of singularities as the radius is dialed from small to large-$L$.  We refer to this phenomenon as Borel flow. 
 Understanding the exact nature of the Borel flow would amount to solution of the mass gap problem on $\R^2$, which is an open problem.

\section{Zero dimensional prototype for resurgence and Lefschetz thimbles}
\label{sec:examples}

In this section, we consider a zero dimensional integral using steepest descent methods. 
as a prototype of the semi-classical approach in path integrals. 
In fact,  the zero dimensional model is related to the 2d QFT by dimensional reduction.  Compactifying the 2d QFT on small $\R \times S^1$ with twisted boundary condition on $S^1$, we land on a quantum mechanical problem with periodic potential. Further compactifying the QM problem and going to the small  $ S^1 \times S^1$ regime, the  integral over the zeroth Kaluza-Klein mode 
reduce to our 0d prototype. 

Perturbative expansion of the  finite-dimensional integral already exhibits non-Borel summability,  Stokes phenomena and cancellation of ambiguities upon Borel-Ecalle (BE)resummation that also take place in path-integral of PCM, and hence 
  provides a useful playground in which we can show many properties very explicitly.   \footnote{ The discussion of this section streamlines the analytic continuation ideas of \cite{Pham1983,Witten:2010cx}
  and aims to make the relation to Ecalle's theory of resurgence \cite{Ecalle:1981}
 as simple as possible.  The material in this section is already known, however we find it useful to detail it since 
 we use a parallel approach for path integrals later in the paper.}

Consider the zero dimensional partition function  $Z (\lambda)$
\begin{align}
Z (\lambda)&= \int_{-{\pi \over 2 \sqrt \lambda }}^{{\pi \over 2 \sqrt \lambda }} dy \, e^{-\frac{1}{2 \lambda} \sin^2(\sqrt \lambda y)} =  {1 \over \sqrt \lambda}   \int_{-{\pi \over 2  }}^{{\pi \over 2  }} dx \, e^{-\frac{1}{2 \lambda} \sin^2(x)}  \\
&= \frac{\pi \, e^{\frac{-1}{4 \lambda }}}{\sqrt{\lambda }}  I_0\left(\frac{1}{4 \lambda }\right) ,\nonumber
\label{part}
\end{align}
where $I_0$ is the modified Bessel function of the first kind.  $Z(\lambda)$ is an integral of a real function over a real domain on a finite interval  $ I = \left[- \frac{\pi}{2}, \frac{\pi}{2} \right]$,  hence the result is manifestly real for real $\lambda$. 
In order to demonstrate the use of some of the resurgence technology that we will use in QFT, we would like to study this integral by using the steepest descent expansion, which is the counter-part of the semi-classical expansion in our QFT example. 
The fundamental idea of the analysis is that to understand the behavior of the $Z(\lambda)$ for $\lambda \in \mathbb{R}^+$ one should understand the behavior of the analytic continuation of $Z(\lambda)$ when $\lambda \in \mathbb{C}$.

Our analysis will proceed as follows:
\begin{itemize}
\item[\bf 1)]{Identify all critical points.}
\item[\bf  2)]{Allow $\lambda$ to move off $\mathbb{R}^+$ into $\mathbb{C}$, and analytically continue $Z(\lambda)$ by rewriting the original integration cycle as a sum over steepest descent paths, which are called Lefschetz thimbles in general. 
}
\item[\bf  3)]{Develop perturbation theory around the P and NP saddles, and derive the respective asymptotic expansions.  This is the counterpart of the semi-classical approximation in QFT.}
\item[\bf  4)]{Show that the  action of the NP saddle governs the growth of late terms in the perturbative series around P-saddle, and that sub-leading corrections to the late terms in the perturbative series around the P-saddle are governed by early terms of the perturbative expansion around the NP-saddle and vice versa. }
\item[\bf  5)]{Show the cancellation of ambiguities and the reality of the trans-series representation of $Z(\lambda)$ on the $\lambda \in \mathbb{R}^+$ Stokes line.}
\end{itemize}

\begin{figure}[tbp]
\centering
\includegraphics[width=0.45\textwidth]{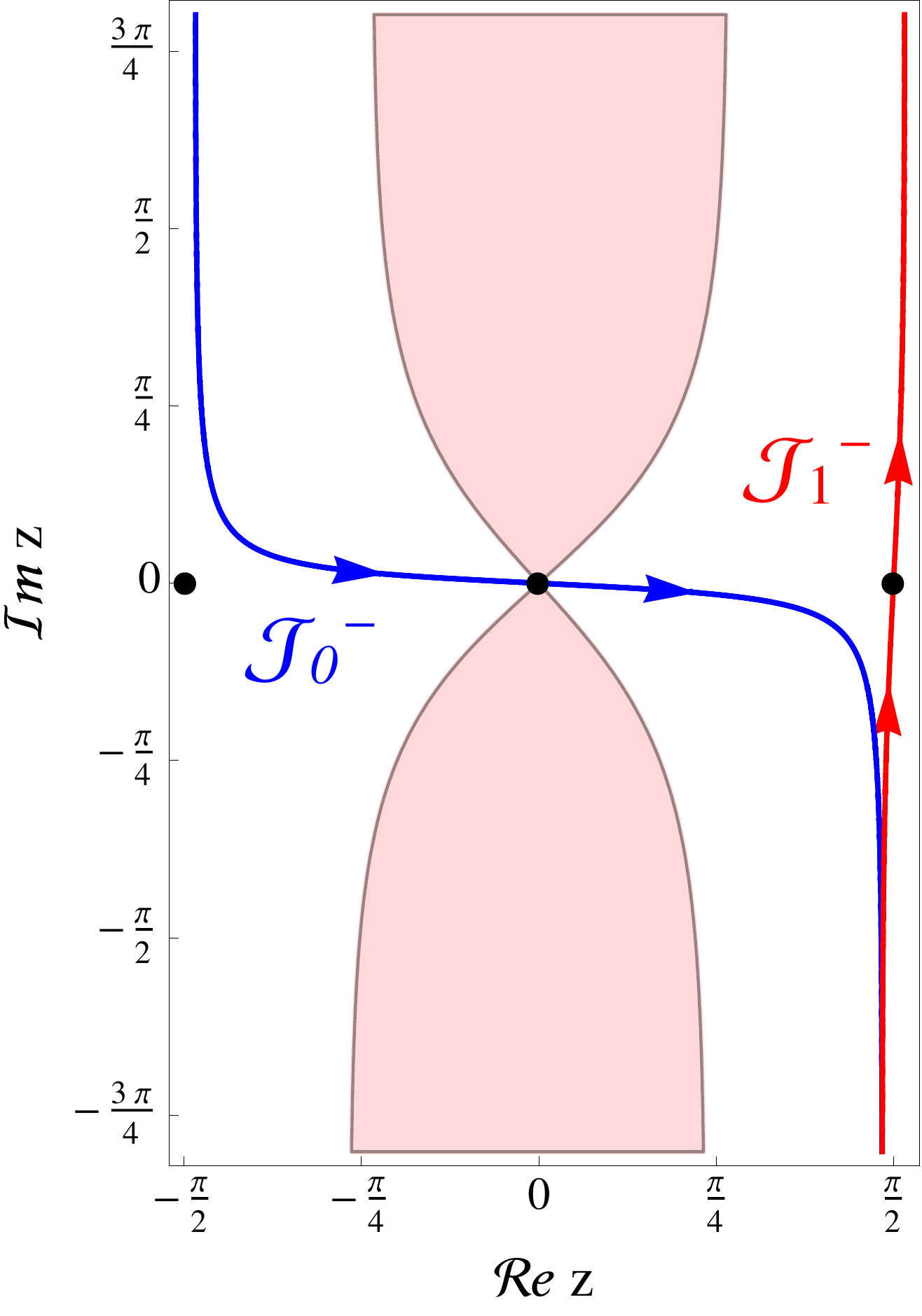}
\hspace{1cm}
\includegraphics[width=0.45\textwidth]{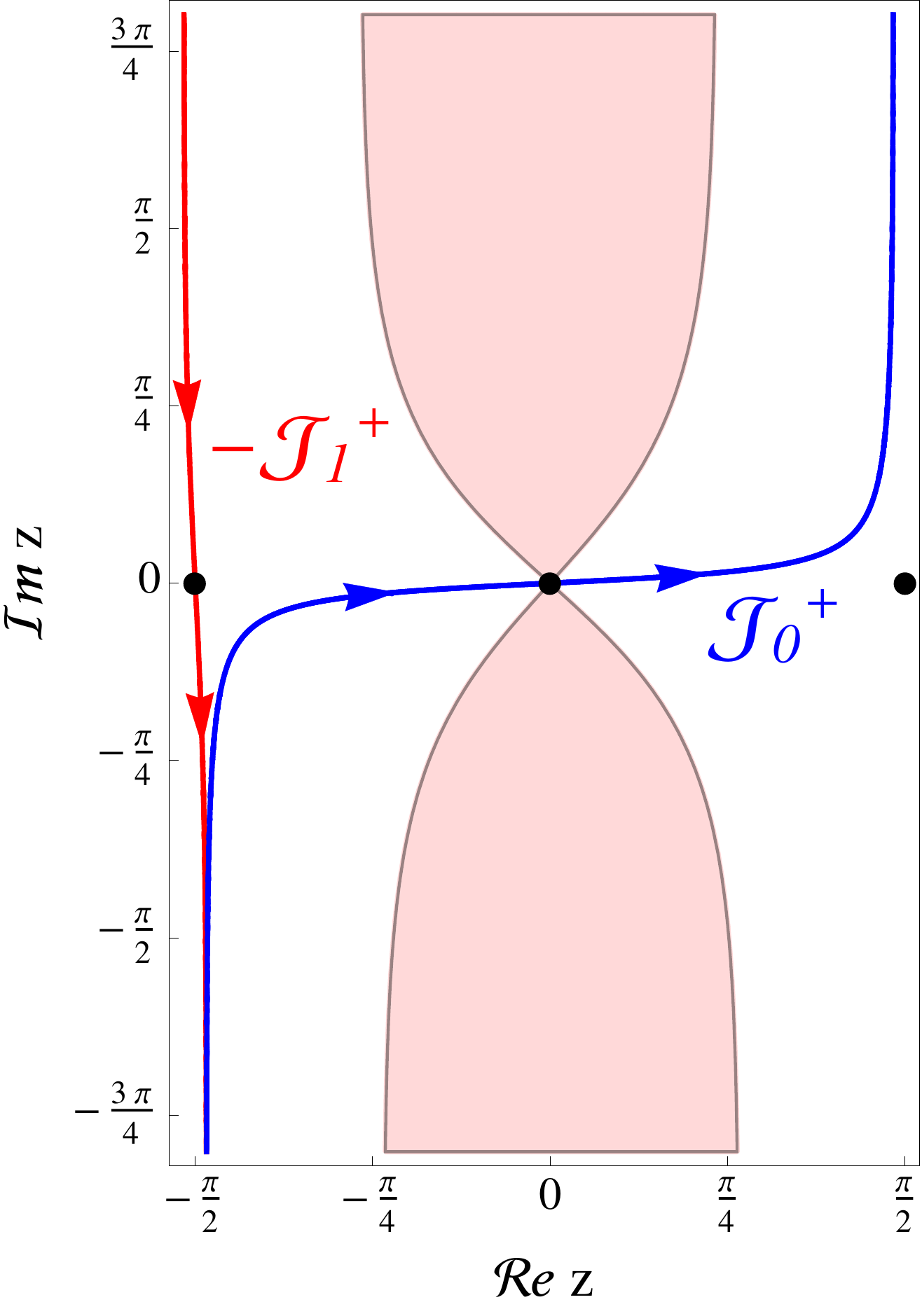}
\caption{{\bf Left:}  Lefschetz thimbles at $\lambda=e^{i\theta}$ with $\theta=0^-$:  $ {\cal J}_0 + {\cal J}_1$. 
{\bf Right:} At $\theta=0^{+}$.   $ {\cal J}_0 - {\cal J}_1. $   We take $\theta=\mp 0.1$  to ease visualization.
   }
\label{fig:thimbles}
\end{figure}

\noindent 
We first  view the action as a meromorphic function $S(z)$. This leads to a more natural description of 
steepest descent method and the semiclassical expansion both in the present zero-dimensional example and in QFT.  It is also the natural way to study the properties of partition functions under analytic continuation. 
 In fact, a judicious analysis of the semi-classical expansion urges us to view {\it all} actions as meromorphic functions of the fields as we will see very explicitly.  
So we now change perspective on the integration cycle $I$ as 
\begin{align}
I \subset \mathbb R \longrightarrow   \Sigma \subset  \mathbb C,  \qquad \Sigma= I \qquad {\rm for\,}   \theta \equiv \arg(\lambda)=0
\end{align}
 where $\Sigma$ has real dimension one for general $\arg(\lambda)$ \footnote{More generally, we generalize $
 I \subset {\mathbb R}^N \longrightarrow   \Sigma \subset  {\mathbb C}^N$, 
where $\Sigma$ has  real dimension $N$.}.   We must now address the question of how the integration cycle in $Z(\lambda)$ changes once $\theta \neq 0$.

There are two non-degenerate critical points, call them $z_0$ and $z_1$, obtained by extremizing  the action 
\begin{align} 
\frac {d S}{dz} = 0 \Longrightarrow  \text {critical points:} \;\; \{z_0, z_1\} = 
\left \{0,  \frac{\pi}{2}  \right \}\,.
\end{align}
We call the first one the P-saddle (perturbative vacuum)  since it has zero action, and call the latter the  NP-saddle since it has a positive action: 
\be
S(z_0)=0, \qquad S(z_1)=\frac{1}{2 \lambda} \qquad S_{10}=S_1 -S_0= \frac{1}{2 \lambda} \,.
\ee
We have also defined the ``relative action" $S_{10}$ (called the ``singulant"  by Dingle\cite{Dingle_asymptotics}), which plays an important role in asymptotic analysis.\footnote{
\label{fn6}
One might naively think that a singulant  is the equivalent of an  instanton  (which is a non-trivial saddle in the path integral formulation) in quantum mechanics or QFT, since both are nontrivial saddle points.  However, in QM, or QFT, there is in general a charge (topological or perhaps emergent, as we will see here) associated with instantons, while the perturbative vacuum is neutral under this charge. Thus, the role that  a singulant plays in the large-order behavior of perturbative series  in ordinary integrals 
 is actually played by instanton-anti-instanton  $[\I \bar \I]$ saddles in QM and/or the weak coupling realization of IR-renormalons in  QFT. So, in passing from $d=0$ examples to $d \geq 1$, the mapping is roughly  ${\rm singulant}   \longleftrightarrow  [\I \bar \I] \;\; {\rm saddle}$. }

Associated with each critical point $z_i$,  there is a  unique integration cycle  ${\cal J}_i$, called a  Lefschetz thimble or a steepest descent path, along which $\Re S$ has a downward gradient flow and the phase $\Im S$ remains stationary.  (See \cite{Witten:2010cx} for a detailed discussion of the construction using Morse theory.)  Indeed, a thimble  $\mathcal{J}_i$ is defined to obey 
\begin{align}
\frac{d z}{dt} = - \overline{\partial_{z} S(z)}
\end{align}
where $t $ is a coordinate along the thimble,  with the initial condition $z(t \to -\infty) = z_i$.  This definition implies that $\Im S$ is stationary, meaning that $\frac{d\, \Im S}{dt} = 0$, and
\begin{align} 
\label{spc}
{\rm Im} S(z)|_{\mathcal{J}_i} =  {\rm Im} S(z_i),  \qquad i=0,1\,.
\end{align}
For each critical point $z_i$ one can also define paths with upward gradient flow $\mathcal{K}_i$, and it can be shown that there is a one-to-one correspondence 
  \be 
  z_i \leftrightarrow  {\cal J}_i ,    {\cal K}_i  
  \ee
between the critical points and  Lefschetz thimbles. 
  The set of the Lefschetz thimbles may be seen as forming a linearly independent  and complete basis of integration cycles for integrals of $e^{-S(z)}$.  
   In general, the contours of integration deform smoothly as $\arg(\lambda)$ is varied, and pass through only the associated saddle. Exactly at the Stokes lines,  these contours also pass through a subset of other saddles.  
    Lefschetz thimbles are the natural geometric surfaces (lines in our example) which can be used to describe the analytic   continuation of $Z(\lambda)$ to complex $\lambda$.

 The Lefschetz thimbles ${\cal J}_i$  are generally unbounded, even when the original integration cycle is bounded, as illustrated in Fig.~\ref{fig:thimbles}. 
  Therefore, we must address the issue of the convergence of the integration over a thimble.  
In doing so, we divide the complex $z$-plane into ``good"  and ``bad"  regions \cite{Witten:2010cx}.
A good region corresponds  to $\Re( S(z) ) >0$ such that $e^{- S(z)} \rightarrow 0$ as $|z| \rightarrow \infty$.  A bad region is the complement,  one in which  $\Re( S(z)) <0$.  
 An admissible contour   for which the integral is finite by construction  is the one which connects two good regions.  
  In Fig.~\ref{fig:thimbles}, the white regions are good and the red regions are bad.   The ${\cal J}_i$ cycles start and end  in the good regions, while the ${\cal K}_i$ cycles, which are not shown in the figure, start and end in the bad regions. 
  
\begin{figure}[tbp]
\includegraphics[width=0.9\textwidth]{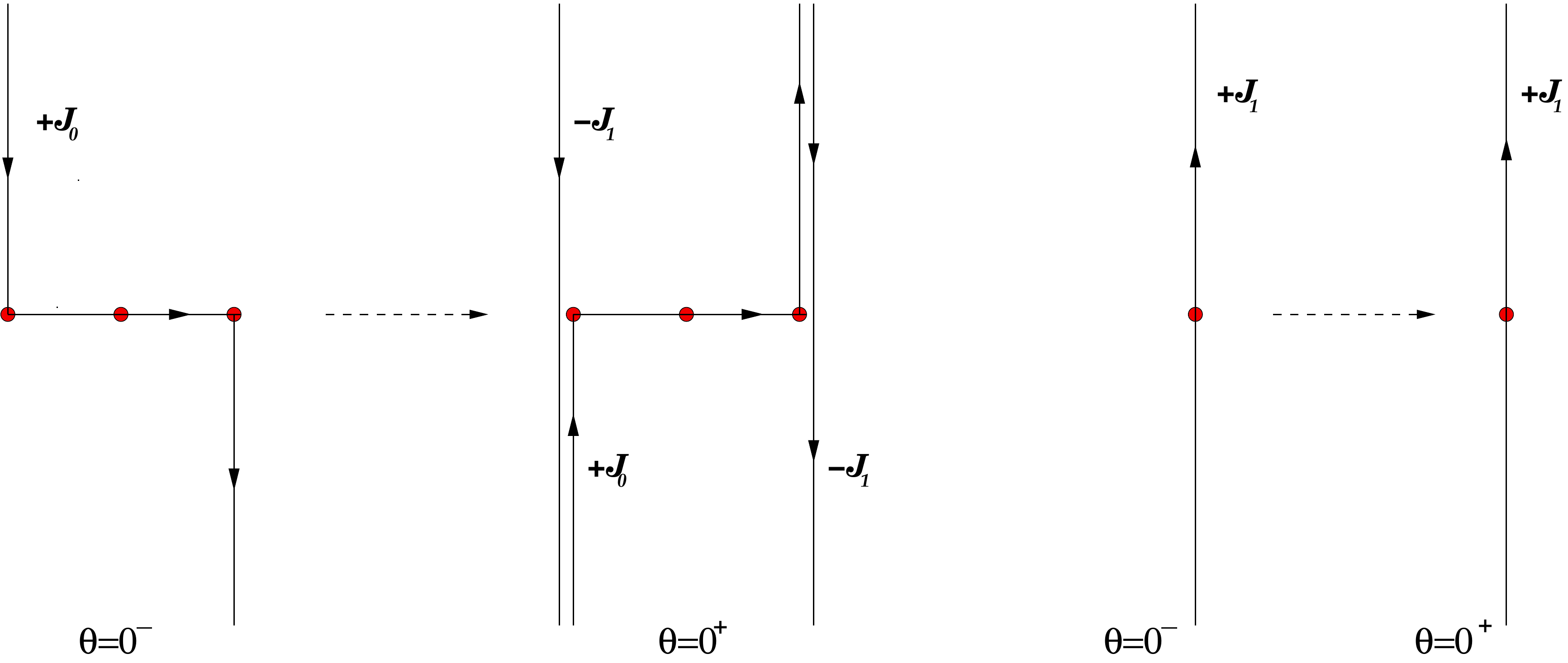}
\hspace{1cm}
\caption{ Stokes phenomenon (wall-crossing) 
at  $\theta=0$:  $ {\cal J}_0  \rightarrow   {\cal J}_0 - 2 {\cal J}_1$, 
while   $ {\cal J}_1 \rightarrow {\cal J}_1 $.   There is also a Stokes line at $\theta=\pi$ where 
 $ {\cal J}_1 $ jumps and    $ {\cal J}_0 $ does not. }
\label{fig:jump2}
\end{figure}

 A general integration cycle  $\Sigma(\theta)$ on which the integral converges  can be written as a sum over the  critical point cycles:
  \be 
  \Sigma(\theta) = \sum_{ i= 0,1}  n_i {\cal J}_i, \qquad   n_i \in \Z,
  \label{eq:CycleDecomposition}
  \ee
The coefficients $n_i$ are piece-wise constant, but have jumps when $\theta$ crosses Stokes lines.  To see an illustration of this, note that our original integration cycle $I=\left[- \frac{\pi}{2}, \frac{\pi}{2} \right]  $ can  actually be written  
  in {\it two} different ways, 
  depending on how one approaches to $\theta = \arg(\lambda)=0$ Stokes line: 
   \begin{align}
   \label{cycle-zero-one}
I = \left[- \frac{\pi}{2}, \frac{\pi}{2} \right]   \longrightarrow \Sigma = 
 \left\{ \begin{array}{l} 
   {\cal J}_0(0^{-})   + {\cal J}_1 (0^{-}) \cr
     {\cal J}_0 (0^{+} )- {\cal J}_1( 0^{+} ) 
\end{array} \right. 
\end{align}
This is illustrated in Fig.~\ref{fig:jump2}.

\begin{figure}[tbp]
\centering
\includegraphics[width=0.9\textwidth]{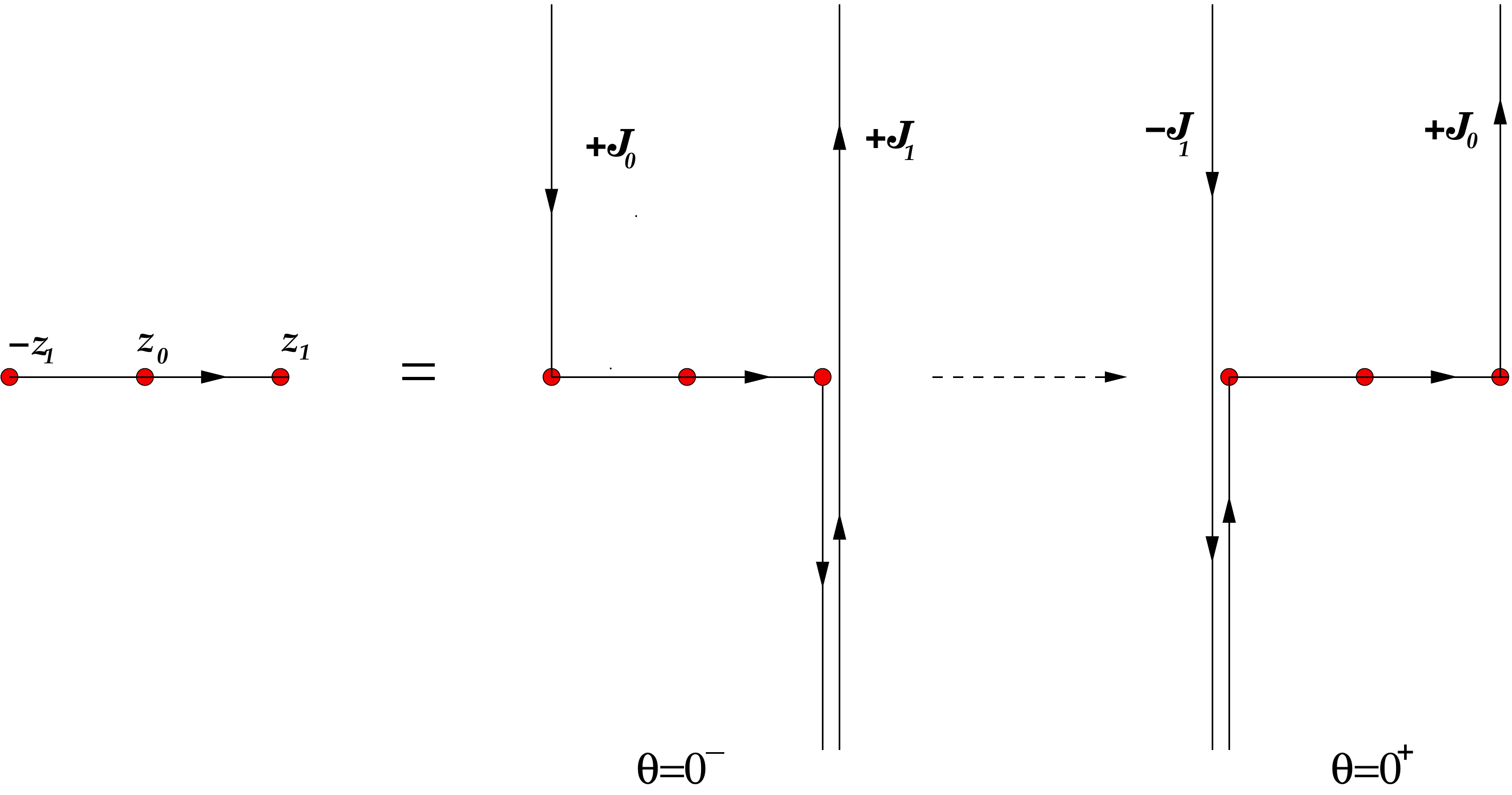}
\caption{The original integration cycle as a linear combination of Lefschetz thimbles at $\theta=0^-$ and   $\theta=0^+$. $\theta=0$ is a Stokes line. 
   }
 \label{fig:thimbles2}
\end{figure}

The relation between the two choices of the integration cycles, the notion of  of cancellation of ambiguities and  BE-summability is discussed in Section~\ref{subsec:cancel}.
Note that despite the fact that $I$ is a finite interval in $\R$,    
the critical point  cycles  ${\cal J}_0$  and  ${\cal J}_1$ are infinite in $\mathbb C$. Moreover, while ${\cal J}_0(0^{-})$ and   $ {\cal J}_0(0^{+})$ coincide on the real  axis, their ``tails"  in the imaginary direction have a relative sign flip.  There is also a 
 flip of the  sign of the coefficient of ${\cal J}_1$ at $\theta=0$  (and  ${\cal J}_0$  at $\theta=\pi$).  These sign flips are realizations of the Stokes phenomenon  \cite{Dingle_asymptotics}.  They are responsible for the cancellation of the imaginary ``tail" contributions to integrals running over either of the cycles in Eq.~\eqref{cycle-zero-one}~\footnote{Recall that $z_1$ coincides with $-z_1$ due to the periodicity of the action.}, so that the value of the integral for $\theta=0^{\pm}$ coincides with its value on the real integration cycle $\Sigma(\theta=0)=I$.  This is an elementary but useful perspective on ambiguity cancellation, which is realized in Fig.~\ref{fig:thimbles2} geometrically. 
 
More generally, note that as $\arg(\lambda)$ changes, the cycles ${\cal J}_i$ deform in a smooth manner, except for Stokes lines, where they undergo jumps.  
The two  Stokes lines are at  $\theta=0$ and $\theta=\pi$, and the respective jumps are:
\begin{align} 
\theta= 0: \qquad  &{\cal J}_1  \longrightarrow {\cal J}_1, \qquad \qquad  \qquad   {\cal J}_0  \longrightarrow {\cal J}_0-2  {\cal J}_1,    \cr 
\theta= \pi: \qquad  &{\cal J}_1  \longrightarrow {\cal J}_1 + 2{\cal J}_0 , \qquad    \;\;\; {\cal J}_0 \longrightarrow {\cal J}_0.
\label{S-jump}  
\end{align}
To see where these relations come from, note for example that when $\theta=0^-$, ${\cal J}_0$ goes from one good domain to another good domain by passing through saddle $z_0$ and being arbitrarily close to saddle-$z_1$.  However, at  $\theta=0^+$,  there is no  single path which can achieve this. To start and end in the same places when  $\theta=0^+$ as when  $\theta=0^-$, one first needs to take the 
$-{\cal J}_1$ thimble, then the ${\cal J}_0$ thimble, then again go along $-{\cal J}_1$. So $ {\cal J}_0  \longrightarrow {\cal J}_0-2  {\cal J}_1$, as is illustrated in Fig.~\ref{fig:jump2}.  
Similar jumps in contours can be found in monodromy problems associated with certain Picard-Fuchs equations \cite{Gulden:2013wya}.

In summary, the expressions for the analytic continuation of the integration cycle from $\arg(\lambda)=0$ to arbitrary $\arg(\lambda)$ must take into account the Stokes phenomena Eq.~\eqref{S-jump} and it depends on the Stokes chamber:  \begin{align}
\label{lt-0}
\Sigma (\theta)  = \left\{ \begin{array}{lc} 
   +{\cal J}_0(\theta) + {\cal J}_1 (\theta)    \qquad& -\pi <  \theta< 0 \,,\cr
     +{\cal J}_0(\theta) - {\cal J}_1 (\theta)  \qquad  & 0 < \theta < \pi  \,,\cr
   - {\cal J}_0(\theta) - {\cal J}_1 (\theta)  \qquad  & \pi< \theta < 2\pi\,.  
\end{array}
 \right. 
\end{align}
This illustrates the reason that Lefschetz  thimbles provide the natural basis for semi-classical expansions and the analytic continuation of Feynman path integral.


{\bf Perturbative expansions around P and NP saddles:} So far we have made no approximations in our analysis.  If one could evaluate the integrals along the cycles Eq.~\eqref{lt-0} exactly, one would obtain an exact result for $Z(\lambda)$.  Usually, however, this is not possible, and the best one can do is evaluate the integrals using perturbative series.   Hence we now find the perturbative expansion around each of the two saddles.
The formal asymptotic expansion around the $z_0$  P-saddle is given by 
\begin{align} 
\frak{Z}_{0} (\lambda) & \equiv   
e^{-\frac{S_0}{\lambda}}  \Phi_{0}(\lambda) = 
 \sum_{n=0}^\infty a_n \lambda^n \equiv
  \sqrt{2\pi}  \sum_{n=0}^\infty  \frac{\Gamma\left(n+\frac{1}{2}\right)^2}{n!\, \Gamma\left(\frac{1}{2}\right)^2}  (2 \lambda)^n  \,.
\label{0-series}
\end{align}
This is a non-alternating asymptotic series for $\th=0$. The late terms grow  as $a_n \sim \frac{n!}{(S_{10})^n}$. The series is non-Borel-summable in the $\th=0$ direction, but it is Borel-summable in the $\th = \pi$ direction.  This formal series is a perturbative representation of the contribution of the integral along the $\mathcal{J}_0$ thimble.

The perturbative expansion  around the NP-saddle (including the NP-factor) is given by 
\begin{align} 
\frak{Z}_{1} (\lambda) &=  e^{- \frac{S_1}{\lambda}}  \Phi_{1}(\lambda) =  
  \sum_{n=0}^\infty   (-1)^n a_n \lambda^n  \equiv  e^{- \frac{1}{2 \lambda}}
 \Phi_{1}(\lambda)\,. 
\label{1-series}
\end{align}
This series is an alternating asymptotic series for  $\th=0$. The late terms grow  as $(-1)^n \frac{n!}{(S_{10})^n}$, and the series is Borel-summable in the $\th=0$ direction. 
On the other hand, it is not Borel-summable in the $\theta=\pi$ direction. This formal series is a perturbative representation of the contribution of the integral along the $\mathcal{J}_1$ thimble.


The semiclassical expansion for $Z(\lambda)$ can be written as a two term trans-series
\begin{align}
   \frak{Z}( \lambda, \s_0, \s_1) =  \s_0 \frak{Z}_{0} (\lambda) + \s_1 \frak{Z}_{1} (\lambda) =  \s_0   \Phi_{0}(\lambda)   + \s_1  e^{- \frac{1}{2 \lambda}}
 \Phi_{1}(\lambda)\,,
 \label{tansseries}
\end{align}
where $\s_i$ are trans-series parameters.  Here the $\s_i$ parameters have exactly the same role as the $n_i$ coefficients of the Lefshetz thimbles in Eq.~\eqref{eq:CycleDecomposition}.  The trans-series is an algebraic representation of the geometric information in Eq.~\eqref{eq:CycleDecomposition}.  The analytic continuation of our original integral to complex $\lambda$ involves different linear combinations of $\frak{Z}_{0} (\lambda)$ and $\frak{Z}_{1} (\lambda)$ in different Stokes wedges.  The value of trans-series representation $\frak{Z}( \lambda, \s_0, \s_1)$ of $Z(\lambda)$ is that because it is an algebraic representation, it is well-suited for direct calculations using standard perturbative methods, while the value of the geometric Lefshetz thimble representation of $Z(\lambda)$ is its ease of visualization.

\subsection{Borel analysis and Stokes Phenomena} 
\label{sec:Borel-Stokes}

The  Borel transforms of the formal series in Eq.~\eqref{0-series} and Eq.~\eqref{1-series} are  given by 
\begin{align}
& \hat \Phi_{0}(t)  \equiv  B[ \Phi_{0}](t) =  \sum_{n=0}^{\infty} a_n \frac{t^n}{n!}  =   \sqrt{2\pi}\sum_{n=0}^{\infty}  \frac{ \Gamma\left(n+\frac{1}{2}\right)^2 }{\Gamma(n+1) 
 \Gamma\left(\frac{1}{2}\right)^2   
   }    \frac{(2t)^n}{n!} \,,  \\
   & \hat \Phi_{1}(t)  \equiv  B[ \Phi_{1}](t) =  \sum_{n=0}^{\infty} (-1)^n a_n \frac{t^n}{n!}  \,,
\end{align}
and since the coefficients $a_n$ grow like $n!$ both $\hat \Phi_{0}(t)$ and $ \hat \Phi_{1}(t)$ define two germs of analytic functions at $t=0$. 
In this simple example we have the luxury of having closed-form expressions for all of the terms of the Borel transforms of the perturbative series, and indeed there are also closed-form expressions for the analytic continuations of these germs into ${\mathbb C}_{s}$ in terms of hypergeometric functions:
\begin{align}
B[ \Phi_{0}](t) = \sqrt{2\pi} ~_2F_1\left(\frac{1}{2},\frac{1}{2}, 1; 2t \right) \;,  \qquad 
 B[ \Phi_{1}](t) =\sqrt{2\pi} ~_2F_1\left(\frac{1}{2},\frac{1}{2}, 1;  -2t\right) 
\end{align}
with singularity (branch point)  at  $t=1/2$  for $B[ \Phi_{0}](t)$ and   at  $t=-1/2$  for $B[ \Phi_{1}](t)$. In more complicated examples, one might only have closed-form expressions for the low-order and high-order terms in perturbative series.  However, this is still enough to determine the positions of the singularities of the Borel transform, since they are determined by the asymptotic behavior of the high-order terms of the perturbative series.

The sectorial (directional) Borel resummations for $ \Phi_i, \; i=0,1$ are given by 
\begin{align}
{\cal S}_\th  \Phi_{i} (\lambda) =& 
\frac{1}{\lambda}  \int_0^{  e^{i \th}\infty}  dt\,e^{- t/\lambda}    \; B[ \Phi_{i}](t)   \,.
 \label{lateral}
\end{align}
These are well-defined  holomorphic functions of $\lambda$ in $\Re( e^{i \th} / \lambda) >0$, where $\theta$ is now parametrizing a generic direction in the complex $t$-plane.  For real coupling, $\arg \lambda = 0$, we cannot directly work with $\mathcal{S}_0$ due to the presence along the line of integration of a singularity of the Borel transform $B[ \Phi_{0}]({t}) $. This is associated with the Stokes phenomenon in the complex $\lambda$- plane: the series Eq.~\eqref{0-series} becomes non-alternating when $\theta = 0$ and hence $\Phi_{0}(\lambda)$ is not Borel summable  when $\th= 0$. 

$\Phi_{0}(\lambda)$ is, however,   {\it right and left  Borel-summable}.
For the right summation  one integrates along a contour which avoids the singularity in such a way that the singularity remains to the right of the contour ($\theta= 0^+$ in Eq.~\eqref{lateral}),  and analogously, for left summation, the singularity stays on the left of the contour ($\theta= 0^-$ in Eq.~\eqref{lateral}).

\begin{figure}[tbp]
\centering
\includegraphics[width=1\textwidth]{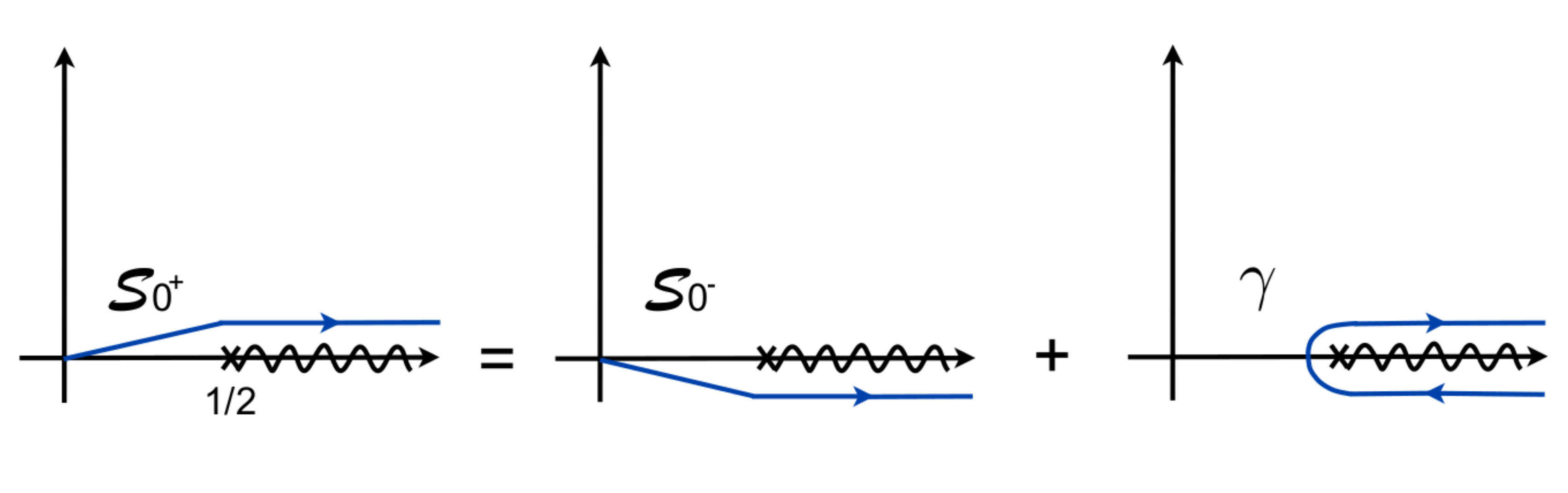}
\caption{The right Borel resummation can be rewritten as the sum of the left Borel resummation plus the contribution coming from the Hankel contour $\gamma$, coming from $t\to-\infty$, circling around the branch cut starting at $t=1/2$ and going back to $+\infty$.}
\label{fig:Hankel}
\end{figure}

As illustrated in Fig.~\ref{fig:Hankel}, the difference of the right and left Borel resummation  can be written as an integral over the Hankel contour $\gamma$ which starts at $\infty$ below the imaginary axis, then circles  the singular point at $t=1/2$, and then goes back to $\infty$ above the imaginary axis:
\begin{align}
({\cal S}_{0^+}   - {\cal S}_{ 0^-} ) \Phi_{0} (\lambda) & = 
 \frac{\sqrt{2\pi}}{\lambda} \int_\gamma dt\,
  e^{- t/\lambda }   ~_2F_1\left(\frac{1}{2},\frac{1}{2}, 1 ;  2t \right)  \cr
& = \frac{\sqrt{2\pi}}{\lambda}   \int_{1/2}^\infty dt\,e^{-t/\lambda }\left[~_2F_1\left(\frac{1}{2},\frac{1}{2}, 1, 2t +i\varepsilon\right)-~_2F_1\left(\frac{1}{2},\frac{1}{2}, 1, 2t -i\varepsilon \right)\right] \cr
&= \frac{\sqrt{2\pi}}{\lambda}   \int_{1/2}^\infty dt \, e^{-  t/\lambda}\, 2 i  ~_2F_1\left(\frac{1}{2},\frac{1}{2}, 1, 1-2t \right) \cr
&= 2i\sqrt{2\pi} e^{-1/(2 \lambda) }   \frac{1}{\lambda}   \int_{0}^\infty dt \, e^{-  t/\lambda}\,   ~_2F_1\left(\frac{1}{2},\frac{1}{2}, 1, -2t \right) \cr
&= 2i e^{-1/(2 \lambda) }  {\cal S}_{0}  \Phi_{1} (\lambda) \,.
\label{disc3} 
\end{align}
To obtain third line in  Eq.~\eqref{disc3} we used the known discontinuity property of the hypergeometric function \cite{NIST}:
\begin{eqnarray}
~_2F_1\left(a, b,c \Big|  t+i\varepsilon\right)-~_2F_1\left(a,b,c \Big| t-i\varepsilon\right)=  \frac{2 \pi i\,\Gamma(c)}{\Gamma(a) \Gamma(b)} ~_2F_1\left(c-a, c-b,1 \Big| 1-t\right)\,, \qquad
\label{hyper-disc}
\end{eqnarray}
valid for $a+b=c$, and the last line is the (unambiguous) Borel resummation of the $ \Phi_{1} (\lambda)$ series in the 
$\theta=0$ direction.  If we only had access to the asymptotic expressions for the high-order behavior of the coefficients of the perturbative series, in the last line we would have obtained a perturbative series expression for $ {\cal S}_{0}  \Phi_{1} (\lambda)$, rather than a closed-form expression for $ {\cal S}_{0}  \Phi_{1} (\lambda) $ itself.  The factor of $2i$ on the right-hand-side of Eq.~\eqref{Stokes} is called a  Stokes constant (or ``analytic invariant" in \'Ecalle's terminology): 
\begin{align}
\label{s-c}
{\rm s} = 2i\,.
\end{align}
Stokes constants are non-vanishing only at singular points on Stokes line and are zero otherwise. 

In fact,  we can obtain the same result by changing the integration variable from the ``field variable" $z$ to the ``action variable" 
$u = S(z)$.  This puts the integral into the form of a Borel sum\footnote{Note that in this case the Borel transform one sees inside the Borel-sum expression are associated with sending $\sum a_n \lambda^n \to \sum_n \frac{a_n}{(n+1/2)!}t^n$, rather than $\sum a_n \lambda^n \to \sum_n \frac{a_n}{n!}t^n$ as we had written above.  This highlights the point that the Borel transform of a given power series is not really unique due to the freedom to divide $a_n$ by $(n+\alpha)!$ for arbitrary fixed $\alpha$ when defining the transform.  However, there is also a corresponding freedom, depending on $\alpha$, in the definition of the Borel sum.  One can easily show that the value of the Borel sum is independent of $\alpha$. }.   Then the integrals over the ${\cal J}_0 (0^{\pm})$ cycles are given by
  \begin{align}
 \frac{1}{\sqrt \lambda} \int_{{\cal J}_0 (0^\mp) } e^{-\frac{1}{2 \lambda} \sin^2(z)}  
  &= \frac{2}{\sqrt \lambda} \int_0^{1/2} du \frac{e^{-u/\lambda}}{\sqrt {2 u (1-2u)}}  
 \mp \frac{2i}{\sqrt \lambda} \int_{1/2}^{\infty}  du \frac{e^{-u/\lambda}}{\sqrt {2 u (2u-1)}} \cr
&= \frac{2}{\sqrt \lambda} \int_0^{1/2} du \frac{e^{-u/\lambda}}{\sqrt {2 u (1-2u)}}  
 \mp i e^{- \frac{1}{2 \lambda} } \frac{2}{\sqrt \lambda} \int_{0}^{\infty} du \frac{e^{-u/\lambda}}{\sqrt {(2 u+1) 2u}}   \cr 
 &=  {\rm Re} {\cal S}_{0}    {\Phi}_0   \mp i e^{- \frac{1}{2 \lambda} } {\cal S}_{0}    {\Phi}_1 
 \end{align}
where  ${\rm Re} {\cal S}_{0}    {\Phi}_0 $ is unambiguous.\footnote{The integral  that we identify with 
${\rm Re} {\cal S}_{0}    {\Phi}_0 $ is dominated by $ u \lesssim \lambda$ in the small $\lambda$ regime.  The procedure to obtain the perturbative expansion  ${\Phi}_0$ from this expression involves two steps.  First, one should extend the integration domain to $[0, \infty)$. Next, one Taylor expands 
$\frac{1}{\sqrt {(1-2u)}}  $ around the origin. Performing the integral will give the divergent asymptotic expansion  ${\Phi}_0$. The reason for the divergence is the use of the Taylor expansion beyond its radius of convergence. One can also obtain the (convergent)  strong coupling expansion from the integral representation in the  $\lambda \gg 1$ regime, by expanding the exponential into a power series in $\frac{1}{\lambda}$ and performing order by order integration.}
 More importantly,  in the integral over the 
Lefschetz thimble associated with the P-saddle, ${{\cal J}_0 (0^\mp) }$, we immediately see the imprint of the NP physics! Moreover, we see all of the data associated with the NP saddle, both its NP weight and the perturbative fluctuations around it: it is encoded in the imaginary part of the integral along ${{\cal J}_0 (0^\mp) }$. 

Clearly, the imaginary part is 
ambiguous for $\theta$ exactly zero, since the result depends on how one approaches the Stokes line. 
This is a reflection of the non-Borel-summability of the perturbative series.  In the geometric perspective we are following here, the flip in the imaginary part of the integral over ${{\cal J}_0 (0^\mp) }$ is due to the flip of the infinite ``tail" of the integration cycle that takes place when crossing the Stokes line $\th=0$. See Fig.~\ref{fig:thimbles2} and Fig.~\ref{fig:jump2}.
 This is the geometric realization of non-Borel-summability.  Happily, 
this is not the whole story, because the integration over the interval $I$ is actually the linear combination of thimbles as seen in Eq.~\eqref{cycle-zero-one}.  
We will come back to the story of how including the contribution from $\mathcal{J}_1$ cures this problem in Section \ref{subsec:cancel}. First, however, it is useful to understand how this happens from the algebraic trans-series point of view.

\subsection{Stokes automorphism and alien derivative}
To understand how ambiguity cancellation works in the trans-series representation, it is useful to introduce the notions of Stokes automorphisms and alien derivatives from resurgence theory.  To keep the presentation more streamlined, Appendix~\ref{sec:ResurgenceAppendix} summarizes some of the results and definitions that we are going to use in what follows. 

Distinct  sectorial  solutions  on two different sides of a Stokes line are ``connected" through the   Stokes automorphism, $\underline{\frak{S}}_\theta$:
\begin{align}
{\cal S}_{\theta^+} = {\cal S}_{\theta^-} \circ \underline{\frak{S}}_\theta \equiv {\cal S}_{\theta^-} \circ \left( \mathbf{1} - {\rm Disc}_{\theta^-} \right),
\end{align}
where  ${\rm Disc}_{\theta^-}$ denotes  is the discontinuity arising on crossing the Stokes line, thus, 
\begin{align} 
\label{boreldisc}
{\cal S}_{\theta^+} - {\cal S}_{\theta^-} = - {\cal S}_{\theta^-} \circ {\rm Disc}_{\theta^-}.
\end{align}
In our example Eq.~\eqref{disc3},  the difference of the right and left summations in the $\theta=0$-direction of $\Phi_0 (\lambda) $  is an exponentially small imaginary term given by 
\begin{align}
 & {\cal S}_{0^{+}}  \Phi_{0} (\lambda)  - {\cal S}_{0^{-}}  \Phi_{0} (\lambda)  =     
2i e^{-1/(2 \lambda) }  {\cal S}_{0}  \Phi_{1} (\lambda) = -
{\cal S}_{0} \circ {\rm Disc}_{0}  \Phi_{0} (\lambda) \,,  \cr
&\Rightarrow 
{\rm Disc}_{0}  \Phi_{0} (\lambda)  = - 2i e^{-1/(2 \lambda) }  \Phi_{1} (\lambda)\,.
\label{disc2} 
\end{align}
The Stokes automorphisms  connecting different sectorial Borel sums are non-trivial only at Stokes lines. The Stokes lines are $\theta=0$ for $\Phi_{0}$ and  $\theta=\pi$ for $\Phi_{1}$, and the resulting 
 Stokes automorphisms  are
\begin{eqnarray}
\underline {\frak S}_0  \Phi_{0} (\lambda) & =  \Phi_{0} (\lambda) + 2 i e^{-{1 \over 2 \lambda}}  \Phi_{1} (\lambda),  \quad \quad   \underline {\frak S}_0  \Phi_{1} (\lambda) & =  \Phi_{1}(\lambda)\,,
  \cr
\underline {\frak S}_\pi  \Phi_{1} (\lambda) & =  \Phi_{1} (\lambda) + 2 i e^{+{1 \over 2 \lambda}}  \Phi_{0} (\lambda), \quad  \quad \underline {\frak S}_\pi  \Phi_{0} (\lambda) & =  \Phi_{0} (\lambda)\,.
\label{Stokes} 
\end{eqnarray}
These equations encode the beautiful structure of resurgence:
the P-saddle carries complete information about the NP saddle which is decoded using its Stokes automorphism in the $\theta=0$ direction.  At the same time,  the NP-saddle carries complete information about the P-saddle, which is decoded by its own Stokes automorphism, but in the $\theta=\pi$ direction.

The set of all formal series appearing in our problem form a closed algebra under the action of the singularity derivative, also called alien derivative, which acts as 
\begin{align}  
&\Delta_{+{1 \over 2}}  \Phi_{0}  = 2 i  \Phi_{1}\,,  \qquad  \Delta_{+{1 \over 2}}  \Phi_{1}  = 0\,, \cr
&  \Delta_{-{1 \over 2}}  \Phi_{1}  =  2 i  \Phi_{0}\,, \qquad 
\Delta_{-{1 \over 2}}  \Phi_{0}  =0\,.
\end{align} 
The first one of these relations means that the action of the singularity derivative at $ \omega=+{1 \over 2}$ on $\Phi_0$ is just $\Phi_1$ times the Stokes constant Eq.~\eqref{s-c}.   The action of the singularity derivative 
at any other point on  $\Phi_0$ just gives zero, because  the Borel transform of $\Phi_0$  does not have any other singularities, i.e.,  $\Delta_{\omega}  \Phi_0=0 $ for $\omega \neq {1 \over 2}$.


  \subsection{Reality of resurgent trans-series for real $\lambda$ and BE-summability}
  \label{summability}
   As stated earlier,  for $\arg(\lambda) =0$, the partition function defined in Eq.~\eqref{part}  is manifestly real. On the other hand, 
the formal  first sum   $ \Phi_0 (\lambda)$ in   Eq.~\eqref{a-c}    is non-Borel summable in the singular direction $\theta=0$, and hence it has an ambiguity, which is of order $ i e^{ - \frac{1}{ 2 \lambda}} $.  
On the other hand, for $\Phi_1$, the singular direction is $\theta=\pi$, and hence it can be Borel resummed in the $\theta=0$ direction.  

For example, for $\theta=0^{-}$ (approaching the real line either from  below), 
   \begin{align}
    \Phi_0 (\lambda)  +   i  e^{ - \frac{1}{2\lambda}}  {\Phi}_1 (\lambda)   \xrightarrow{ \rm BE-summation}    & {\cal S}_{0^-}    {\Phi}_0   +  i e^{ - \frac{1}{2\lambda  }}    {\cal S}_{0^{-}}  {\Phi}_1 \cr 
   = &\left( {\rm Re} {\cal S}_{0}    {\Phi}_0  + i   \, {\rm Im} {\cal S}_{0^{-}}    {\Phi}_0  \right)  + i  e^{ - \frac{1}{ 2\lambda }}    {\cal S}_{0}  {\Phi}_1  \cr   
  = & {\rm Re} {\cal S}_{0}    {\Phi}_0  + i   \left(  {\rm Im} {\cal S}_{0^{-}}    {\Phi}_0    +   e^{ - \frac{1}{ 2\lambda }}    {\cal S}_{0}  {\Phi}_1 \right)  \cr
     = & {\rm Re} {\cal S}_{0}    {\Phi}_0
\label{a-c-2}
\end{align}
and we have similar cancellation for $\theta=0^{+}$, approaching the real line from  above. 
The non-Borel summability of  the perturbative expansion  $ \Phi_0$ leads to two-fold  purely imaginary ambiguity. But exactly at the Stokes line, the integration path is also two-fold ambiguous, 
  ${\cal J}_0 \pm {\cal J}_1 $, for $\theta=0^{\mp}$.  This maps to a two-fold ambiguity of the coefficient of the  NP-term in the trans-series.  The observable $Z(\lambda)$ is the combination of the two contributions, and the ambiguities cancel in the appropriate combinations in any of the Stokes chambers, leading 
 to the real (physical) result on positive real axis in coupling constant plane.
   This is an example  of  median resummation and Borel-\'Ecalle summability.  \footnote{In the QM and QFT examples, the cancellation mechanism of ambiguities is essentially the same, but 
   unlike the 0d example where the NP-term completely disappears, there are  (infinitely many)  
   real  non-perturbative contributions that also contribute to observables along with the perturbative contribution.}

Consequently, the trans-series expansions for the analytic continuation of the partition function Eq.~\eqref{part} in different Stokes chambers are given by
\begin{align}
  Z (\lambda)=  \left\{ \begin{array}{lc}  
   \frak{Z}_{0} (\lambda) +  i \frak{Z}_{1} (\lambda)  = 
   \Phi_0(\lambda)  +  i  e^{ - \frac{1}{ 2 \lambda}}  {\Phi}_1 (\lambda)     \qquad& -\pi <  \theta< 0\,, \cr
     \frak{Z}_{0} (\lambda) -  i \frak{Z}_{1} (\lambda)  =    \Phi_0(\lambda)  -  i  e^{ - \frac{1}{ 2 \lambda}}  {\Phi}_1 (\lambda)     \qquad  & 0 < \theta < \pi  \,,
\end{array} \right.
\label{a-c}
\end{align}
with a Stokes jump at   $\arg(\lambda)=0$.   
Approaching the $\lambda \in \mathbb{R}^+$ line from above or below, we observe that the real solution for a trans-series (\ref{tansseries}) is given by 
\begin{align}
\frak{Z}_{\mathbb R} (\lambda, 1, 0) =  {\cal S}_{0^{-} }\frak{Z}\left( z, 1,   + \frac{1}{2} {\rm s} \right) =
 {\cal S}_{0^{+} }\frak{Z}\left( z, 1,  -\frac{1}{2} {\rm s}  \right)\,,
 \end{align}
 where $s=2i$ is once again the Stokes constant.
This is a very simple example of the fact that cancellation of nonperturbative ambiguities leads to median resummation of trans-series\cite{Marino:2008ya}.  This also seems to be valid for non-linear systems with infinitely many Borel plane singularities \cite{Aniceto:2013fka}. We comment on the generalization of this formula to QFT in Sec.~\ref{sec:Resurgence}.

 \subsection{Lefschetz thimbles and  geometrization  of ambiguity cancellation}
 \label{subsec:cancel}
We now return to the picture offered by the Lefshetz thimble decomposition of the integration cycle to see the geometrization of ambiguity cancellation on the Stokes line  $\arg(\lambda)= 0$ which we have already seen in the algebraic trans-series representation Eq.~\eqref{a-c-2}.  The integral over the P-thimble ${\cal J}_0$ at $\theta=0^{-} $ and 
  at $\theta=0^{+} $  can be written schematically as 
 \begin{align}
 \int_{{\cal J}_0 (0^-) } = \int_{- {\pi \over 2 } + i 
 \infty}^{- {\pi \over 2 } }  +    \int_{- {\pi \over 2 } } ^{+{\pi \over 2 } } + \int_{+ {\pi \over 2 } } ^{+{\pi \over 2 }  -i\infty}  \underbrace{\longrightarrow }_{\footnotesize{  \textsf {  {\rm periodicity 
} }}} \int_{- {\pi \over 2 } } ^{+{\pi \over 2 } }  + \int_{+ {\pi \over 2 } + i \infty } ^{+{\pi \over 2 }  -i\infty} =
  Z  -  \int_{{\cal J}_1 (0) } \qquad  \cr
 \int_{{\cal J}_0 (0^+) } = \int_{- {\pi \over 2 } - i 
 \infty}^{- {\pi \over 2 } }  +    \int_{- {\pi \over 2 } } ^{+{\pi \over 2 } } + \int_{+ {\pi \over 2 } } ^{+{\pi \over 2 }  +i\infty}  \underbrace{\longrightarrow }_{\footnotesize{  \textsf {  {\rm periodicity 
} }}} \int_{- {\pi \over 2 } } ^{+{\pi \over 2 } }  + \int_{+ {\pi \over 2 } - i \infty } ^{+{\pi \over 2 }  +i\infty} =
  Z   +  \int_{{\cal J}_1 (0) } \qquad    
   \end{align}
where in the second  step, we cut   the segment  
 $[- {\pi \over 2 } + i \infty, {- {\pi \over 2 } }]$  and glued it   to $[ {\pi \over 2 } + i \infty, { {\pi \over 2 } }]$.
  Because of  the  $\pi$ periodicity of the integrand, the integral remains  unchanged. 
  $ Z$ is the original real valued  partition function Eq.~\eqref{part}, and  $\int_{{\cal J}_1 (0) } $ is purely imaginary and is of order $e^{ - \frac{1}{2\lambda  }}$. This formula makes it manifest that the integral over the P-thimble $\mathcal{J}_0$ has an imaginary part and is \emph{not} equivalent to the original partition function $ Z$.


For $\th=0^-$ the addition of  $+ \int_{{\cal J}_1 (0) } $   kills the (undesired) imaginary part of the $\mathcal{J}_0$ integral, and 
the combination $\mathcal{J}_0 (0^-)   +    \mathcal{J}_1 (0^{-}) $ is the linear combination of thimbles associated with   $Z$  at   $\theta=0^{-}$, namely
 \begin{align}
\frac{1}{\sqrt{\lambda} } \int_{{\cal J}_0 (0^-) + {\cal J}_1 (0^-) } e^{-\frac{1}{2\lambda}\sin^2 (z) }=    Z  \,.
   \end{align}
   Upon the Stokes jump at $\theta= 0$  ${\cal J}_1  \longrightarrow {\cal J}_1, \;  {\cal J}_0  \longrightarrow {\cal J}_0-2  {\cal J}_1$    given in  Eq.~\eqref{S-jump}, we obtain 
    \begin{align}
  \frac{1}{\sqrt{\lambda} } \int_{{\cal J}_0 (0^+) - {\cal J}_1 (0^+) }    e^{-\frac{1}{2\lambda}\sin^2 (z) }    = Z  \,,
   \end{align}
  where the combination of ${\cal J}_0 (0^+), {\cal J}_1 (0^+)$ is simply the unique linear combination which cancels the imaginary part exactly at 
    $\theta=0^{+}$.   It is also  important to note that except for $\arg(\lambda)=0$, (and  $\arg(\lambda)=\pi$), there is never an exact cancellation between the contribution of the two saddles, and both saddles contribute! In fact, at the anti-Stokes lines $\theta=\pm {\pi \over 2}$, the modulus of the two contributions is the same, and there is an exchange of dominance.

 Thus, the ambiguity in the imaginary part of the integration  $\int_{{\cal J}_0 (0^\mp) }$ is cancelled exactly by the  ambiguity in the prefactor of the $  \int_{{\cal J}_1 (0) } $ integral. 
   This is a simple geometric realization of the cancellation of ambiguities on the Stokes line.  Stated another way, one can observe that on approaching the Stokes line from above $\theta=0^{+}$ or   from   below $ \theta=0^{-}$, the ``amplitude" associated with the NP-saddle $[z_1]_{\theta=0^{\pm}}$ is given by a two-fold ambiguous result:
\begin{align} 
[\rho_1]_{\theta=0^{\pm}} =  \pm i e^{ - \frac{1}{2\lambda  }}    {\cal S}_{0}  {\Phi}_1 \,.
\label{two-fold-1}
\end{align}
 This is the counter-part of the ambiguous structure of instanton-anti-instanton-type amplitudes in QM and QFT examples, where the associated amplitude  (which may naively be expected to be real) actually possess an unambiguous real part and a  two-fold ambiguous imaginary part:
 \begin{align} 
[\I \bar \I]_{\theta=0^{\pm}}  \sim  e^{ - 2S_I  }    \pm i  \pi e^{ - 2S_I  }  \,.
\label{two-fold-2}
\end{align}
This is taking place in QFT for the same reason as in ordinary integration.  Of course, in semi-classically calculable regimes of QFTs and in QM,  there are infinitely many saddle points and Lefschetz thimbles. The thimbles are infinite dimensional algebraic varieties and when the theory is regularized on a finite lattice with a finite size, they become finite dimensional  algebraic varieties.
 The saddle points are the perturbative vacuum and also instantons, bi-instantons, topologically neutral instanton molecules, etc, with appropriate terminological modifications in theories without the topology to support instantons, as we explore in what follows. Let 
  $\rho_0$  denote the perturbative vacuum and   $\rho_n$ denote various NP-sectors,  with action $S_n$, 
  that can communicate with the P-sector according to the structure of  graded  resurgence triangle explained in e.g. Sec~\ref{sec:Resurgence}. The evidence gathered so far suggests that the integration over ${\cal J}_n, n \neq 0$ yields both real and imaginary parts in QFT, while, in the 0d example in this section,  ${\cal J}_1$  yields only a purely imaginary contribution for $\arg(\lambda)=0$.
    The cancellation of the imaginary parts in path integral examples is essentially the same as for ordinary integrals.  However, in QFT there are also real unambiguous contributions to observables from NP-saddles.

\subsection{NP-data in late terms  of P-expansion}
Resurgence and the Stokes automorphism allow  one to extract the structure of late terms in both P and NP sectors. The asymptotic large order behavior of the perturbative expansion can be deduced by using Cauchy's theorem. Taking $z= \frac{1}{\lambda}$, we can write
\begin{align}
F(z) = \frac{1}{2 \pi i} \sum_{a} \int_0^{e^{i \theta_a} \infty}  d \omega \frac{ {\rm Disc}_{\theta_a}  F(\omega)}  {\omega - z}  +
 \frac{1}{2 \pi i} \oint_{C^\infty} \frac{F(\omega)}  {\omega - z} \,,
 \label{dispersion}
\end{align}
where summation over $a$ is over the singular directions in Borel plane, or Stokes lines in the physical coupling plane (for simplicity we omitted the contributions coming from simple poles), and ${C^\infty}$ is a closed loop at infinity. 

In the present problem, the singular directions are $\theta= {0, \pi}$.  For example, consider the large order behavior of $\Phi_0 $. The only discontinuity of $\Phi_0 $ is in the $\theta=0$ direction and using 
Eq.~\eqref{Stokes} and  Eq.~\eqref{dispersion},  we obtain 
\begin{align}
a_n^{(0)} \sim \frac{\rm s}{2 \pi i} \frac{\Gamma(n)}{(S_{10})^n} \left[ a_0^{(1)}  +  a_1^{(1)}  \frac{S_{10}}{(n-1)} + a_2^{(1)} \frac{ (S_{10})^2 }{(n-1) (n-2)}  +  \ldots  \right]  .
\label{late-1}
\end{align}
So the leading large order behavior  of  the asymptotic expansion around the P-saddle is determined by 
the relative action with respect to the NP-saddle. The corrections to the leading behavior are governed by the early terms in the perturbative expansion around the NP-saddle.  This is a very explicit realization of the idea of resurgence  stated just after Eq.~\eqref{Stokes}: the information in the series expansion around the NP-saddle surges up, in a disguised form, in the expansion around the P-saddle and vice versa \cite{Ecalle:1981}.

  \section{Review of the Principal Chiral Model on $\R^2$}
  \label{basic}
  This section briefly summarizes some basic aspects of  the principal chiral model (PCM) with and without fermions. 
The bosonic   PCM in $d=2$ dimensions is an asymptotically free matrix field theory.
  The  classical action  is given by 
\begin{align}
\label{eq:BosonicSigmaModel}
S_v = \frac{1}{2g^2} \int_{M} dt dx \, \tr \partial_{\mu} U \partial^{\mu} U^{\dag}\,, \qquad  \frac{1}{g^2}= \frac{N}{\lambda}
\end{align}
where $U(t,x) \in SU(N)$, $M$ is a two-dimensional manifold with $\mu$ running over $t,x$, and $\lambda = g^2 N$ is a dimensionless coupling constant which must be held fixed when taking the large-$N$ limit. This action is invariant under the global symmetry group $SU(N)_L\times SU(N)_R$ acting as $U\rightarrow g_L\, U\, {g_R}^\dagger$, with $g_L\in SU(N)_L$ and $g_R\in SU(N)_R$.  

Classically, the theory is scale invariant,  and  the non-linear wave solutions for the $U$-field propagate at the speed of light.   This means the classical theory has   $N^2-1$  gapless  degrees of freedom.
This is similar to classical Yang-Mills theory, which also has 
$N^2-1$ massless (gapless) gluons. 

In the quantum theory, the situation is believed to be both qualitatively and quantitatively
different. In particular, a macroscopic observer in a hypothetical $\R^{1,1}$ universe
would not see the $N^2-1$ non-linear $U$-field waves, just as we do not see non-linear Yang-Mills waves. Indeed, numerical lattice simulations indicate that the theory is gapped, see e.g. \cite{Rossi:1993zc}.
The mass gap is expected to be of the order of the
 dynamically generated strong scale $\Lambda$
\begin{align}
\Lambda^{\beta_0}= \mu^{\beta_0} e^{-\frac{4 \pi}{g^2(\mu)}}, \qquad \beta_0=N
\label{beta0}
\end{align} 
where $\mu$ is a cut-off scale, and $\beta_0$ is the leading coefficient of renormalization group beta-function. 

This model is  interesting because it possesses  a matrix-like large-$N$ limit which is dominated by planar diagrams when $N \to \infty$ with $\lambda$ fixed, just like Yang-Mills theory.  However, 
it has only received scant attention as a useful toy model for Yang-Mills. One of the primary reasons behind this is an  apparent dissimilarity to Yang-Mills: the PCM does not have instantons while Yang-Mills theory does. 
We will come to the conclusion that this difference from Yang-Mills is only superficial, because under suitable conditions, there is an infinite class  of the NP-saddles in PCM model as well.

As it happens, some NP saddles in the PCM, the unitons, have been discovered some time ago by Uhlenbeck \cite{uhlenbeck1989harmonic}.  Unitons are harmonic maps from $S^2$  to $SU(N)$~\footnote{Here, $S^2$ should be viewed as one point compactification of $\R^2$ by including a point at infinity.}.
 Unlike instantons, there is no homotopy  argument for the stability of a uniton. However, (and sounding almost contrary to the previous statement), the uniton action is quantized in units of $\frac{8 \pi}{g^2}$~\cite{Valli1988129,Piette:1987qp,Wood01051989,Ward:1990vc,Dunne:1994uy}~\footnote{We are grateful to N. S. Manton for suggesting an elegant way to think about the quantization of the uniton actions.  The key observation is that while $\pi_2[SU(N)] = 0$, $\pi_3[SU(N)] = Z$.  The relevance of $\pi_3$ comes from considering the homotopy groups of the \emph{infinite-dimesional} manifold of field configurations $\mathcal{M} = \mathrm{Maps}: S^2 \to SU(N)$,$\pi_{n}[\mathcal{M}] = \pi_{n+2}(SU(N))$.  This perspective suggests that the vanishing of $\pi_2[SU(N)]$ is just the statement that there is only one connected component in $\mathcal{M}$ \cite{Manton:2004tk},  so the unitons must lie in the same conjugacy class as the vacuum, on the other hand the non-triviality of $\pi_{3}[SU(N)]$ means that there are non-contractible one-cycles in $\mathcal{M}$, which appear to be realized by unitons, with the integrality of $\pi_3[SU(N)]$ related to the integrality of the uniton action. }. Thus the uniton amplitude is parametrically of the form:
\begin{align}
{\cal U} \sim  e^{-\frac{8 \pi}{g^2(\mu)}}  \qquad { \rm and }   \qquad   
\Lambda^{2 \beta_0} = \mu^{2 \beta_0}  {\cal U}\,.
\label{Uniton}
\end{align} 
The structure of the moduli of a uniton, its zero and quasi-zero modes are so far not completely understood.

The bosonic principal chiral matrix  theory is expected to posses the following properties: 
\begin{itemize}
\item[\bf 1)] Mass gap
\item[\bf 2)]  Confinement:  free energy $\mathcal{F}/N^2$ approaching zero in the {\it low} temperature regime.  
\item[\bf 3)]  Deconfinement: $O(N^2)$ free energy in the {\it  high} temperature regime.
\end{itemize}
A few remarks are in order about these properties: 
 {\bf 1)} In the integrability studies of the PCM where the model is viewed as solvable, the existence of the mass gap is an {\it assumption}. This assumption is something that we would like to derive in our framework.  
 {\bf 2)}
We refer to the low temperature  regime as a confined regime because there the microscopic $\mathcal{O}(N^2)$ degrees of freedom are not present in the physical Hilbert space. 
 {\bf 3)} We refer to the high temperature regime as deconfined because of the  liberation of the $\mathcal{O}(N^2)$ microscopic degrees of freedom.

\vspace{3mm}
\noindent{\bf  Fermions}
\vspace{3mm}

 \noindent It will also be useful to add Lie algebra valued Majorana fermions to the bosonic Lagrangian, since that makes it easier to draw analogies with QCD with fermions in adjoint representation.  
The action Eq.~\eqref{eq:BosonicSigmaModel} is replaced by
\begin{equation}
S_{\rm f} = S_b +\frac{1}{2g^2} \int_M d^2 x \,  \tr\left[ i {\bar{\psi}}_i \gamma^{\mu}( \partial_{\mu} \psi^i+ \frac{1}{2} [U^{\dag} \partial_{\mu} U,\psi^i])
 - \frac{1}{16} \{{\bar{\psi}}_i, \gamma_5 \psi^i \} \{{\bar{\psi}}_j, \gamma_5 \psi^j \} \right]\,,
\label{eq:BFSigmaModel}
\end{equation}
 where $\psi^i$, with $i, \ldots, N_f$, are $\mathfrak{su}(N)$ valued Majorana fermions and $\gamma_5$ is the chirality matrix in $2d$. 
The fermions are two-index matrix valued fields. Under $SU(N)_L\times SU(N)_R$ they transform as $\psi^i\to g_R\,\psi^i g_R^\dagger$, and despite the asymmetric-looking form of the fermion transformations the theory still has an $SU(N)_L \times SU(N)_R$ symmetry for any $N_f$.

\vspace{5mm}

{${\bf N_f =1 \; or \;  \N=(1,1)}$:} The model with $N_f=1$ has $\mathcal{N}=(1,1)$ supersymmetry, the minimal non-chiral supersymmetry in 2D, see e.~g.~\cite{DiVecchia:1984ep,Evans:1996ah}.    
On top of the three properties of the bosonic model  mentioned above, this fermionic model  is also  believed to possess the property
\begin{itemize}
\item[\bf 4)] Discrete chiral symmetry breaking and two isolated vacua (for any $N$).
\end{itemize}
The $\mathcal{N}=(1,1)$  model has a $\Z_2$ discrete chiral symmetry:
\begin{align}
\Z_2: \psi \rightarrow \gamma_5 \psi, \qquad {\rm or}  \qquad \Z_2:  \tr  \bar \psi \psi  \rightarrow -
 \tr  \bar \psi \psi
\end{align}
which is believed to be dynamically broken by the formation of fermion bilinear condensate, $ \langle \frac{1}{N} \tr  \bar \psi \psi \rangle  = \pm  \Lambda $. 

The existence of the two isolated vacua can also be backed up by the following independent argument, for the case of the $\mathcal{N}=(1,1)$ $SU(2)$ PCM. 
The group manifold for the $SU(2)$ PCM  is three-sphere $S^3$,  same as  the $O(4)$ sigma model. The supersymmetric index for the $O(4)$ model  (or generally for $O(N)$ models) was calculated a long time ago and it is equal to $I_S =2$ \cite{Witten:1982df}. This is indeed compatible with the discrete chiral symmetry breaking and the existence of two isolated vacua.  We claim that the supersymmetric Witten index for the PCM model for arbitrary $N$ must also be equal to two:
\begin{align}
I_S (SU(N)) =2 \,.
\end{align}

\vspace{5mm}
{$\bf N_f >1$:} The PCM model with  $N_f$ Weyl Majorana fermions has an $SU(N_f)$ continuous 
global symmetry and $ \Z_2$ discrete chiral symmetry. A continuous global symmetry in a finite $N$ theory cannot be broken on $\R^2$ due to the Mermin-Wagner-Coleman theorem. We expect that for sufficiently low $N_f $ the discrete chiral symmetry should be broken, and at large-$N$ the $SU(N_f)$ global symmetry may be broken as well. 

The inclusion of an arbitrary number of fermion flavors does not modify the one loop beta function given in Eq.~\eqref{beta0}.
Unlike in four dimensions,  two-dimensional non-linear sigma models remain 
asymptotically free even at arbitrarily large-$N_f$. 
Whether these theories are confining or exhibit IR conformal behavior should depend on the number of fermionic flavors, $N_f$. The value $N_f^{\star}$ at  which these theories move from gapped (for $U$-fluctuations) behavior to gapless behavior is currently unknown.   We expect that for $N_f <N_f^{\star}$, the discrete chiral symmetry should be broken  and these theories possess two isolated vacua.

\subsection{Perturbation theory  on $\R^2$ and IR-renormalons}
\label{sec:IR-ren}
The structure of perturbation theory in PCM is similar to other asymptotically free matrix field theory. 
On general grounds, we expect two types of factorial divergences in the perturbative series associated with the P saddle point. These are  
\begin{enumerate}
\item Combinatorial $n!$ growth in the number of  Feynman diagrams
\item Phase space  $n!$  contribution coming from the integration of high and low momenta from a fixed class of diagrams.  These contributions are referred to as the UV and IR renormalons, respectively. 
\end{enumerate} 
Usually, in theories with instantons,  the combinatorial growth is associated with instanton-anti-instanton $[\I  \bar \I]$ pairs. That is, the ambiguity of Borel resummation of perturbation theory is cancelled against the ambiguity in the $[\I  \bar \I]$-amplitude. However, in the PCM, there are no instantons to begin with. In \cite{Cherman:2013yfa}, we argued that the uniton saddle ought to substitute the  $[\I  \bar \I]$-saddle and its ambiguity, yielding an ambiguity of order $\pm i e^{- \frac{8 \pi}{g^2 (Q)} }$.

 On $\mathbb{R}^2$, Fateev, Kazakov, and Wiegmann\cite{Fateev:1994dp,Fateev:1994ai} used integrability techniques to show that there exists a much larger ambiguity.  It is guessed there that this is related to the leading renormalon ambiguity:
\begin{align}
\text{ Leading IR renormalon on} \; \mathbb{R}^2: \pm i e^{-\frac{8\pi}{g^2[Q] N}} \sim   \pm i  \frac{\Lambda^2}{Q^2},
\end{align}
where $Q$ is a large Euclidean momentum. 

This is a sensible guess in the light of the operator product expansion (OPE). In the PCM, 
the leading non-trivial condensate is the  dimension two operator $\tr \partial_{\mu} {U}  \partial^{\mu} {U}^{\dag}$.  Since  $\arg \lambda=0$ is a Stokes line in PCM, the expectation value of generic operators  that can mix with perturbation theory are  two-fold ambiguous\cite{David:1983gz}. The condensate evaluated at $ \theta=  0^{\pm}$ must give 
\begin{align}
\left \langle { \frac{1}{N}} \tr \partial_{\mu} {U}  \partial^{\mu} {U}^{\dag} \right \rangle_{ \theta = 0^{\pm}}   Q^{-2} \sim  c_1 \frac{\Lambda^2}{Q^2} 
\pm i c_2 \frac{\Lambda^2}{Q^2} 
\label{condensate}
\end{align}
where $c_1$ and $c_2$ are pure numbers.  
Of course, one also expects more-suppressed IR renormalon ambiguities of the order  $\Lambda^{2m}/ Q^{2m}, m>1$. 

On $\R^2$, there is no semi-classical interpretation for IR renormalons. However, at large $Q^2$ there is a useful hierarchy of non-perturbative scales associated with the IR renormalons
\begin{align}
&\underbrace{ e^{- \frac{8 \pi}{g^2 N  (Q)} } }_{\small \rm 1^{st}  \;  IR-renormalon}   \gg 
\underbrace{ e^{- \frac{16 \pi}{g^2 N  (Q)} } }_{\small \rm 2^{nd}  \;  IR-renormalon} \gg  \ldots 
 \gg  \underbrace{ e^{- \frac{8 \pi}{g^2   (Q)} } }_{\small \rm  1-uniton}  \gg     \ldots  \cr
& {\rm or \;  equivalently } \qquad  \frac{\Lambda^2}{Q^2} \gg \frac{\Lambda^4}{Q^4} \gg  \ldots \gg  \frac{\Lambda^{2N}}{Q^{2N}}  \gg \ldots 
\end{align}
which is exploited in the OPE approach.
It is worth 
emphasizing that the leading IR renormalon ambiguity is parametrically much  larger than the 
uniton ambiguity at large Euclidean momentum $Q$.  It is actually the $N^{\rm th}$ root of it. 

In  theories with fermions, if  $N_f  < N_f^{\star}$, 
we expect only minor and relatively  unimportant changes in the Borel plane structure. 
If $N_f  > N_f^{\star}$,  one does not expect IR-renormalons to exist, since the coupling does not diverge in the infrared. 
This drastic change at $N_f^{\star}$ is consistent with a speculation by  't Hooft, in the context of gauge theories,    that  
IR-renormalons are connected with  the mass gap and confinement \cite{tHooft:1977am}.

\section{Compactification to $\R \times S^1$ and   adiabatic continuity}
\label{sec:SmallLLimit}
Usually, one can only hope to get analytic insights into the non-perturbative physics of a field theory when it is weakly coupled.  This is definitely not the case for the PCM on $\mathbb{R}^2$, where the IR physics is strongly coupled in terms of $\lambda$. When the PCM is compactified on $\mathbb{R} \times S^1$, however, at small enough circle size $L$ the coupling will become set at the scale $1/L$, and the theory is guaranteed to become weakly coupled thanks to   asymptotic  freedom.  

In recent years it has become clear that in QCD-like gauge theories on $\mathbb{R}^3\times S^1$, the appearance of a weak-coupling regime at small-$L$  can occur in two dramatically different ways:
\begin{itemize}
\item[i)] If the $S^1$ circle is thermal,  as $L$ is dialed from large to small,  the gauge theory goes through a phase transition, or a rapid cross-over, in such a way that the physics in the small-$L$ theory is not adiabatically connected to the physics in the large-$L$ decompactification limit.  
\item[ii)] If the $S^1$ circle is spatial in theories with e.g. massless adjoint fermions or if a center-stabilizing deformation is used,  then as $L$ is dialed from large to small,  
one finds that the small-$L$ limit of the gauge theory is adiabatically connected to the large-$L$ regime without any phase transition or rapid cross-over. 
\end{itemize}
In gauge theory, thermal Yang-Mills theory and thermal ${\cal N}=1$ Super-Yang-Mills (SYM) yield examples of the first type of limit, as is well known from thermal field theory \cite{Gross:1980br}.  In this case, the long distance\footnote{Here by ``long distance" we mean physics dominated by contributions from distance $\ell \gg (g_3^{2} N)^{-1} =1/(g_{4}^{2} NT)$.} physics is strongly coupled and incalculable. 
  Some examples of the second type are deformed Yang-Mills and spatially compactified ${\cal N}=1$ SYM \cite{Unsal:2007jx,Unsal:2008ch}.  In this case, the long distance physics remains weakly  coupled and is non-perturbatively calculable.

\begin{figure}[tbp]
\centering
\includegraphics[width=0.7\textwidth]{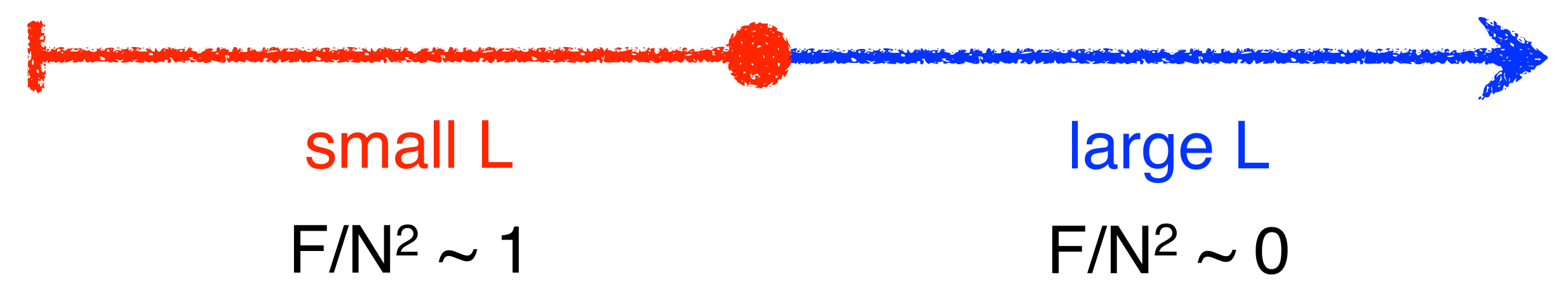}
\includegraphics[width=0.7\textwidth]{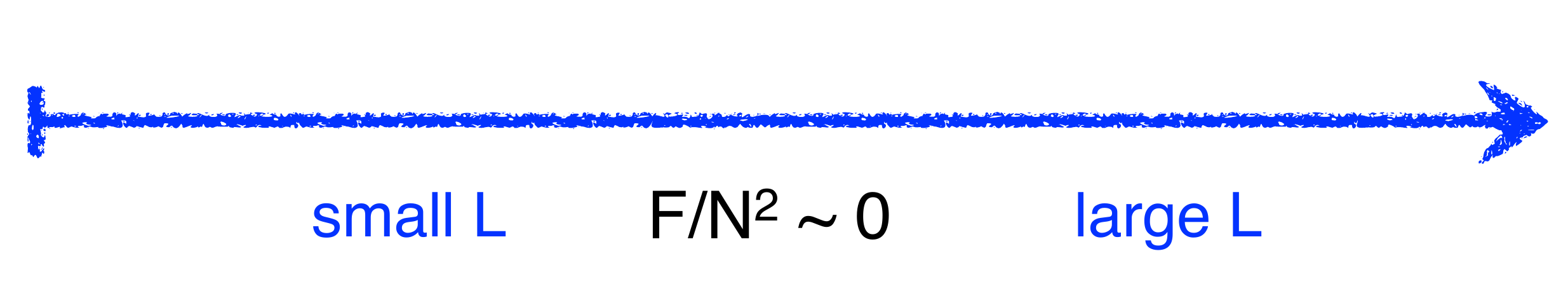}
\caption{{\bf Top:} With a thermal compactification on $\R\times S^1_\beta$,  there is a rapid-crossover from an $F/N^2 \sim O(1)$ 
(deconfined)  behavior of the free energy to  $F/N^2 = 0 $ confined regime, which becomes a genuine phase transition at $N=\infty$. Even at finite-$N$, however,  the quantitative behavior of the theory changes dramatically between these two regimes, despite the fact that there is no sharp phase transition in a finite volume.
  \newline
{\bf Bottom:} By using spatial compactification,  we find a unique small-$L$ limit in which ``free energy"  scales as $F/N^2 = 0 $. The behavior of the theory does not change dramatically from small-$L$ to large-$L$. This is the idea of {\it adiabatic continuity}. Since the small-$L$ theory is weakly coupled thanks to asymptotic freedom, it is NP-calculable, and the knowledge gained therein is continuously connected to the physics of decompactified theory on $\R^2$.
   }
\label{fig:adiabatic-continuity}
\end{figure}

 Our goal in this section is to discuss the realization of both of these classes of small-$L$ limits in the PCM by a careful analysis of the compactification procedure, as depicted in Fig. \ref{fig:adiabatic-continuity}.  
The theory defined by Eq.~\eqref{eq:BosonicSigmaModel} has a global $SU(N)_L\times SU(N)_R$ symmetry acting as $U \to g_L\, U\, g_R^{\dag}$.   Defining the theory on an Euclidean base space manifold $\mathbb{R}\times S^1$ requires a choice of boundary conditions on the circle.  We will consider a family of compactified theories labeled by a choice of boundary conditions
\begin{align}
\label{eq:BosonicBCs}
U(t, x+L) &= e^{i H_L} U(t, x) e^{-i H_R}, \;   \\
\psi_i(t, x+L) & = (\pm)  e^{i H_R} \psi_i(t,x) e^{-i H_R}\,,
\end{align}
where $H_L, H_R \in \mathfrak{su}(N)$.    We refer to the $(-)$ boundary conditions for the fermions as  thermal,  since they reduce to the standard anti-periodic thermal boundary conditions when $H_R=H_L=0$, and we refer to $(+)$ boundary conditions as spatial, since they become purely periodic BCs when $H_R=H_L=0$.  For generic $H_L, H_R$, these are non-thermal compactifications.  For the supersymmetric  ${\cal N}=(1,1)$ theory, the spatial boundary conditions respect supersymmetry. 

The question is which theory, in this family, is in the same `phase' at small-$L$ as the theory on $\mathbb{R}^2$~\footnote{We are working with a two-dimensional theory, for which the Coleman-Mermin-Wagner theorem implies the lack of any local order parameters which could distinguish distinct phases at finite $N$. Nevertheless, there can still be rapid cross-overs in the physical properties as a function of $L$ at finite $N$ (which become a sharp phase transition at $N=\infty$) and as a result small-$L$ physics can be very different from large-$L$ physics.}.  With a thermal compactification, there is a rapid cross-over/phase transition at finite/large $N$ as $L$ is dialed from large to small values.  Our goal is to construct a special small-$L$ limit in which this does not happen, so that the changes between large-$L$ and small-$L$ are `adiabatic'.  One of the two observables we will use in our construction will sharpen into a proper order parameter in the large-$N$ limit, where the Coleman-Mermin-Wagner theorem no longer forbids such a notion.

To proceed with the analysis we find it convenient to switch variables to $\tilde{U}, \tilde{\psi}$:
\begin{align}
\Ut &=  e^{-i H_L \frac{x}{L}} U\, e^{ i H_R \frac{x}{L}}\,,\\
\tilde{\psi} &=  e^{-i H_R \frac{x}{L}} \psi\, e^{i H_R \frac{x}{L}} \,,
\end{align}
so that $\tilde{U}, \tilde{\psi}$ are periodic on $S^1$.  In terms of $\tilde{U}, \tilde{\psi}$, the action becomes
\begin{align}
\label{eq:GaugedAction}
S[H_L, H_R] 
&= \frac{N}{2\lambda} \int_{\mathbb{R}\times S^1} dt dx  \tr D_{\mu}\tilde{U}  D^{\mu} \tilde{U}^{\dag}\, \\
&+  \frac{N}{2\lambda} \int_{\mathbb{R}\times S^1} dt dx \,  \tr\left[ i \tilde{\bar{\psi}}_i \gamma^{\mu}( D_{\mu}\psit^i + \frac{1}{2} [\tilde{U}^{\dag} D_{\mu} \tilde{U}, \tilde{\psi}^i])
 - \frac{1}{16} \{\tilde{\bar{\psi}}_i, \gamma_5 \tilde{\psi}^i \} \{\tilde{\bar{\psi}}_j, \gamma_5 \tilde{\psi}^j \} \right]\,, \nonumber
\end{align}
where 
\begin{align}
 \label{Dmu}
D_{\mu} \tilde{U} &\notag= \partial_{\mu} \tilde{U} - i \frac{\delta_{\mu,x}}{L}\, \left(  H_L\tilde{U} -  \tilde{U} H_R \right)  = \partial_{\mu} \tilde{U}  - i \delta_{\mu,x}\, \left( [H_V, \tilde{U}] +  \{H_A, \tilde{U}\} \right)\,, \\
D_{\mu} \tilde{\psi}^i &= \partial_{\mu} \tilde{\psi}^i - i \frac{\delta_{\mu,x}}{L}\,[H_R, \tilde{\psi}^i ]\,,
\end{align}
and $H_{V,A} = \frac{1}{2L} (H_L \pm H_R)$.  We can interpret $\frac{1}{L} H_{L}$ and $\frac{1}{L} H_{R}$ as background gauge fields for the global symmetry group $SU(N)_L\times SU(N)_R$.  The symmetry of the original action Eq.~\eqref{eq:BosonicSigmaModel} under $U\to g_L\,U\,g_R^\dagger$ becomes a symmetry of Eq.~\eqref{eq:GaugedAction} under $\Ut \to g_L\,\Ut\,g_R^\dagger$ together with $H_L\to g_L \,H_L\, g_L^\dagger$ and $H_R\to g_R \,H_R\,g_R^\dagger$, which are effectively global `gauge' transformations. A theory with the twisted boundary conditions (BCs) of Eq.~\eqref{eq:BosonicBCs} can be equivalently viewed as a theory with periodic BCs for $\tilde{U}, \tilde{\psi}$ with the constant background gauge fields $\frac{1}{L}H_{L/R}$~\footnote{Note that these constant background fields $H_{L,R}$ are \emph{not} standard chemical potentials for the conserved currents
\begin{align}
J^L_{\mu} =& i\,U^\dagger \partial_\mu U \,,\\
J^R_{\mu}=& i\,\partial_\mu U \, U^\dagger\,.
\end{align}
A chemical potential in Minkowski space enters the action as a constant background \emph{time} component of a gauge field.  In Euclidean space, a chemical potentials enters the action as a constant imaginary Euclidean-time component of a background gauge field. In contrast, we interpret the background gauge fields $H_V, H_A$ as the \emph{space} components of the gauge fields associated to the $SU(N)_{L,R}$ symmetries, and they are real in \emph{both} Minkowski and Euclidean space. }.  From here onward we will work with the latter picture, and drop the tilde on $U, \psi$ for simplicity.

\subsection{Large-$L$ expectations}
\label{LargeLexp}

To figure out the right choice for $H_V, H_A$, we have to decide what properties of the large-$L$ theory we want to capture in the small-$L$ limit. From our perspective, the most essential features of the bosonic PCM on $\R^2$ and large $S^1 \times \R $ are  
the existence of a mass gap and the  order $N^0$ free energy density   in the large-$N$ limit.  


We want to find a small-$L$ limit  which allows both of these expected properties to persist.  To translate these heuristic expectations into sharp conditions on the theory, we first observe that the existence of mass gap  implies that when $L\to \infty$, the dependence on the boundary conditions must vanish, since in a theory with a mass gap we expect finite-volume effects to vanish as $e^{-\Delta L}$, where $\Delta$ is the mass gap. Let   $Z(L; H_L, H_R)$ be  the partition function of the theory with the background fields turned on,  and let us define  as $\mathcal{F}(L; H_L, H_R) =\mathcal{V}^{-1}\log Z(L; H_L, H_R)$ as the ``twist free energy", where  $\mathcal{V}^{-1}$ is the volume of the space-time manifold. Then it is clear that in the large-$L$ confined phase, the free energy must be independent of the boundary conditions at leading order in $N$.   At small-$L$, it is not possible to have complete independence of the theory from $L$\footnote{If it were so, the theory would still be strongly coupled at small-$L$.}.  The highest degree of independence from $L$ one can demand at small-$L$ is 
\begin{align}
\label{eq:NoPersistentCurrents}
\frac{\partial \mathcal{F} }{\partial H_i} &= \langle J^i_x\rangle_{H_V, H_A} =0 \qquad i= V, A
\end{align}
where $J^{V/A}_{\mu}=\frac{1}{2} (J^L_{\mu}\pm J^R_{\mu})$, and the subscript on $\langle \cdot \rangle_{H_V, H_A} $ is a reminder that the expectation value is to be taken with background fields turned on.   Heuristically, this means that we demand that changing boundary conditions should not result in persistent currents in the compactified direction.   
Equation~\eqref{eq:NoPersistentCurrents} automatically holds in the decompactification limit in the large-$L$ theory, but it becomes a non-trivial constraint at small-$L$.  

Next,  to have any chance that the small-$L$ theory is confined according to the count of the degrees of freedom contributing to the physical Hilbert space,
the ``twist free energy", i.e, the free energy of the system normalized to $N^2$ at a given value of the the boundary conditions 
must zero. For instance, with purely `thermal' boundary conditions, with $H_V = H_A = 0$, when $L$ can be interpreted as an inverse temperature $1/T$, at small-$L$ the theory becomes weakly coupled, but one  expects the theory to be `deconfined', with all $N^2-1$ components of the matrix $U$  liberated and contributing equally to free energy density, so that $\mathcal{F}/N^2 \sim \mathcal{O}(1)$.  (We show this explicitly in Sec.~\ref{sec:highT}.) 
 This is in sharp contrast to what happens in the  low temperature decompactification limit, where we expect the theory to be in a `confining' phase, in the sense that $\lim_{N\to \infty} \mathcal{F}/N^2 = 0$.  That this is the case can be seen from lattice simulations (see e.g. ~\cite{Rossi:1993zc}) as well as from the exact solution for the spectrum using integrability\cite{Polyakov:1983tt,Wiegmann:1984ec,Fateev:1994dp,Fateev:1994ai}.  Indeed, as shown in \cite{Fateev:1994dp,Fateev:1994ai}, the spectrum consists of $k$ particle modes with masses $m_k$
\begin{align}
m_k = m \frac{\sin \left(\frac{\pi k}{N}\right)}{\sin(\pi/N)}
\end{align}
where $m = \Lambda$ is the strong scale, with degeneracies given by $d_k = N!/[k!(N-k)!]$.  Then the free energy can be written as
\begin{align}
\mathcal{F} = -T\sum_{k=1}^{N-1} d_k \int \frac{dp}{2\pi} \log \left(1-e^{-\omega_k/T }\right)^{-1} 
\end{align}
where $\omega^2_k = p^2 + m_k^2$.  If $T \ll m$, the $m_1$ mode will dominate the thermodynamics so long as $N e^{-m/T} \ll 1$, and as a result\footnote{We are very grateful to V. Kazakov for explaining this argument to us, and correcting a mistake about the scaling of the free energy in an earlier version of the paper.} 
\begin{align}
\mathcal{F} &\to N T^2 \left(\frac{m}{2\pi T}\right)^{1/2} e^{-m/T}
\end{align}
So as claimed above $\lim_{N\to \infty} \mathcal{F}/N^2 = 0$ for small $T$.
 
  At large-$N$, the value of $\frac{1}{N^2} \mathcal{F}$ becomes a bona-fide order parameter:  it is zero in the decompactification limit, but may become order one at small-$L$, depending on the values of $H_V, H_A$.  Hence we will demand that $\mathcal{F}/N^2$ (or more precisely the natural dimensionless quantities $L^2 \mathcal{F}$ or $\Lambda^{-2} \mathcal{F}$ associated to $\mathcal{F}$) must go to zero at large $N$ for any $L$.  This expectation is believed to be automatically met in the decompactification limit based on previous studies using integrability\cite{Polyakov:1983tt,Wiegmann:1984ec} or lattice Monte Carlo simulations, see e.g.  But demanding $L^2 \mathcal{F}/N^2 \to 0$ poses a non-trivial constraint at small-$L$. 

To summarize, our \emph{adiabaticity conditions} are
\begin{samepage}
\begin{enumerate}
 \item We demand an insensitivity to changes in boundary conditions, Eq.(\ref{eq:NoPersistentCurrents}).
 \item We demand the free energy normalized to $N^2$ goes to zero at large $N$.
 \item For the supersymmetric $N_f=1$ PCM, we impose susy preserving boundary conditions. For the non-susy $N_f>1$ case we impose the same supersymmetric boundary conditions for all the fermions\footnote{While it is clear that this is a sensible demand for the $N_f=1$ theory, the reason to demand this for general $N_f \ge 1$ is more subtle.  The reason we do so is that we expect the large-$N$ PCM with $N_f$ `adjoint' fermions to have an emergent fermionic symmetry at large-$N$ due to arguments similar to the ones recently given for QCD[Adj] in \cite{Basar:2013sza}.}.    
\end{enumerate}
\end{samepage}
We emphasize that we are \emph{not} assuming that the theory is gapped at small-$L$.  That is a dynamical question about the theory, which should be settled by a calculation at small-$L$.  What makes the construction interesting is that the unique small-$L$ theory selected by the conditions above \emph{does} turn out to have a non-perturbatively-generated mass gap!  This is consistent with the notion that the small-$L$ limit we construct really is adiabatic.

\subsection{Choosing the right small-$L$ limit}

Our goal for the rest of this section is to see which choice of $H_V, H_A$ results in a  $Z(L; H_L, H_R)$ which reproduces the large-$L$ expectations encoded in Eq.~\eqref{eq:NoPersistentCurrents} and $L^2 \mathcal{F}/N^2 \to 0$ even at small-$L$.  When $L \Lambda$ is small enough, the theory will become weakly coupled in terms of $U$.  The precise meaning of  `small enough' is subtle, and we address it below.  If $U$ is parametrized as
\begin{align}
U = e^{i W}, \; W\in \mathfrak{su}(N)
\label{algebra-par}
\end{align}
we expect that at weak coupling the dominant contribution to $Z$ will come from fluctuations which are quadratic in $W$.

We first determine the form of $H_A$ consistent with Eq.~\eqref{eq:NoPersistentCurrents}.  To do this, we perform a global axial `gauge' transformation using $g \in SU(N)_A$, so that $H_A\to g \,H_A\,g^\dagger$ is diagonalized and lies within the Cartan subalgebra of $\mathfrak{su}(N)$ $H_A = a_{\alpha} t^{\alpha}$, with $\{t^{\alpha}\}$ being the Cartan generators normalized as $\mbox{Tr} (t^\alpha t^\beta) = \delta^{\alpha\beta}/2$.  The small $W$ (i.e., perturbative) action then becomes 
\begin{align}
S = \frac{N}{2\lambda} \int_{\mathbb{R}\times S^1} dt dx\left( - 2 \sum_{\alpha=1}^{N-1} (a_\alpha)^2 +\mathcal{O}(W^2)  \right)\,.
\label{eq:AxialAction}
\end{align} 
This constitutes a \emph{tree-level} potential for the eigenvalues of $H_A$, which is extremized when $H_A=0$.   Hence to satisfy  Eq.~\eqref{eq:NoPersistentCurrents}, we must set{\footnote{
Note that the partition function actually depends on the conjugacy classes of $H_L,\,H_R$: $Z[H_L,H_R]=Z[g_L H_L g_L^\dagger, g_R H_R g_R^\dagger]$ with $g_{L/R}\in SU(N)_{L/R}$. Hence once can trade a purely axial background $H_V=0,H_A\neq0$ (as in \cite{Fateev:1994ai,Fateev:1994dp}) for a purely vectorial one, provided one modifies Eq.~\eqref{algebra-par} accordingly. We thank V.~Kazakov and Z.~Bajnok for pointing this out to us.}} $H_A = 0$.  So long as the theory is weakly coupled, quantum effects cannot  change the extremum of the potential energy as a function of $H_A$, whether or not there are fermions in the theory.   The reason is the presence of the \emph{non-vanishing} tree-level potential above. Hence from here onward, we will set $H_A = 0$.

Now we can focus on working out the effective potential for $H_V$.  When $H_A=0$ we can use a global vectorial `gauge' transformation to diagonalize $H_V$, so that it lies in the Cartan subalgebra.  The energy is independent of $H_V$ at tree-level, so now we must do a one-loop analysis to compute the first non-trivial contributions to the twist free energy.  

Before presenting the result for the twist free energies with the two classes (thermal and spatial) of BCs we are considering, it is useful to introduce a Wilson loop operator associated to $H_V$, which will allow us to write the one loop twist free energy in a more illuminating form:
\begin{align}
\Omega&= e^{i \oint d x_2 \, H_V} = e^{ i \,L\, H_V}\, \qquad {\rm where} \qquad 
(H_V)_{jk} = 2 \pi \mu^{j} L^{-1} \delta_{jk},\; \tr H_V = 0\,.
\end{align}
Note that the eigenvalues of $\Omega$ are $SU(N)_V$ gauge-invariant, and $\tr \Omega$ transforms non-trivially under the $\mathbb{Z}_N$ center symmetry of $SU(N)_V$ acting as $\Omega \to \omega \Omega, \omega \in \mathbb{Z}_N$.  The expression for the twist free energy of the PCM in the presence of the background gauge field $H_V$ can now be rewritten in terms of the $\Omega$~\footnote{We use the $S_N$ Weyl symmetry of $\mathfrak{su}(N)_V$ to arrange the eigenvalues of $\Omega$ in a canonical order in the case where they are not coincident.}:
\begin{align}
\Omega = \begin{pmatrix}
e^{2\pi i\mu_1} &0&\dots &0 \cr
0& e^{2\pi i\mu_2}&\dots &0\cr
\vdots\cr
0&0&\dots & e^{2\pi i\mu_N}
\end{pmatrix}
\quad, \quad 0\leq \mu_1\leq \mu_2 \leq \dots \leq\mu_N<1 .
\label{twist2}
\end{align}
One can now calculate the potential for the background holonomy $\Omega$ by integrating out  the weakly coupled  KK-modes at one-loop level. The result is
 \begin{align}
&L^2 V_{-} ( \Omega) =   
\label{tpotential}  \frac{1}{ \pi}  \sum_{n=1}^{\infty} \frac{1}{n^2}  (-1+ (-1)^n N_f)  (|\tr \, \Omega^n |^2-1)  \qquad \;\;\; (\rm thermal) \,, \\
&L^2 V_{+} (\Omega)  =   (N_f -1) \frac{1}{ \pi}  \sum_{n=1}^{\infty} \frac{1}{n^2}   
 (|\tr \, \Omega^n |^2-1)  \qquad \qquad  \qquad  (\rm spatial) \,.
 \label{spotential}
\end{align} 

An intuitive way to derive Eq.~\eqref{spotential} is as follows.  (We only detail the derivation for the spatial case; the thermal analysis is very similar.)  At sufficiently small-$L$, the Kaluza-Klein tower of modes in the principal chiral model can be viewed as a collection of simple harmonic  bosonic and fermionic  oscillators. The strategy of the derivation follows L\"uscher and  van Baal \cite{Luscher:1982ma,vanBaal:1988va}.  However, the idea of adiabatic continuity, which is instrumental for our purpose, 
did not appear in these earlier works. 
   
In spatial compactification, i.e., with periodic boundary conditions for fermions,  the KK modes of bosons and fermions are degenerate at the classical level. (This may or may not be lifted quantum mechanically, depending on the theory.)
 There are  $(N^2-1)$ physical 
bosonic fluctuations  
and $N_f (N^2-1)$ fermionic fluctuations  for each Kaluza-Klein level $k \in Z$.
The vacuum energy density associated with the background Eq.~\eqref{twist2} is  the sum of ground state energies of corresponding bosonic and fermionic harmonic oscillators, and it is equal to: 
  \begin{eqnarray}
L {\cal E}[ \mu_{ij}]&&= \sum_{\rm bosons}  {\textstyle \frac{1}{2}} \omega^{b} -  \sum_{\rm fermions}  
  \textstyle \frac{1}{2} \omega^f  \;    \cr
  &&= 
   (1-N_f) \frac{1}{2L} \sum_{i ,j} \sum_{ k \in \Z}  | \mu_i-\mu_j + 2 \pi  k | \,.
\label{SHO} 
\end{eqnarray}
Since the vacuum energy density  Eq.~\eqref{SHO} is periodic in   $ \mu_{ij} \equiv \mu_i-\mu_j  $ with period $2 \pi$, 
it can be Fourier transformed, 
 \begin{eqnarray}
{\cal E} [\mu_{ij}]  = \sum_{ij} \sum_{ n \in\, \Z \setminus  \{0\}} P_{n}  \; 
e^{i  \mu^{ij} n}  \,,
\end{eqnarray}
which is actually the Poisson resummation of the original formula.  Here,  
$P_{ n} =\frac{(1-N_f)}{2L^{2}} I_{n} $. 
The advantage of this form is that 
$\sum_{ij}     e^{i  n \mu^{ij}}  =|\tr \Omega^n |^2  $, and we can express the result in terms of the background holonomy.  The summation over KK modes $k \in \Z$ thus turns into a summation over the winding number of the line operator. The prefactor is: 
\begin{eqnarray}
I_{ n}  =&& \frac{1}{2 \pi} \int_{0}^{2 \pi} d \mu 
\sum_{k \in \Z}  | \mu + 2 \pi \vec k | \;  e^{i n \mu}  \cr
  =&&  \frac{1}{2 \pi} \int_{-\infty}^{\infty} d \mu  
    |\mu | \;  e^{i n \mu} =  \frac{1}{2 \pi} 
      \frac{1}{\Gamma(-\frac{1}{2})}   \int d\mu 
    \int_0^\infty \frac{d  \alpha}{\alpha^{3/2}}   
  e^{- \alpha 
   \mu^2}  e^{i n \mu}   \cr
   =&&   \frac{1}{2 \pi} \frac{1}{\Gamma(-\frac{1}{2})} \int_0^\infty \frac{d  \alpha}{\alpha^{3/2}}   
 \left(  \frac{\pi}{\alpha}\right)^{1/2} 
    e^{- \frac{  n^2} {4 \alpha} }   = -  \frac{1}{\pi} \frac{1}{  n^2}\; , 
 \end{eqnarray}
 resulting in  vacuum energy density
 \begin{eqnarray}
 {\cal E} [\Omega]  = V_{+}[\Omega]= 
    \frac{1}{2 \pi L^2 }
\sum_{ n \in \,\Z \setminus \{ 0\}} 
       (N_f-1) 
    \frac
    {\left|\tr \Omega^{n}    \right|^2}
    {n^2}
    \,,
\label{smallT3b}
 \end{eqnarray}
 which is identical to Eq.~\eqref{spotential}.

  
Our adiabaticity conditions now tell us that we must find the extrema of the twist free energy, and compute the large-$N$ scaling of $\mathcal{Z} = L^2 V$ at the extremum.  Note that the expressions for the twist free energy is very similar to the effective potential for the eigenvalues of Polyakov loops in $SU(N)$ gauge theories on $\mathbb{R}^3 \times S^1$\cite{Kovtun:2007py,Argyres:2012ka} with $N_f$ adjoint  Weyl  fermions.   
The minimization problem  for the potential  is same as  the one in  QCD(Adj) on 
$\R^3 \times S^1$~\footnote{This analysis can easily be generalized to PCM with  
other classical Lie groups such as  $SO(N), Sp(N), \ldots$ along the same lines as  \cite{Argyres:2012ka}.}.  It turns out that there are two extrema that we must deal with in general.

 \subsection{Thermal  compactification and non-adiabaticity}
 \label{sec:highT}
 There are $N$ `thermal' extrema of the thermal or spatial twist free energy at
\begin{equation}
\Omega^{\rm thermal}  = e^{i \frac{2 \pi  k}{N}}  \left ( \begin{array}{ccccc}
1 & &&& \cr
  & 1 &&&  \cr
  && \ddots &&  \cr
      &&&&1
  \end{array} \right)  \; ,   
    \label{eq:MinThermal}
\end{equation}
 where $k$ labels the center-position of the lump of eigenvalues.  For these extrema $H_V = 0$, and if we use anti-periodic BCs for fermions this is precisely the standard Euclidean thermal compactification.  Note that $\tr \Omega^{\rm thermal}$ transforms non-trivially under all non-trivial elements in the center subgroup $\mathbb{Z}_N \subset SU(N)_V$. 
  
In the thermal case, the free energy density is 
  \begin{eqnarray}
&&{\cal F} = V_{-} [ \Omega=1] =   (N^2-1) \frac{\pi}{6} \left(1 + \frac{N_f}{2}\right) T^2\,.
\label{SB} 
\end{eqnarray}
This gives a nice check of our calculations, because this is precisely the expected Stefan-Boltzmann law: 
 $(N^2-1)$ is the number of bosonic degrees of freedom and   $\frac{\pi}{6}T^2$ is the free energy per boson, while  $(N^2-1)N_f$ is the number of fermionic  degrees of freedom, with  $\frac{\pi}{12}T^2$ being the free energy per fermion. The factor of two difference follows from statistics, Bose-Einstein versus Fermi-Dirac. 

This is the deconfined regime shown in Fig. \ref{fig:adiabatic-continuity} where the $O(N^2)$ degrees of freedom are liberated, not adiabatically connected to the large-$L$ confined regime.

 \subsection{Spatial compactification and adiabaticity}
 The only other extremum of the spatial twist free energy for $N_f=0, N_f>1$ is  
 \begin{equation}
  \Omega^{\rm spatial}  =   e^{i \frac{\pi}{N} \nu }  \left ( \begin{array}{ccccc}
1 & &&& \cr
  & e^{i \frac{2\pi}{N} } &&&  \cr
  && \ddots &&  \cr
      &&&& e^{ i \frac{ 2\pi (N-1)}{N}}
  \end{array} \right) \; ,  \qquad \;   \nu=0,1\; \textrm{for} \; N=\textrm{odd, even}.
  \label{eq:MinSpatial}
\end{equation}
This extremum of the twist free energy is \emph{unique}, and $\tr \Omega$ is neutral under the $\Z_N$ center symmetry $\mathbb{Z}_N \subset SU(N)_V$, with $\tr \Omega^n$ vanishing for $n \mod N \neq 0$.   This is in sharp contrast with the thermal holonomy in Eq.~\eqref{eq:MinThermal}, for which $\tr \Omega^n \neq 0$ for all $n$.

\begin{figure}[tbp]
\centering
\includegraphics[width=0.6\textwidth]{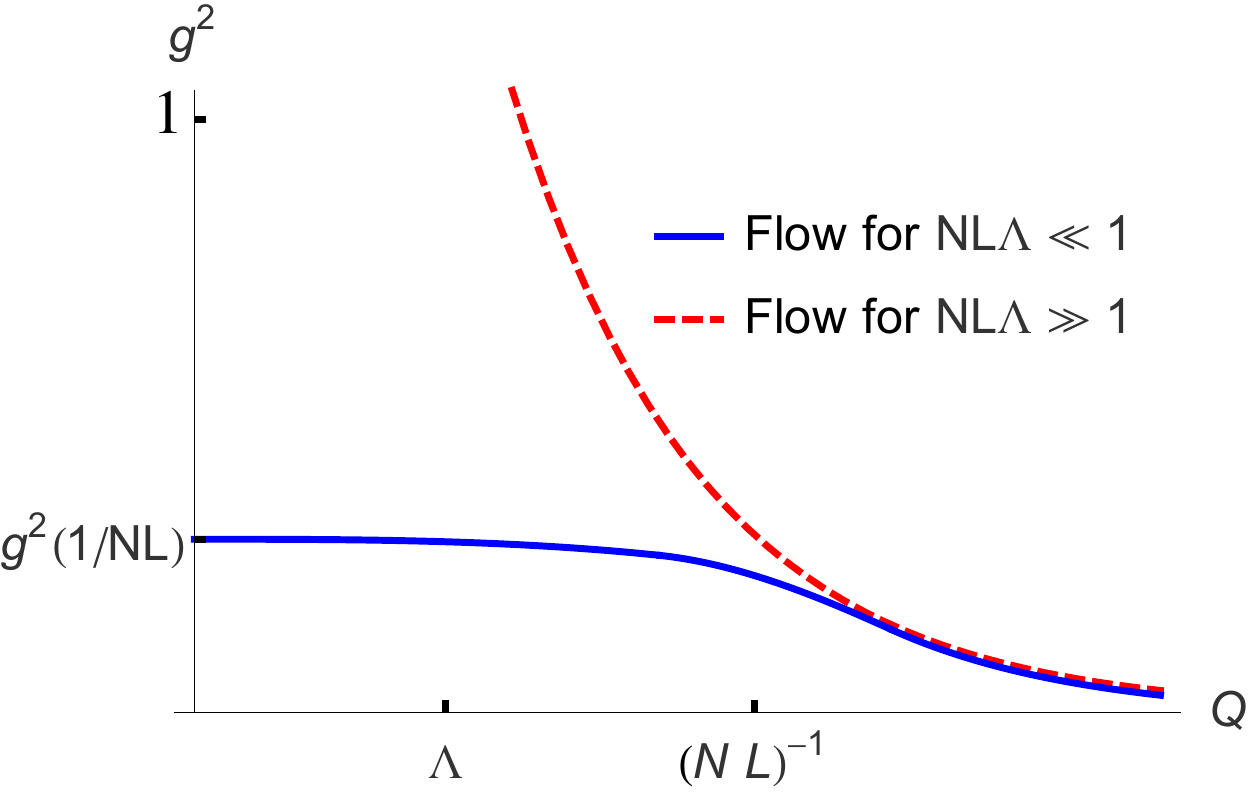}
\caption{A sketch of the flow of the coupling constant $g^2$ in the two regimes $\frac{N L \Lambda}{2\pi} \ll 1$ and $\frac{N L \Lambda}{2\pi} \gg 1$ in the principal chiral model with a  $\mathbb{Z}_N$-symmetric background holonomy.   If $N$ is large with $L$ fixed, large-$N$ volume independence holds in the $N L \Lambda \gg 1$ regime.  The physics of the theory on $\mathbb{R} \times S^1$ exactly coincides with the physics of the theory on $\mathbb{R}^2$, up to $1/N$ corrections, and the theory is strongly coupled at large distances and not tractable using semiclassical methods.   On the other hand, if $N L \Lambda \ll 1$, the physics of the theory is continuously connected to the physics on $\mathbb{R}^2$, but volume independence does not hold.  Here the coupling constant flow freezes at some small value, and the theory is semi-classically tractable.   }
\label{fig:CouplingSketch}
\end{figure}

For $N_f=1$, the theory has ${\cal N}=(1,1)$ supersymmetry. The fermion  boundary conditions associated with $V^{(+)}$ also respect supersymmetry. Consequently, the vanishing of the one-loop contribution actually extends to all orders in perturbation theory.   
At the non-perturbative level, we expect to find non-vanishing contributions to $V^{(+)}$, and for the background in Eq.~\eqref{eq:MinSpatial} to be a non-trivial extremum,  as is the case in 4D $\mathcal{N}=1$ SYM \cite{Davies:2000nw}  and the 2D $\mathcal{N}=(2,2)$ $\mathbb{CP}^{N-1}$ model.  However, we leave an explicit verification of this for the PCM to future work.

  At the unique center-symmetric extremum characterized by Eq.~\eqref{eq:MinSpatial}, the twist free energy is 
  \begin{align}
  V_{+}( \Omega^{\rm spatial}) = \frac{(N_f -1)}{\pi L^2} \times \frac{\pi^2}{6}.  
  \end{align}
 Note that it remains $O(N^0)$ at arbitrarily small-$L$, 
  as opposed to the thermal case, where the associated free energy is $O(N^2)$, as anticipated in Fig. \ref{fig:adiabatic-continuity}.   Hence $L^2 \mathcal{F}/N^2 \to 0$, and we refer to this center-symmetric small-$L$ regime as confined.  We also observe that \emph{only} the theory associated to $V_{+}$ with $\Omega$ satisfying Eq.~\eqref{eq:MinSpatial} satisfies \emph{all} of our conditions of Section \ref{LargeLexp} for continuity between small-$L$ and large-$L$ regimes.  Hence we have found a \emph{unique} choice of boundary conditions for the principal chiral model for which one can expect that the physics at small-$L$ should be smoothly connected to physics at large-$L$.   We conjecture that with the $\mathbb{Z}_N$-symmetric background holonomy,  the PCM on $\mathbb{R} \times S^1_{L}$ has an adiabatic small-$L$ limit.  If $L$ is small enough, the theory remains weakly coupled at long distances, as illustrated in Fig.~\ref{fig:CouplingSketch}.  This is the counter-part of the adjoint-Higgs/Wilson line branch in gauge theories.

\section{Structure of the perturbative series on small $\R \times S^1_L$ }
 \label{sec:pert-th}


Now that we have devised a weak-coupling limit on $\R \times S^1$ adiabatically connected to 
the theory on  $\R^2$,  we can expect long-distance observables to be calculable using semiclassical methods, with a trans-series  representation  of the form
\begin{align}
\langle \mathcal{O}[\lambda] \rangle = \sum_{n=0}^{\infty} p_{0,n}  \lambda^n + \sum_{c} e^{-S_c/\lambda} \sum_{n=0}^\infty \sum_{k=0}^n p_{c, n,k} \lambda^{n} \left[\log(\lambda^{-1})\right]^{k} ,
\label{eq:ResurgentTransSeries}
\end{align}
where $\lambda = \lambda(1/NL)$. The reason for the inclusion of the $\log \lambda^{-1}$ terms will become apparent in Section~\ref{sec:Resurgence}.

As we mentioned in Section \ref{sec:IR-ren} the leading singularity, along the positive real axis of the Borel plane, for the $\mathbb{R}^2$ theory is the IR-renormalon singularity and it is located at $1/N$ of the uniton action.   On $\mathbb{R}^2$, the theory is strongly interacting, and there is no semi-classical interpretation of this singularity.  Once we put the theory on $\mathbb{R}\times S^1$ and go to the adiabatic small-circle limit described in Sec.~\ref{sec:SmallLLimit}, the Borel singularity positions should move in a smooth way relative to their locations in the $\mathbb{R}^2$ limit\footnote{With a thermal compactification, one can show that the renormalon singularities simply disappear, consistent with our claim that the high temperature limit is not smoothly connected to the theory on $\mathbb{R}^2$.}.  Since the adiabatic small-$L$ limit we constructed in Sec.~\ref{sec:SmallLLimit} is weakly coupled, however, we should be able to see renormalon ambiguities in the large-order behavior of perturbation theory in $\lambda[1/NL]$, and reproduce at least the factor of $N$ in the expression above.  We now demonstrate that this is indeed the case. 

The quickest route to see the semi-classical realization of the IR renormalons involves working out the perturbative series describing the contributions  to the ground state energy $E(\lambda)$ from fluctuations around the $P$ saddle:
\begin{align}
\mathcal{E}(\lambda) = E(\lambda) \xi^{-1}= \sum_{n=0}^{\infty} p_n \lambda^n\,,
\end{align}
where we have chosen to write the ground state energy in the natural units of the problem, which turn out to be $\xi = 2\pi/(NL)$.   The ground state energy will receive contributions from modes with momenta $Q > 1/(NL)$, as well as modes with $Q < 1/(NL)$. 

The effective coupling constant for the UV modes with $Q > 1/(NL)$ is $\lambda(Q)$, and runs logarithmically, as can be seen in Fig.~\ref{fig:CouplingSketch}.   Such modes will produce factorially-growing and \emph{sign-alternating} contributions to the perturbative coefficients $p_n$.  More precisely, the sign-alternating part of the perturbative series is
\begin{align}
p_n g^{2n} \sim (-1)^n \frac{n!}{\left(\frac{8\pi}{g^2N}\right)^n} .
\end{align} 
In the phase space integration over high momenta, the dominant 
contribution aries from the UV scale 
 \begin{align} 
  \textbf{UV-renormalon support:}\;\; 
k_{\rm UV}^* \sim Q  e^{n} \gg \Lambda \;\; \iff \;\;  \ell_{\rm UV}^*  \sim \frac{1} {k_{\rm UV}^*} 
\sim LN  e^{-n}\ll \Lambda^{-1}
\end{align}
 where  $Q$ is some large external momentum.  
 These are associated with  singularities  on the negative real axis in the Borel plane, and are called the UV renormalons\cite{Beneke:1998ui}.  The UV renormalons can not be affected by compactification, because   $\ell_{\rm UV}^* \sim    LN  e^{-n} \ll LN$.  In other words, UV renormalons probe only the high-momentum behavior of a theory and do not care if the space is compactified or not. So it is natural to see them appear unchanged in the compactified theory.  Their presence is not related to the particular regularization used to compute the phase space integration over high momenta and the renormalons can be recovered also on the latticized theory \cite{Bauer:2011ws,Bali:2013pla,Bali:2014fea}.

The contributions from modes with $Q < 1/(NL)$ are a different and more subtle matter.  As illustrated in Fig.~\ref{fig:CouplingSketch}, the relevant coupling for these modes is effectively $\lambda(1/NL)$, which 
ceases to run at energies lower than $1/LN$. {\it Naively}, the absence of running would mean that 
the IR-renormalon should disappear. For example, if we take the PCM on $\R^2$ with $N_f >N_f^{\star}$,  where $ N_f^{\star}$ is the value of $N_f$ above which the theory is conformal in the IR, the coupling stops running in the IR as well, and this was our reason for asserting in Section~\ref{sec:IR-ren} that  the IR-renormalons must disappear when $N_f> N_f^{\star}$. 
So why is it that the freezing of the coupling at a small value in the infrared, which leads to the absence of IR-renormalon singularities when it happens on $\R^2$, does not lead to the same effect on $\R \times S^1$?   Are we not running into a contradiction?

In fact, this is precisely at the heart of our adiabatic continuity idea. First, on $\R^2$, 
 the dominant contribution to the renormalon singularities  comes from the scale 
\begin{align} 
  \textbf{IR-renormalon support:}\;\;   k_{\rm IR}^* \sim Q e^{-n} \ll \Lambda \;\; \iff \;\; 
    \ell_{\rm IR}^*  \sim \frac{1} {k_{\rm IR}^*} \sim Q^{-1}  e^{n} \gg \Lambda^{-1}
\end{align}
which is exactly the scale where the theory becomes strongly coupled and perturbation theory becomes unreliable.  Studying the Borel resummation of perturbation theory, as briefly summarized in Section~\ref{sec:IR-ren}, one finds the leading-order  ambiguity
$ \pm i  {\Lambda^2 Q^{-2}} \sim \pm i e^{-\frac{8 \pi}{g^2N(Q)}} $  for processes involving a large external momentum $Q$. Note that this is of order $ [e^{-\frac{8 \pi}{g^2(Q)}}]^{1/N}$ where $e^{-\frac{8 \pi}{g^2(Q)}} \sim \Lambda^{2N} Q^{-2N}$ is the uniton amplitude at momentum scale $Q$. 

Because of the strong coupling problem, there is no semi-classical interpretation for the IR renormalons on $\mathbb{R}^2$.  And indeed, until very recently it was widely believed that there is {\it never} a semi-classical interpretation for IR renormalons. However, 
the more useful/refined question to pose is actually the following.  
\begin{itemize}
\item Start with an asymptotically free  theory on $\R^d$ which has IR renormalon singularities.  
Compactify on   $\R^{d-1} \times S^1$ and dial radius of $S^1$ to a small value.  What happens to the renormalon singularities if we are able to work with a compactification which adiabatically connects the theory on $\mathbb{R}^2$ to a  regime where the long distance dynamics becomes weakly coupled?\footnote{It is conceivable that there may be alternative setups to compactification in which an analogous question can be posed.}
\end{itemize}
Below, we show that in the adiabatic small-$L$ limit of the PCM, we can describe the long-distance physics using a small-$L$ effective field theory, which is a quantum mechanical system. This theory has ordinary 1d instantons  whose actions are $\frac{1}{N}$ that of the unitons on $\R^2$.  We refer to these 1d instantons as fractons, since they are the fractionalized constituents of the uniton saddle from a 2d point of view.  

We then show that the  IR renormalon singularities on $\R^2$ which arise from phase space integration and which are located at $\sim  \frac{S_{\rm uniton}} {N}$ 
transmute into  semi-classical correlated fracton-anti-fracton saddles on small $\R \times S^1_L$ which are also located at $\sim  \frac{S_{\rm uniton}} {N}$.   The crucial factor here is the appearance of $\frac{1}{N}$, 
which is the characteristic of renormalons,  but the exact location of the singularities will generally change mildly between the semi-classical and non-semi-classical regimes. 
This is what is guaranteed to hold via  adiabaticity.  So we have reached the same conclusion in PCM  as was found in deformed YM theory, QCD(adj), and the
$\mathbb {CP}^{N-1}$ model\cite{Argyres:2012vv,Argyres:2012ka,Dunne:2012ae}, but in a theory which has no topologically-stable stable points!  

\subsection{Covering $ SU(N)$ with $SU(2)$s} 

\noindent Before discussing the construction of the leading term in the small-$L$ effective field theory, we make some remarks on the organization of perturbation theory more generally.  The most common way to set up a perturbative calculation would be to write $U = e^{i W}, W \in \mathfrak{su}(N)$, expand the action in powers of $W$, and then do perturbation theory in terms of $W$.  Indeed, this was how we organized our calculation in Section~\ref{sec:SmallLLimit}.  From here onward, however, we find it advantageous to organize our perturbative calculations in a different way.  We will examine the contributions to perturbative observables from field fluctuations within $SU(2)$ subgroups of $SU(N)$, and then sum over the various $SU(2)$ subgroups.  \footnote{This  is the process by which which gauge configurations are generated in lattice Monte Carlo simulations for $SU(N)$ Yang-Mills theory\cite{Cabibbo:1982zn}, and is known to cover the whole  $SU(N)$ manifold.}

 Consider the embedding of $SU(2)$ into $SU(N)$ as two by two diagonal blocks:
\begin{equation}
  U_ {j}^{(1)} =    \left ( \begin{array}{cccccc }
1 & &&&  &\cr
  & \ddots  &&& &  \cr
  &&  
   \left ( \begin{array}{ll}
U_{j,j} &   U_{j,j+1}\cr
U_{j+1,j} & U_{j+1, j+1}
    \end{array} \right) 
   &&  & \cr 
      &&&& \ddots & \cr
      &&&& &1
  \end{array} \right) = \left ( \begin{array}{cccccc }
1 & &&&  &\cr
  & \ddots  &&& &  \cr
  &&  
   \left ( \begin{array}{cc}
z_1 &   i z_2 \cr
  i \bar z_2 & \bar z_1
    \end{array} \right) 
   &&  & \cr 
      &&&& \ddots & \cr
      &&&& &1
  \end{array} \right) \,,
  \label{eq:USimple}
\end{equation}
 and
 \begin{equation}
  U_{N}^{(1)}=      
   \left ( \begin{array}{ccccc}
U_{1,1}   &  & & &  U_{1,N} \cr
 &  1  &  &&  \cr
 &  & \ddots  &  &  \cr 
  &  &  & 1 &  \cr 
U_{N, 1} & & && U_{N,N}
    \end{array} \right)  = 
     \left ( \begin{array}{ccccc}
\bar z_1 &  & & &  i  \bar  z_2 \cr
 &  1  &  &&  \cr
 &  & \ddots  &  &  \cr 
  &  &  & 1 &  \cr 
  i  z_2 & & &&   z_1
    \end{array} \right) \,,
  \label{eq:UAffine}
\end{equation}
where $z_1, z_2 \in \mathbb{C}$, and $ |z_1|^2 + |z_2|^2 =1$.   

One can associated  the affine root system of the associated  $\mathfrak{su}(N)$ Lie algebra with each of these embeddings,   $U_ {j}^{(1)} , j=1, \ldots, N$, as the root associated with the corresponding $SU(2)$.  The affine root system consists of  the $N-1$ simple roots $\alpha_{i}, i=1, \ldots, N-1$ along with the affine root  $\alpha_{N}$:
\begin{align}
\alpha_{j} =   \left ( \begin{array}{ccccccc}
0_1 &  0_2 & \ldots & 1_{j} & -1_{j+1} &\ldots &0_N
    \end{array} \right),  \;\; 
      \qquad   
    \alpha_{N} =   \left ( \begin{array}{ccccccc}
-1_1 &  0_2 & \ldots & 0 & 0 &\ldots &1_N
    \end{array} \right) \,.
\end{align}
While the $N$ distinguished $SU(2)$ subgroups above will be the most important for our analysis, note that the generic $SU(2)$ subgroup  looks like
 \begin{equation}
  U_{j}^{(k)}=      
   \left ( \begin{array}{cccccc }
1 & &&&  &\cr
  & \ddots  &&& &  \cr
  &&  
   \left ( \begin{array}{ccccc}
U_{j,j}   &  & & &  U_{j,j+k} \cr
 &  1  &  &&  \cr
 &  & \ddots  &  &  \cr 
  &  &  & 1 &  \cr 
U_{j+k, j} & & && U_{j+k,j+k}
    \end{array} \right)
   &&  & \cr 
      &&&& \ddots & \cr
      &&&& &1
  \end{array} \right)
  = \left ( \begin{array}{cccccc }
1 & &&&  &\cr
  & \ddots  &&& &  \cr
  &&  
   \left ( \begin{array}{ccccc}
z_1 &  & & &  i z_2 \cr
 &  1  &  &&  \cr
 &  & \ddots  &  &  \cr 
  &  &  & 1 &  \cr 
  i \bar z_2 & & &&  \bar z_1
    \end{array} \right)
   &&  & \cr 
      &&&& \ddots & \cr
      &&&& &1
  \end{array} \right) \,.
  \label{eq:GeneralSU2Embedding}
\end{equation}
Each one of these $SU(2)$ subgroups can be associated with sums of roots in the affine Lie algebra which are themselves roots.

To write down the actions describing the fluctuations within the $U_{i}^{(k)}$ $SU(2)$ sub manifolds, we need to choose a parametrization of $SU(2) \simeq S^3$.   We use the  
 Hopf coordinate parametrization:
\begin{align}
U \equiv    \left ( \begin{array}{cc}
z_1 &   i z_2 \cr
  i \bar z_2 & \bar z_1
    \end{array} \right) 
\qquad \text{where}  \qquad 
\begin{array}{ll}
z_1& = r_1  e^{i \phi_1}   =   \cos \theta  e^{i \phi_1}\,,  \cr
z_2&= r_2  e^{i \phi_2}   =  \sin  \theta  e^{i \phi_2} \,,
\end{array} 
\label{eq:HopfCoordinates}
\end{align}
and we take the angular variables to have the ranges $\theta \in [0, \pi], \phi_1 \in [0, \pi], \phi_2\in [0, 2\pi]$.  For fixed $\theta$, the variables $\phi_1, \phi_2$ parametrize a torus. At the degeneration points,   $\theta =0, \pi/2, \pi$, the torus  shrinks to a circle.   With this parametrization, for each one of our $SU(2)$ embeddings, using the bosonic part of Eq.~\eqref{eq:GaugedAction}, we find that the fluctuations are described by 
\begin{align}
 S_{i}^{(k)} &=
\frac{1}{ g^2} \int_{\R \times  \mathbb S^1}  dtdx\,\left[ (\partial_\mu \theta)^2 + \cos^2 \theta   (\partial_\mu \phi_1)^2 +  
\sin^2 \theta   (\partial_\mu \phi_2 +  \xi_{i,k} \delta_{\mu,x})^2\right]\,,
\end{align} 
where 
\begin{align}
\xi_{i,k} = \frac{ 2 \pi \mu_{i,i+k}} {L} = \frac{ 2 \pi ( \mu_{i+k} - \mu_i)}{L}\,.
\label{eq:xik}
\end{align}
  For the $\Z_N$-symmetric background, $\xi_{i,k} = 2\pi k /(NL)$ for all $i=1,...,N$ (using affine roots when needed)\footnote{
This action is identical to the one  of the $\mathbb C \mathbb P^{1} $ model given in \cite{Dunne:2012ae}, except for an overall factor of $2$, if one forgets about the $\phi_1$ coordinate and sets $k=1$. 
This follows from viewing $S^3$ through the lens of the Hopf fibration 
$S^1 \rightarrow S^3 \rightarrow S^2 \sim \mathbb C \mathbb P^{1}$
where $S^3$ is the total space and $S^1$ is fibered over $S^2$.}.

\subsection{Large order behavior and  Stokes phenomenon}
\label{sec:HighOrderBehavior}
To efficiently deduce the large-order behavior of the perturbative contributions to $\cal{E}$ when $NL\Lambda/2
\pi \ll 1 $, we will use a small-$L$ effective field theory.  
The IR properties of the theory for small enough $L$ can be calculated from a theory where one integrates out modes with energies larger than the scale $(NL)^{-1}$, up to a subtlety which we comment on below.   Therefore, the effective field theory is {\it forgetful} of UV-renormalon singularities, which can only be extracted from the microscopic theory. 
The leading term of the action of the  small-$L$ effective field theory is 
\begin{align}
S&=\frac{L}{2g^2}\int dt \tr\left( \partial_{t} U \partial_{t} U^{\dag} +  [H_V, U] [H_V, U^{\dag}] \right) \\
&\to \frac{L}{g^2}\int dt\, \Big[  \dot \theta ^2 +   \cos^2 \theta \dot \phi_1^2  +    \sin^2  \theta    \dot \phi_2^2
+ {\xi_{i,k}}^2 \sin^2 \theta     \Big]\,,
  \label{eq:actionQM}
\end{align} 
where $ \xi_{i,k}$ is the same as Eq.~\eqref{eq:xik}.
  The expression in the second line comes from a restriction to a generic  $SU(2)$ subgroup of $SU(N)$, given in Eq.~\eqref{eq:GeneralSU2Embedding}.
   This leading term of the small-$L$ EFT is  a $0+1$ dimensional field theory, which is just quantum mechanics.  
 
An important subtlety with the derivation of Eq.~\eqref{eq:actionQM} is that it is not quite true that all KK-momentum carrying states decouple at low energies.  The issue is that states that carry non-zero \emph{winding number} can contribute to the low-energy dynamics on the same footing as states that carry zero winding number.  To see this, note that if we look at the contribution from $U_{i}^{(N-1)}$ and allow $\phi_2$ to carry $-1$ units of winding number, we find
\begin{align}
S &= \frac{L}{g^2}\int dt\, \Big[  \dot \theta ^2 +   \cos^2 \theta \dot \phi_1^2  +    \sin^2  \theta    \dot \phi_2^2
+ \left(\frac{-2\pi}{L} +\xi(N-1)\right)^2 \sin^2 \theta     \Big]\\
&= \frac{L}{g^2}\int dt\, \Big[  \dot \theta ^2 +   \cos^2 \theta \dot \phi_1^2  +    \sin^2  \theta    \dot \phi_2^2
+ \xi^2 \sin^2 \theta     \Big]\,,
\end{align} 
where $\xi = 2\pi/NL$.  This contribution is in fact associated precisely with the affine root of the $\mathfrak{su}(N)$ Lie algebra, and its relevance is due to the compactness of the $\Z_N$-symmetric background holonomy.  So there are contributions  to the low-energy physics from some configurations with winding number $-1$ which are of the same magnitude as contributions from field configurations that carry winding number $0$.  This is an illustration of the fact that the adiabatically-compactified theory ``remembers" its two-dimensional nature even when $L$ is very small.  This subtlety will also be important in the analysis of non-perturbative saddle points in the next section.

To understand the high-order behavior of perturbation theory in $g^2[1/NL]$ using the leading order part of the EFT action --- that is, leading order in an $ N L \Lambda/2 \pi \ll 1 $ expansion --- it turns out to be useful to temporarily move back to Minkowski space and use Hamiltonian methods, because it lets us use some known results from the literature. At small-$L$ the Hamiltonian associated with Eq.~\eqref{eq:actionQM} is
\begin{align}
H = \frac{g^2}{4L} P_{\theta}^2 + \frac{L\xi_k^2}{g^2} \sin^2 \theta + \frac{g^2}{4L \sin^2 \theta} P_{\phi_1}^2+ \frac{g^2}{4L \cos^2 \theta} P_{\phi_2}^2\,.
\label{Ham-master}
\end{align}
 We emphasize that this Hamiltonian describes the contributions to the energies of the states from the $N$ $SU(2)$ subgroups $U_{j}^{(1)}, j=1,\ldots, N$. In general, the way we have organized perturbation theory, we must sum over the contributions from all $SU(2)$ subgroups to compute the energies of states in the full $SU(N)$ PCM.  Fortunately, we will see that the $U_{j}^{(1)}$ subgroups make the dominant contribution to e.g. the ground state energy at small-$L$, which simplifies our analysis.   Also, to avoid confusion, we note that with the center-symmetric background gauge fields turned on, each of the $U_{j}^{(1)}$ subgroups will make the same contribution to $\mathcal{E}$.  If the background fields were to be slightly perturbed away from the $\mathbb{Z}_N$ symmetric points this would no longer be the case; the contributions are in principle distinguishable.  
 
Having made the point about the importance of taking into account the various $SU(2)$ subgroups, we present the subsequent calculations with $\xi \equiv \xi_1 = 2\pi/(NL)$ to lessen the notational clutter. The dependence on $\xi_k$ always follows from the replacement  $\xi \to \xi_k$.

The potential for $\theta$ has minima at $\theta=0, \pi$, and the action has a discrete symmetry $P$ which acts as $P: \theta \to \pi -\theta $.  The states of the system will have well-defined eigenvalues $\pm 1$ under $P$.  Moreover,  since $\phi_1, \phi_2$ are cyclic compact coordinates, associated to quantized conserved charges, a state with the quantum numbers of $(\pm, n_1, n_2)$ will have an energy of order 
\begin{align}
E_{\pm, n_1, n_2} \sim g^2 \xi (n_1^2 + n_2^2)
\sim \frac{ g^2}{NL} (n_1^2 + n_2^2)\,.
\end{align}
We expect the ground state and first excited states to have the quantum numbers $(+, 0, 0)$, and $(-,0,0)$, with an excitation energy non-perturbatively  small compared to the scale $1/NL$.  

To compute the energy of the ground state as a function of $g^2$, we  use the Born-Oppenheimer (BO) approximation familiar from atomic physics.  The Born-Oppenheimer approximation states that the lowest excitation energies of a coupled quantum system can be extracted by solving the Schr\"odinger equation for the `light' degrees of freedom (in our case, the dynamics of $\theta$ field) with the `heavy' degrees of freedom (the dynamics of the $\phi_1, \phi_2$ fields) frozen to their energy-minimizing values.   Essentially, one is exploiting a separation in energy scales between the two sets of degrees of freedom to solve for the behavior of the heavy variables classically, and then treating the light variables quantum-mechanically.   In our case, we will see that the separation in energy scales between states with $n_1, n_2 \neq 0$ and $n_1=n_2=0$ is exponential in $g^2$, so the use of the Born-Oppenheimer is parametrically well-justified when $g^2\ll 1$, as is the case when $NL$ is small compared to $\Lambda^{-1}$.  So in the Bohr-Oppenheimer approximation, we only need to compute the ground state energy in the $n_1 = n_2 = 0$ sector of the theory, described by the Hamiltonian 
\begin{align}
\label{eq:BOHamiltonian}
H = \frac{g^2}{4L} P_{\theta}^2 + \frac{L\xi^2}{g^2} \sin^2 \theta \,.
\end{align}
  The Schr\"odinger equation can be written as 
 \begin{align}
\textstyle{ \left[  \frac{d^2}{d \theta^2} + p + \frac{ 2  \xi^2}{g^2} \cos (2 g \theta) \right] \psi =0,  \qquad p= 4E - \frac{ 2\xi^2 }{g^2}}, \qquad \theta \in [0, \pi]\,,
\label{form1}
\end{align}
where we have set $L=1$.
 It is convenient to define  a dimensionless Hamiltonian $\tilde H = H \xi^{-1}$,  and bring the kinetic term into canonical form by the change of variables $\theta= \sqrt \frac{g^2}{2 \xi} \tilde \theta$. Then  the Hamiltonian reads 
 \begin{align}
 \label{canonical}
\textstyle{\tilde H =  - \frac{1}{2}   \frac{d^2}{d \tilde \theta^2} +  \frac{   \xi}{2g^2} \cos \left( \sqrt  \frac{2g^2}{\xi}   \tilde \theta \right)\,. }
 \end{align}
 This form is same as the Hamiltonian in Zinn-Justin's text, Section~41.2 \cite{ZinnJustin:2002ru}  with the identification $g_{\rm ZJ}= \frac{2 g^2 }{\xi}$.  
 
\begin{figure}[tbp]
\centering
\includegraphics[width=0.5\textwidth]{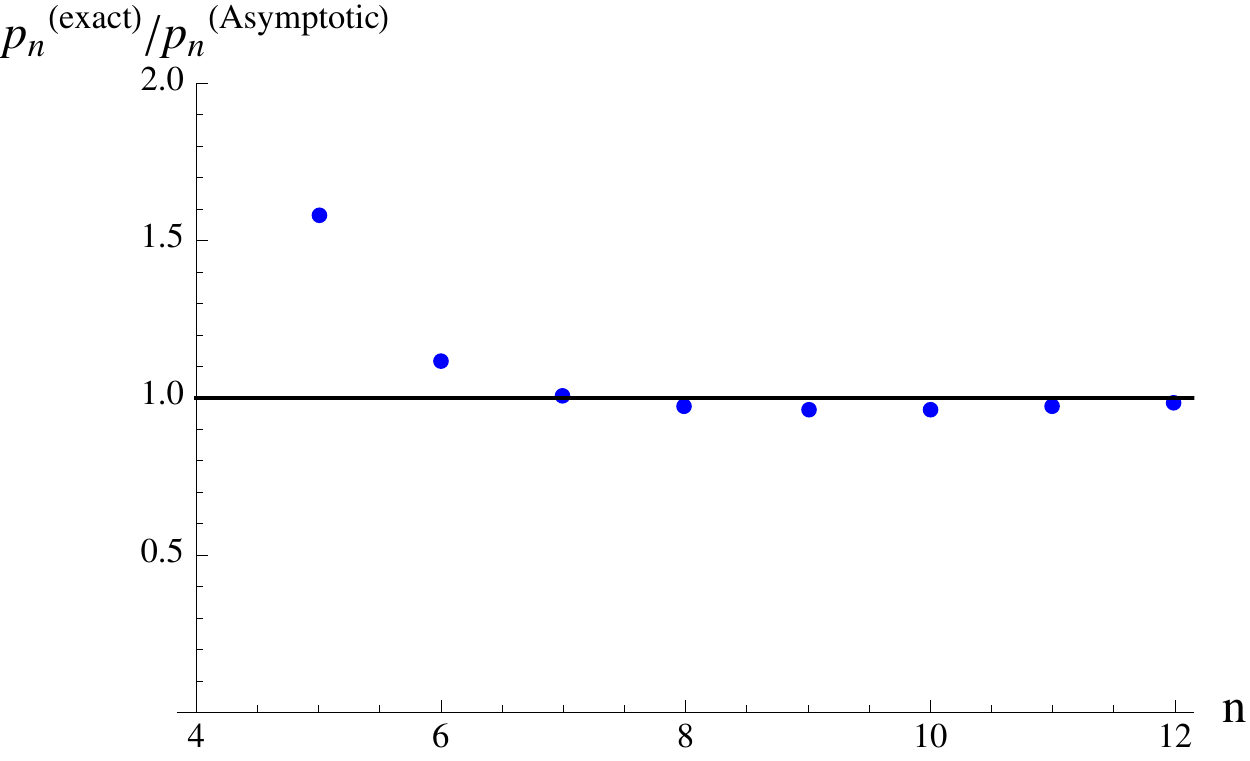}
\caption{ The  ratio of the exact value  of the pre-factors in  perturbation theory  to the asymptotic formula,  $\frac{ p_{n}^{\rm exact}}{p_{n}^{\rm asymptotic}}$.}
\label{fig:coefficient}
\end{figure}

The energy of the ground state now follows from the $P = +$ solution, and 
 the large-order behavior of this solution was already determined by Stone and Reeve\cite{Stone:1977au} using the methods developed by Bender and Wu\cite{Bender:1969si,Bender:1971gu}. From  \cite{Stone:1977au} we see that the contributions of each of the $U_{(i)}^{1}$ in the $SU(2)$ subgroups to the perturbative series for the ground state energy, $\mathcal{E}_i$, behave as
\begin{align}
\label{pt-low}
&\mathcal{E}_i(g)  = E_i(g) \xi^{-1} \\
&=  \textstyle{   \frac{1}{2}-  \frac{1}{2} \left(\frac{g^2}{8 \xi}\right) -  \frac{1}{2}  \left(\frac{g^2}{8 \xi}\right)^2 -  \frac{3}{2} \left(\frac{g^2}{8 \xi}\right)^3 - \frac{53}{8}  \left(\frac{g^2}{8 \xi}\right)^4  - \frac{297}{8} \left(\frac{g^2}{8 \xi}\right)^5    - {3961 \over 16}  \left(\frac{g^2}{8 \xi}\right)^6   - {60727  \over 32}  \left(\frac{g^2}{8 \xi}\right)^7} \cr& \textstyle{- {2095501 \over 128} \left(\frac{g^2}{8 \xi}\right)^8  - {20057205 \over 128} \left(\frac{g^2}{8 \xi}\right)^9  - { 421644859 \over 256} \left(\frac{g^2}{8 \xi}\right)^{10} -  {4835954237 \over 256}  \left(\frac{g^2}{8 \xi}\right)^{11} + \quad \ldots } \nonumber
\end{align}
with an asymptotic form given by 
\begin{align}
\mathcal{E}_i(g) = \sum_{n=0}^{\infty} a_{n}^{[0]} g^{2n}, \; \qquad  p_{n} \sim  -\frac{2}{\pi} \left(\frac{1}{8\xi }\right)^{n} n! 
\left(1- \frac{5}{2n}  - \frac{13}{8 n^2} +  \mathcal{O}(n^{-3})\right)\,.
\label{eq:LargeOrderPT}
\end{align}
We have confirmed the correctness of the asymptotic form (which includes sub-leading  corrections to the leading asymptotic) by numerical analysis. See Fig.~\ref{fig:coefficient} for an illustration of the rapid convergence of the series coefficients to their asymptotic large-order form. 

The expansion parameter of perturbation theory is   $g^2/(8 \xi) = g^2N/(16 \pi)$.  The perturbative  series is non-alternating and has factorially-growing coefficients, of the form $n!  \left[g^2/(8 \xi)\right]^n $. This means that the perturbative expansion is along a Stokes line, and its resummation is ambiguous. 
The analytic continuation of the Borel transform for the leading  order $n!$ divergence is given by 
\begin{align}
B \mathcal{E}_i(t) = - \frac{2}{\pi} \frac{1}{1-\frac{t}{8 \xi}} \,.
\end{align}
Hence the perturbative series will not be Borel summable along $\R^{+} $ in Borel plane, due to the (leading) singularity located at
$t_{*} = 8\xi  = \frac{16 \pi}{N}$.
If the integration contour defining the Borel-resummation of the series is deformed to pass either above or below the real axis near this singularity, so that it is avoided, one can obtain a finite result.   However, the result of the resummation will depend on the choice of contour, leading to a  two-fold ambiguity.  An equivalent way to think about this which we find particularly useful in working with the NP saddles is to analytically continue $g^2 \to g^2e^{i\epsilon}, \epsilon \in (-\pi,\pi) $.  Then so long as $\epsilon \neq 0$, the Borel sum converges, but the analytic continuation back to $\epsilon = 0$ depends on whether $\epsilon$ approaches $0$ from above the real axis or below the real axis.  This matches what one sees with the contour deformations.

Using either of these equivalent approaches we can define the lateral 
Eq.~\eqref{lateral} Borel resummation as in the simple example discussed in Sec.~\ref{sec:Borel-Stokes}.  In particular, the right  ${\cal S}_{0^+} {\mathcal{E}}_i$ and left 
${\cal S}_{0^-} {\mathcal{E}_i} $ Borel resummations yield 
\begin{align}
{\cal S}_{0^\pm} {\mathcal{E}_i}   & = \frac{1}{g^2} \int_{C^{\pm}}  \; dt e^{-t/g^2}  B \mathcal{E}_i(t) 
= P \frac{1}{g^2} \int  \; dt e^{-t/g^2}  B \mathcal{E}_i(t)    \mp  i \frac{16  \xi}{g^2} e^{-\frac{8 \xi}{g^2}}  \cr
&= \Re {\cal S}_{0} {\mathcal{E}_i}  \mp  i \frac{32 \pi}{g^2N} e^{-\frac{16 \pi }{g^2N}} \,,
\label{eq:Borel-res}
\end{align}
where $P$ stands for the (unique) Cauchy principal value.

\vspace{3mm}
\noindent{\bf  Stokes Jumps}
\vspace{3mm}

We can now determine  the action of the Stokes automorphism:
\begin{align}
\label{Stokes-auto}
({\cal S}_{0^+} -{\cal S}_{0^-}  ){\mathcal{E}}   = - i  \frac{32   \xi}{g^2} e^{-\frac{8 \xi}{g^2}}  = - i  \frac{64  \pi }{g^2N } e^{-\frac{16 \pi }{g^2N}}  \,.
\end{align}
Note that this expression is valid for each of the individual $SU(2)$'s embedded into $SU(N)$ as in Eq.~\eqref{eq:USimple}.  There are $N$ such embeddings associated with the $N$ roots of the affine Lie algebra in a center-symmetric background for which $\xi= \mu_{i+1} - \mu_i = \frac{2\pi}{N}$.  So there are $N$ leading-order ambiguities in the Borel resummation of perturbative series describing the fluctuations within the $U_{i}^{(1)}$ subgroups.  These $N$ ambiguities are identical when the background holonomy is $\Z_N$-symmetric, so that the overall leading ambiguity in the $SU(N)$ theory is just $N$ times the expression above\footnote{The ambiguities would become split if the $\Z_N$ symmetry were to be slightly broken.}.

Of course, there are also ambiguities coming from other embeddings of $SU(2)$ into $SU(N)$ such as the one shown in  
Eq.~\eqref{eq:GeneralSU2Embedding} which are associated with roots which can be written as positive linear combination of the simple roots. In this case,  with the $U_{j}^{(k)}$ embeddings in the center-symmetric background, the leading non-perturbative ambiguity is of order 
\begin{align}
\mp  i \frac{16  \xi k  }{g^2} e^{-\frac{8 \xi k }{g^2}}, \qquad k\geq 2\,.
\end{align}
As advertised, this is exponentially small compared to the leading ambiguities coming from fluctuations living within $U_{i}^{(1)}$.



\subsection{NP-data from the late terms  of the P-expansion}
Energy eigenstates in a real periodic potential must be real.  Yet we have just seen that the perturbative series generates a result which has an imaginary part upon Borel resummation. Furthermore, as seen in Eq.~\eqref{eq:Borel-res} the imaginary part is ambiguous.  A non-perturbative completion of this result which gets rid of both of these problems is the resurgent trans-series. See  \cite{Aniceto:2013fka} for a very explicit discussion. 
In fact, the NP-completion of the perturbative results, even in the case of simplest single parameter trans-series, leads to infinitely many NP-saddles.

If we assume that the semiclassical representation of observables in the PCM is resurgent,    the appearance of the ambiguity  in perturbation theory described above  implies that there must   exist some finite-action field configuration with action $S=\frac{16 \pi}{Ng^2}$ 
whose amplitude should be ambiguous in just the right way to cancel this leading ambiguity.   
 Another way to say this is that resurgence theory \emph{predicts} the existence of such non-perturbative saddle-points.  However, as we have mentioned before, naive topological considerations seem to leave no room for stable finite-action solutions in the PCM.    In the next section, we resolve this puzzle by showing the existence of such saddles, which in general are \emph{not}  classified according to the topological structure of the microscopic theory.


In the low energy effective field theory for the PCM on small $\R \times S^1$, there is a sense in which topology is {\it emergent}. Indeed, the periodic potential  Eq.~\eqref{eq:BOHamiltonian} has  one-dimensional instantons, which can be classified according to topology in quantum mechanics. However, in QM  and QFT,  it is well-known that instantons   cannot mix with perturbation theory.  The  first NP-saddles which can  mix with perturbation theory and fix its ambiguity  are of the instanton-anti-instanton form. 
Indeed, we should expect a left/right  $[\F \bar \F]_{\pm} = \Re [\F \bar \F] \pm  i  e^{-\frac{16 \pi}{Ng^2}}  $ correlated amplitude for some event with amplitude $\mathcal{F}$:  this is the only way the leading ambiguities of perturbation theory can be cancelled in the trans-series.  So resurgence theory and perturbation theory in the semi-classically calculable regime of the 2d PCM compactified down to $\R \times S^1$ predicts that there must exist a NP saddle with amplitude 
 \begin{align} 
\F  \sim   e^{ - \frac{4 \xi}{g^2 } }  \sim  e^{ - \frac{8 \pi}{g^2 N} }    \,. 
\label{1-fold}
\end{align}
Note that the action of this NP saddle in the compactified theory is $1/N$ of the uniton action.  Explicit field configurations with such actions are discussed in detail in the following Section.


\section{Non-perturbative saddle points}
\label{sec:FractonsUnitons}
We now begin our exploration of the non-perturbative features of the PCM at small $S^1$, with the boundary conditions/background gauge field corresponding to Eq.~\eqref{eq:MinSpatial}. As we have mentioned in the introduction, there are no stable topological defects in the theory on $\R^2$.  However, on $\mathbb{R} \times S^1$ with a $\Z_N$-symmetric background, the situation is different.   
The background holonomy  amounts to a potential on the $SU(N)$ target space, effectively modifying it to $U(1)^{N-1}$, the maximal torus of $SU(N)$, at energies low compared to $1/(NL)$.  
 This is the counterpart of adjoint Higgsing in gauge theories via a Wilson line, see e.g. 
 \cite{Davies:2000nw,Unsal:2007jx}.

As we briefly mentioned  in Section~\ref{sec:HighOrderBehavior}, this potential has isolated minima  on the group manifold. Consequently, there are tunneling events between these minima, which should admit a semiclassical Euclidean description as \emph{stable} instanton-like field configurations when $L$ is small enough and the theory is weakly coupled.  This happens despite the fact that the microscopic theory has a trivial $\pi_2$ homotopy group and hence has no instantons. 
This is the sense in which there is an emergent topological classification  in the low energy effective field theory. 

 We will give a careful classification of these new saddle points of the PCM path integral.  Moreover, we will see that these small-$L$ field configurations are the constituents of unitons, with the minimal uniton fractionalizing into $N$ constituents at small-$L$.  For this reason, we  refer to these small-$L$ stable saddle points as \emph{fractons}.  This terminology was originally  introduced in \cite{Shifman:1994ce} in the multi-flavor Schwinger model for fractionalized instantons.   The fractons play a critical role in the non-perturbative physics of the PCM on small $S^1$, and are responsible for the mass gap of the theory.  Moreover, they lead us to the resolution of the deep puzzle about the interpretation of renormalon ambiguities in the PCM. 

\subsection{Unitons}
\label{sec:Unitons}
We start with a description of the unitons, which are finite-action solutions of the second-order  Euclidean equations of motion \cite{uhlenbeck1989harmonic}. The reason that one should expect such solutions is that the $\mathbb{CP}^{N-1}$ manifold can be embedded as a totally geodesic submanifold into $SU(N)$ \cite{Eichenherr:1979hz}.  This means that solutions of the $\mathbb{CP}^{N-1}$ field equations can be lifted to solutions of the $SU(N)$ field equations.  However, while finite-action solutions on $\mathbb{CP}^{N-1}$ carry a topological charge, 
and  have no negative modes  while staying within the $\mathbb{CP}^{N-1}$ manifold, 
once the solution can evolve in the full $SU(N)$ target space there is no longer any conserved topological charge. Hence unitons are known to be ``unstable", which just means that the fluctuation operator around a uniton has negative modes \cite{Piette:1987ia}. 

It turns out that in general, not all the solutions of the PCM are obtained directly from the $\mathbb{CP}^{N-1}$ embedding, and in \cite{uhlenbeck1989harmonic} Uhlenbeck gave an exhaustive classifications of all possible solutions to the PCM equation of motion.  However, our interest here will be in what happens to the minimal-action uniton, which \emph{can} be obtained from a $\mathbb{CP}^{N-1}$ embedding, and has the action
\begin{align}
S_{\rm uniton} = \frac{8 \pi}{g^2} =  N  \times \frac{ 8 \pi }{ \lambda}\,.
\label{Suniton}
\end{align} 
This minimal-action uniton solution on $\mathbb{R} \times S^1$ with the boundary conditions determined by Eq.~\eqref{eq:MinSpatial} is associated with the minimal-action instanton of $\mathbb{CP}^{N-1}$, and takes the form
\begin{align}
U_{\rm uniton} = e^{i \pi/N} ( 1 - 2 {\mathbb P}),  \;   \qquad  {\mathbb P}_{ij} = \frac{v_i \, v^{\dagger}_j }{v^{\dagger} \cdot v}\,,
\label{eq:UnitonProj}
\end{align} 
where   ${\mathbb P}$ is a projector, ${\mathbb P}^2 = {\mathbb P}$. Here, we defined $v = \Omega(z) \cdot v_I$
\begin{align}
v_I &= \left(\begin{array}{c}1 \\ \lambda_N \\ \lambda_{N-1} \\
\vdots \\
\lambda_{3} \\
\lambda_2 + e^{2\pi z} \lambda_1 \end{array}\right) 
\end{align}
and 
\begin{align}
\Omega(z) &  =   e^{i \frac{\pi}{N} \nu }  \left ( \begin{array}{ccccc}
1 & &&& \cr
  & e^{ \frac{2\pi}{N} z} &&&  \cr
  && \ddots &&  \cr
      &&&& e^{  \frac{ 2\pi (N-1) z}{N}}
  \end{array} \right) \; ,  \qquad \;   \nu=0,1\; \textrm{for} \; N=\textrm{odd, even},
\end{align}
and $z =( x_1+ i x_2 )/L$, with $x_{1,2}$ coordinates on Euclidean $\mathbb{R}\times S^1$, while $\lambda_i \in \mathbb{C}$ are moduli of the solution. We note that $v(z)$ are precisely the twisted instanton configurations discussed in the context of the $\mathbb{CP}^{N-1}$ model in \cite{Bruckmann:2007zh,  Brendel:2009mp} and \cite{Dunne:2012ae,Dunne:2012zk}, with the correct twisted periodicity in $x_2$.

\begin{figure}[htbp]
\begin{center}
\subfigure{
\includegraphics[width=0.48\textwidth]{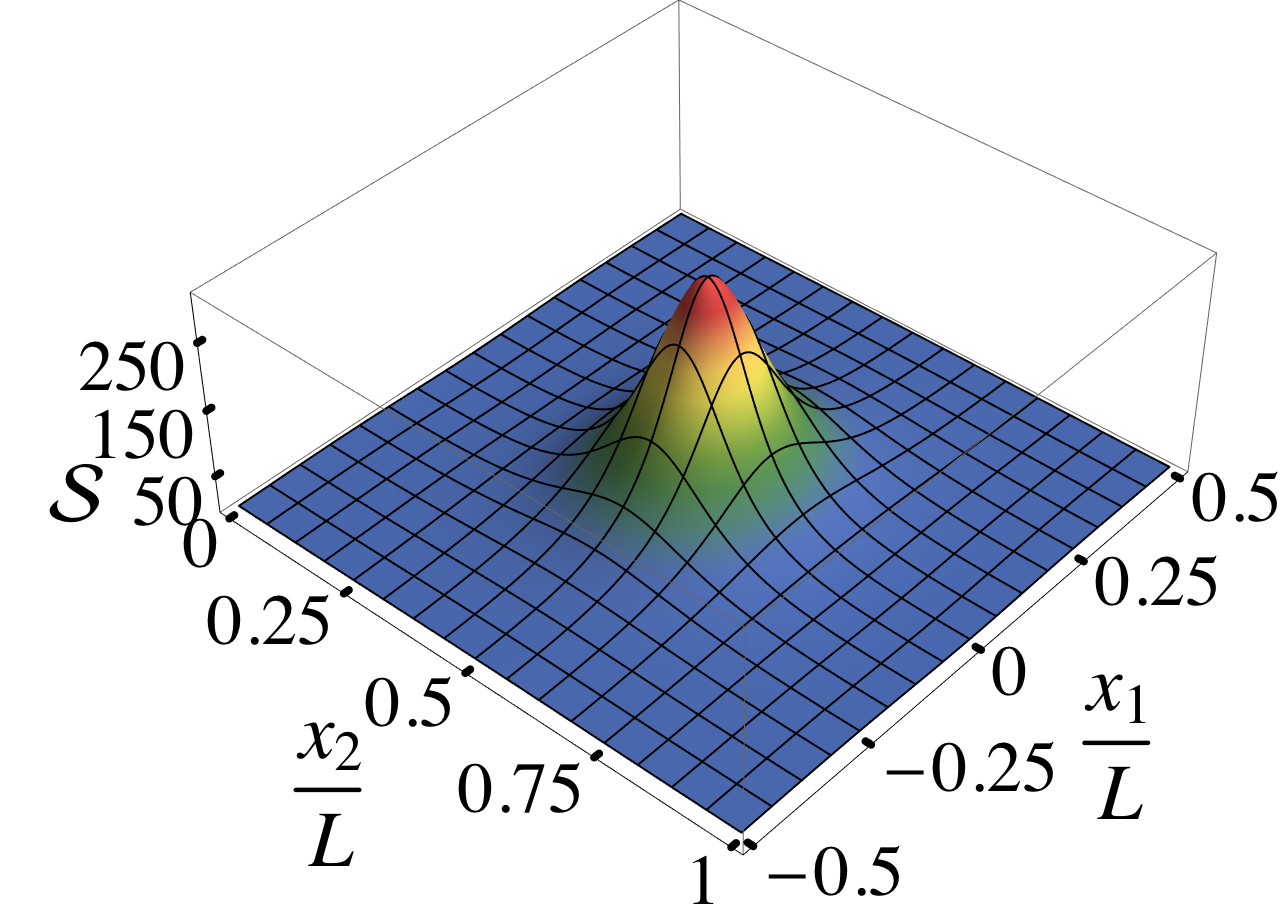}}
\subfigure{
\includegraphics[width=0.48\textwidth]{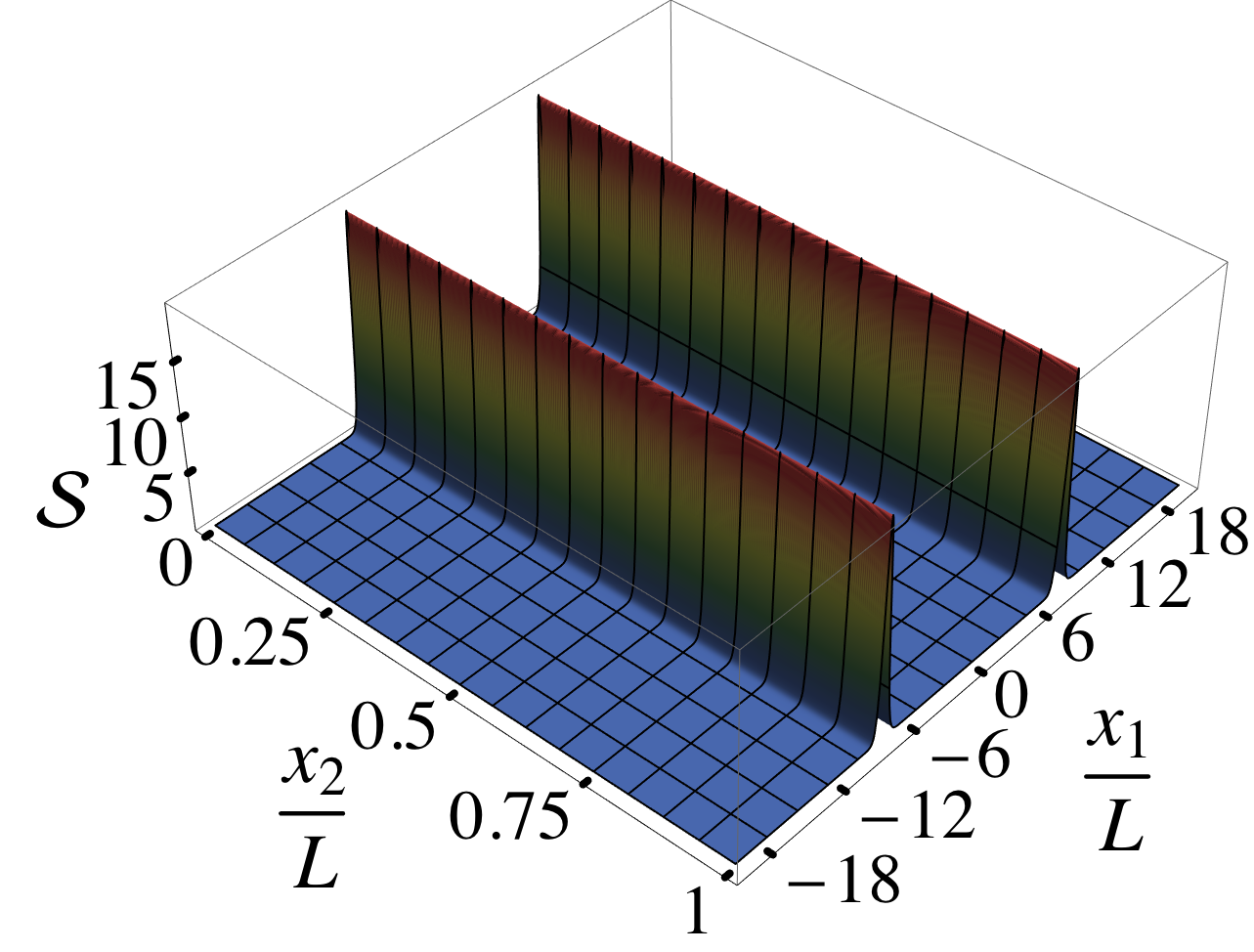}}
\caption{Action densities for the $SU(2)$ unitons: 
Left: small uniton. Right: Large uniton fractionalize to two fracton.
 }
\label{fig:SU2Uniton}
\end{center}
\end{figure}

\begin{figure}[htbp]
\centering
\subfigure{
\includegraphics[width=0.48\textwidth]{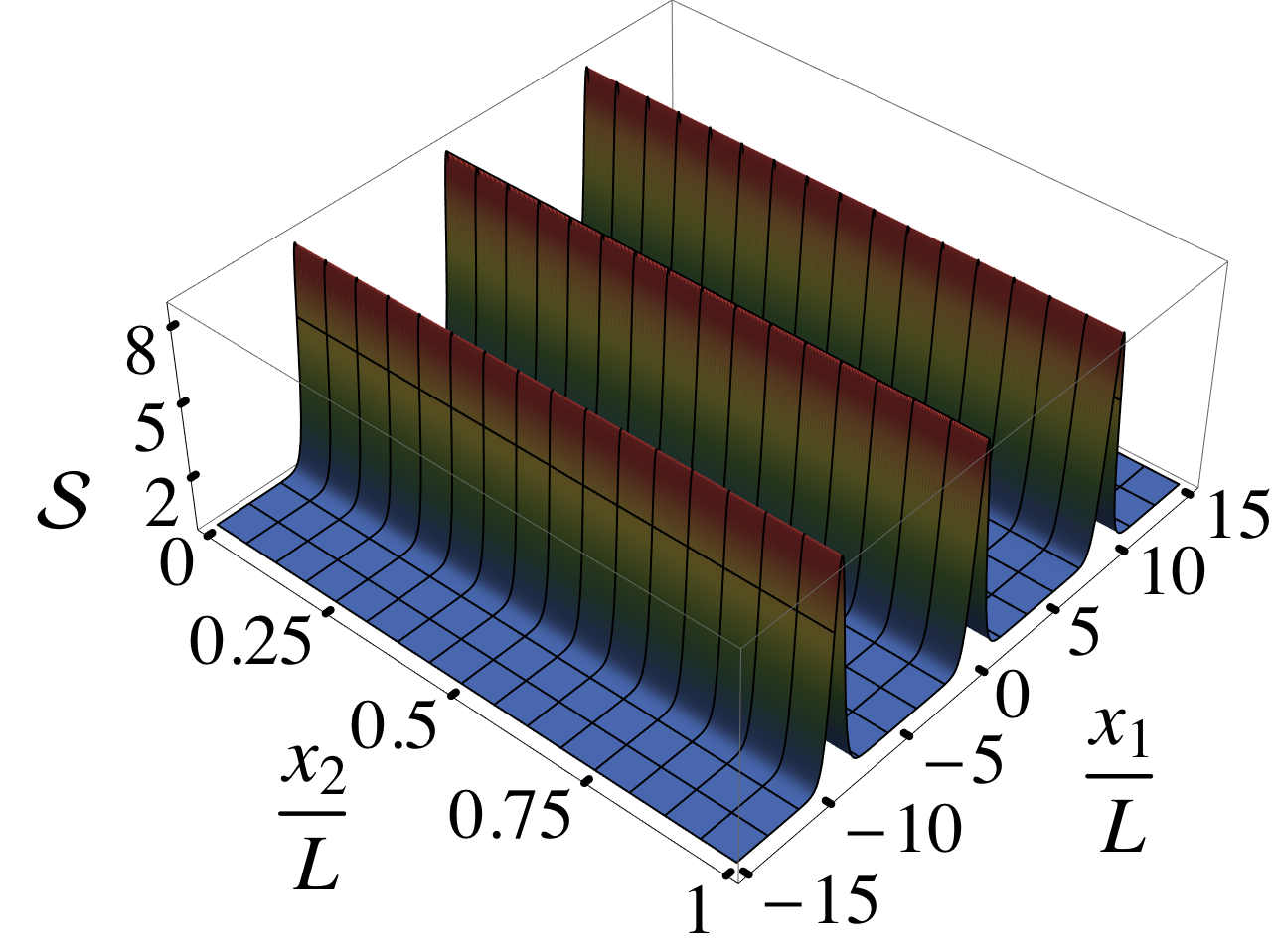}}
\subfigure{
\includegraphics[width=0.48\textwidth]{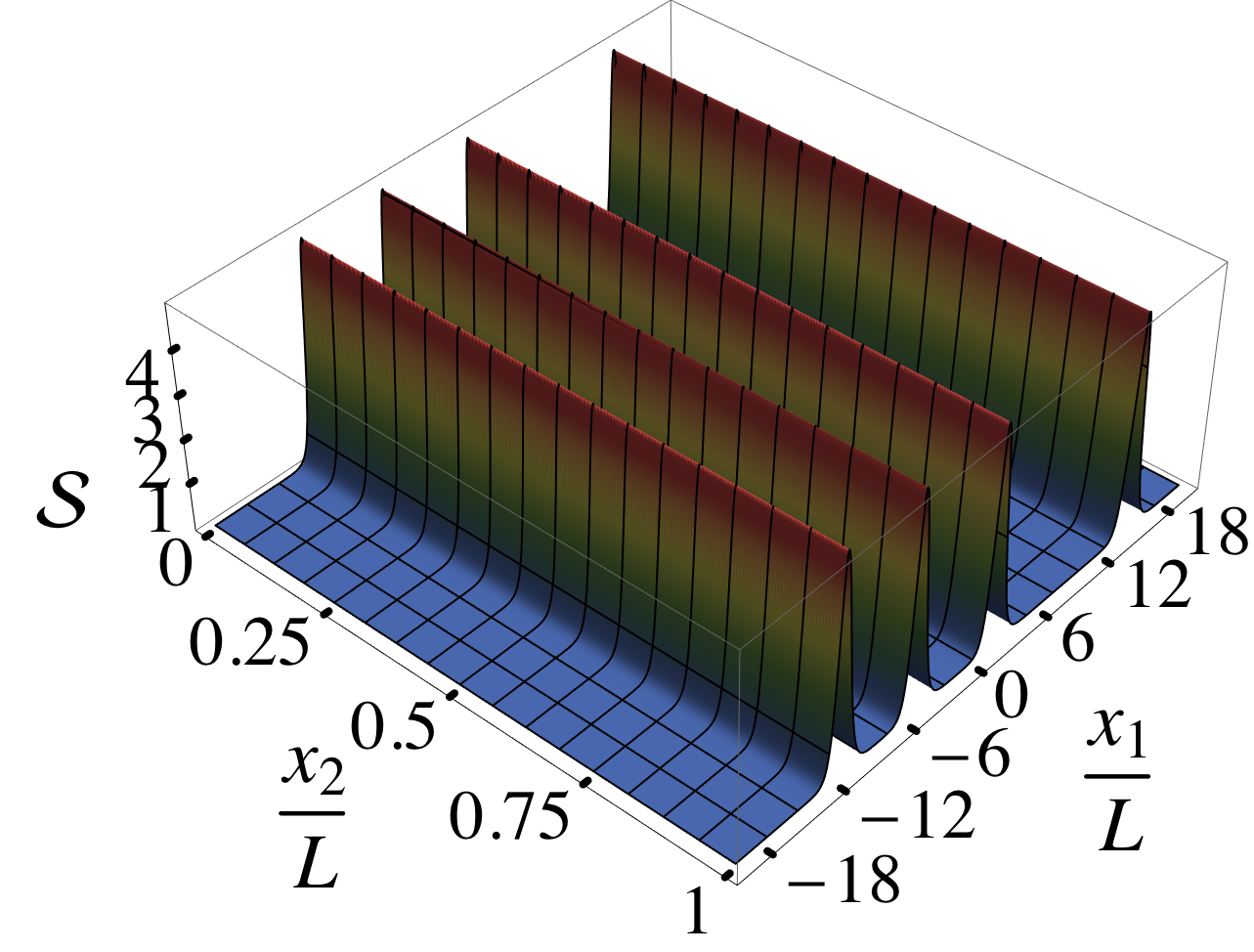}}
\caption{Action densities for the large $SU(3)$  and  $SU(4)$ unitons. They split to three and four fractons, respectively. 
 }
\label{fig:SU3SU4Unitons}
\end{figure}

 Once we work on $\mathbb{R}\times S^1$ with a center-symmetric background holonomy, we expect the emergence of some stable instanton-like field configurations, the fractons.   The fastest way to see  the emergence of the fractons is to  look at some plots of the action densities associated to these solutions as a function of the moduli $\lambda_{i}$ shown above\footnote{The uniton moduli $\lambda_i$ that we show in the text are inherited from the moduli of the $\mathbb{CP}^{N-1}$ instanton, but they are not the only moduli of the uniton solution, since there are also moduli associated with the embedding of $\mathbb{CP}^{N-1}$ into $SU(N)$. The full structure of the zero and quasi-zero modes of a uniton is not known.}.   It turns out that `small' unitons, which is to say ones whose characteristic size is small compared to $L$, resemble the profile of a uniton on $\mathbb{R}^2$.  An example of such a small-uniton configuration is shown on the left in Fig.~\ref{fig:SU2Uniton}, where the minimal uniton looks like a single lump of Euclidean action density, $\mathcal{L}_E$, centered near $x_1 = 0, x_2 = L/2$.  On the other hand, unitons which are large compared to $L$ tend to fractionalize, in the background  Eq.~\eqref{eq:MinSpatial}, 
  into multiple lumps, the fractons. The locations of the different fractons are controlled by the $\lambda_i$, with an example shown on the right of Fig.~\ref{fig:SU2Uniton}.   The number of lumps is at most $N$. \footnote{ These type of plots, in the context of  fractionalization of BPST instantons in the background of non-trivial holonomy,  
  goes back to the work of van Baal et.al. \cite{Kraan:1998pm} in QCD-like theories. For other works  emphasizing similar effects in various QFTs, see e.g. \cite{ Shifman:1994ce,GonzalezArroyo:1995zy,Lee:1997vp,vanBaal:2000zc,Bruckmann:2007zh,Collie:2009iz,Brendel:2009mp,Harland:2009mf,Parnachev:2008fy,Ogilvie:2012is}
   The amusing feature illustrated by the PCM is that the idea of fractionalization even works in the absence of topologically stable instantons!  
A more direct route to study the constituents of instanton appears in 
\cite{Davies:2000nw,Unsal:2008ch,Unsal:2007jx} by studying the theory in a weakly coupled calculable regime, see e.g.\cite{Thomas:2011ee,Nishimura:2011md,GarciaPerez:2009mg}
 }   More examples of fractionalization are shown for the $SU(3)$ and $SU(4)$ cases in Fig.~\ref{fig:SU3SU4Unitons} on the left and right respectively.

From the plots we see that in the fractionalized limit, the lumps are constant along the compact direction $x_2$, and seem to approach a vacuum configuration in between each other.  This strongly suggests that the PCM should have isolated fracton solutions\footnote{In \cite{Bolognesi:2013tya} it was shown that a direct extrapolation of classical field configurations from small-$L$ to large-$L$, using only the leading term in the small-L EFT action, is quite subtle for non-BPS solutions. On the other hand we know already that the PCM, in the decompactified $\mathbb{R}^2$ regime, becomes strongly coupled and one cannot recover the full quantum theory starting from classical field configurations.  Indeed, even before thinking about the strongly-coupled $L\to \infty$ limit, one can see that the small-L EFT action itself gets corrections as $L$ is increased.  In this paper, we confine our attention to the weakly-coupled small-$L$ limit, in which the subtleties explored in \cite{Bolognesi:2013tya} are not relevant. } for small-$L$, and this will be explicitly verified in the next section.

We also note that in the thermal compactification, at high temperatures, where the trivial holonomy of Eq.~\eqref{eq:MinThermal} must be used, the fractionalization of unitons does not occur.

\subsection{$SU(2)$ Fractons and  KK Fractons}
\label{sec:SU2Fractons} 
It is instructive to start our search for the fracton solutions by considering the $SU(2)$ PCM, in which case the group manifold is $S^3$.   The discussion of the fractons becomes most transparent if we parametrize the group element using Hopf coordinates as we did in Section~\ref{sec:pert-th}, 
 see Eq.~\eqref{eq:HopfCoordinates}.  The round metric on the 3-sphere in
 the Hopf coordinates is given by 
 \begin{align}
ds^2 = d\theta ^2 +   \cos^2 \theta d \phi_1^2  +    \sin^2  \theta    d\phi_2^2
\end{align}
and this fixes the kinetic term. From the action Eq.~\eqref{eq:GaugedAction},we see that a non-trivial potential term is induced on the $S^3$ manifold due to the background field determined by Eq.~\eqref{eq:MinSpatial}. The associated Euclidean Lagrangian on  $\R \times S^1$ is given by
\begin{align}
 S &=
\frac{1}{ g^2} \int_{\R \times  \mathbb S^1}  dx_1dx_2\,\left[ (\partial_\mu \theta)^2 + \cos^2 \theta   (\partial_\mu \phi_1)^2 +  
\sin^2 \theta   (\partial_\mu \phi_2 +  \xi \delta_{\mu,x_2})^2\right]\,,
\label{eq:HopfAction}
\end{align} 
where $\xi = 2\pi/NL = \pi/L$.


Given the fact that unitons which are large compared to $L$ split into $N$ configurations which appear to have flat profiles along the $S^1$, it is tempting to start by setting all Kaluza-Klein (KK) momenta in Eq.~\eqref{eq:HopfAction} to zero, and thus obtain an effective one-dimensional theory.  The usual reasoning behind such an approach is that field configurations carrying $n$ units of KK momentum have an energy density $n^2/L^2$, so for studying physics on length scales large compared to $L$ it is sufficient to focus on states with $n=0$, since these are parametrically lighter.  This would certainly be correct if $\xi$ were set to zero.  
However, as we already saw in the perturbative context, this approach is too naive for the PCM in the $\mathbb{Z}_N$-symmetric small-$L$ limit, since the small-$L$ theory actually `remembers' that it is  microscopically a \emph{two-dimensional} theory, and neglecting all configurations with non-zero winding number (and hence KK momentum) is not quite correct. It will turn out to be important to consider finite-action field configurations also carrying non-zero winding number for $\phi_2$.

 \vspace{3mm}
{\bf $\bf SU(2)$ Fractons:}  Let us start by looking for instanton configurations carrying \emph{zero} units of KK momentum in the compact direction $x_2$ for $\theta, \phi_1$ and $\phi_2$.  By setting $\partial_{x_2}  = 0$, the euclidean action for the low-lying modes reduces to 
\begin{align}
S&= \frac{L}{g^2}\int_{-\infty}^{+\infty} dx_1\, \Big[  \dot \theta ^2 +   \cos^2 \theta \dot \phi_1^2  +    \sin^2  \theta    \dot \phi_2^2
+ \xi^2 \sin^2 \theta     \Big] \,,
\end{align} 
where dotted quantities are derived with respect to $x_1$.
As expected,  the KK reduction lands us on a theory  with a non-trivial  potential on the  target space ${\cal T}=S^3$ due to presence of the $\Z_2$-symmetric background holonomy.
 
The Euclidean equations of motions associated with this action are 
\begin{align}
&\ddot \theta - \frac{1}{2}\sin 2 \theta [  (\dot \phi_2)^2 - (\dot \phi_1)^2  + \xi^2 ] =0\,,\\
&\partial_{x_1} \left(\cos^2\theta \dot{\phi}_1\right) =0\,, \\
&\partial_{x_1} \left(\sin^2 \theta \dot{\phi}_2\right) =0\,. 
\label{Eeom}
\end{align} 
Setting  $\phi_{1,2}$ constant in time solves the second and third equations while the first one reduces 
to the equation for a kink-instanton --- the fracton from the 2d viewpoint --- in one dimension:  
\begin{eqnarray}
&&\ddot \theta - \frac{\xi^2}{2} \sin 2 \theta  =0  \,.
\label{kinksecond}
\end{eqnarray} 
The existence of the {\it stable} fracton events  is  the crucial difference compared to  $\R^2$, or  to thermal compactification $\R \times S^1_\beta$  associated with the  trivial background holonomy. 

We can find the action of the instanton configuration by focusing only on the $\theta$ field, and using Bogomolny's method to rewrite the action in the simple form
\begin{align}
&S= \frac{L}{g^2}\int_{-\infty}^{+\infty}dx_1\,  \Big[  (\dot \theta  \mp    \xi  \sin \theta)^2 \pm   2 \xi \dot \theta       \sin \theta   \Big]    \geq   \frac{2L \xi }{g^2}
\int_{0}^{\pi}   d \theta       \sin \theta \geq   \frac{4L \xi }{g^2} \\
&\Rightarrow S_F = \frac{ 4 \pi}{g^2 } = \frac{S_{\rm uniton}}{2} 
\,,
 \label{actioninstanton}
\end{align} 
where  we substituted the $\mathbb{Z}_2$ symmetric value for $\xi=\pi / L $. 
 Note that the action of a fracton $S_F$ is precisely $1/2$ ($1/N$ in the $SU(N)$ case) times the action of a uniton Eq.~\eqref{Suniton}, supporting our claim that the fractons are constituents of unitons.  

The equality in Eq.~\eqref{actioninstanton} holds whenever the fracton/anti-fracton satisfy the first order BPS equation
\begin{align}
\dot{\theta}(x_1) - \xi \sin \theta &=0\,,\\
\dot{\bar{\theta}}(x_1) + \xi \sin \bar{\theta} &=0.
\end{align}
The solutions to these equations take the form
\begin{align}
\theta(x_1; x_1^{(0)}) &= 2 \arccot \left[e^{-\xi (x_1 - x_1^{(0)})} \right] \,,\\
\bar{\theta}(x_1; x_1^{(0)}) &=\pi -  2 \arccot \left[e^{-\xi (x_1 - x_1^{(0)})} \right]\,,
\end{align}
 where $x_1^{(0)}$ is a position modulus. The fracton obeys the boundary conditions $\theta(x_1 \to -\infty) = 0, \; \theta(x_1 \to +\infty) = \pi$, while the anti-fracton goes from $\pi$ to $0$ and these solutions have precisely the BPS-saturated (from the small-$L$ effective field theory point of view) action
 \begin{equation}
 S_{\rm fracton} = \frac{4 \pi}{g^2} =  \frac{S_{\rm uniton}}{2} \,.
 \label{Sfracton2}
 \end{equation}

 \vspace{3mm}
 
{\bf KK Fractons, and Unitons from Fractons:} 
\vspace{0mm} 
\label{sec:KKFractons}
As we have mentioned above, discarding all KK non-zero modes misses some critical information, which is important even when $L$ is small.  In particular, consider doing the dimensional reduction to 1D with the \emph{periodic} field $\phi_2\in[0,2\pi]$ winding $n$ times as we move along the $S^1$ (i.e. $\phi_2 = 2 \pi \,n\,x/L$), rather than just being set to a constant, as we did above.  We will shortly see that configurations with $n=-1$ have a distinguished role. Let us assume that
\begin{align}
\phi_2(x_1,x_2)=\phi_2(x_1)+\frac{2 \pi\,n\,x_2}{L}\,,
\end{align}
this means that $\phi_2$ carries some KK momentum, by way of a non-trivial winding number $n$.  The reduced action becomes
\begin{align}
S&= \frac{L}{g^2}\int_{-\infty }^{+\infty} dx_1\,  \Big[  \dot \theta ^2 +   \cos^2 \theta \dot \phi_1^2  +    \sin^2  \theta    \dot \phi_2^2
+ \left(\frac{2 \pi \,n}{L}+\xi\right)^2 \sin^2 \theta     \Big] \,.  
\end{align} 
The same arguments as before then show that this action gives rise to stable instanton configurations, which we will refer to as KK-fractons\footnote{This construction is analogous to the one done for gauge theories on $\mathbb{R}^3 \times S^1$, where the resulting configurations are known as KK-monopoles\cite{Lee:1997vp,Kraan:1998pm}. }.   These configurations have action
\begin{align}
S^{(n)}&=  \frac{4 | 2\pi\,n+\xi\,L| }{g^2} \,.
\end{align} 
Now observe that when $\xi$ takes on the $\Z_2$ symmetric value $\xi = \pi/ L$ and $n=-1$, this becomes 
\begin{align}
S_{\textrm{KK fracton}}&=  \frac{4\pi}{g^2} = S_{F}  = \frac{S_{\rm uniton}}{2}  \,.
\end{align} 
The action for a generic KK fracton with $n\neq -1$, will be higher than $S_{\rm fracton}$, and hence their contribution will be suppressed  in the semiclassical expansion.  But fractons and the $n=-1$ KK fractons enter the semiclassical expansion on the \emph{same} footing!  

Despite having the same action as the fractons, it is important to emphasize that the $n=-1$ KK fractons and $n=0$ fractons are \emph{distinct} field configurations.  Indeed,  the explicit solutions for the field configurations of the $n=-1$ KK fractons are
\begin{align}
\theta(x_1; x_1^{(0)})_{\rm KK} &= 2 \arccot \left[e^{-(\xi-2\pi/L) (x_1 - x_1^{(0)})} \right]\,, \\
\bar{\theta}(x_1; x_1^{(0)})_{\rm KK} &=\pi -  2 \arccot \left[e^{-(\xi-2\pi/L) (x_1 - x_1^{(0)})} \right]\,.
\end{align}
 These KK fractons behave as $\theta(x_1 \to -\infty)_{\rm KK} \to \pi$ and $\theta(x_1 \to +\infty)_{\rm KK} \to 0$, while  $\bar{\theta}(x_1 \to -\infty)_{\rm KK} \to 0$ and $\bar{\theta}(x_1 \to +\infty)_{\rm KK} \to \pi$, in contrast to the standard fractons. 
 
From the solution for the minimal fractionalized $SU(2)$ uniton (\ref{eq:UnitonProj}), depicted in Fig.~\ref{fig:SU2Uniton}, we can indeed read the exact form of the various component fields $\theta,\,\phi_{1,2}$.
The $\theta$ component can be seen as the gluing of a fracton at $x_1= x_1^{(0)}$, followed by a KK-fracton at some $x_1>x_1^{(0)}$, so that the combined field configurations goes from $\theta = 0$ at $x_1=-\infty$ back to $\theta =0$ at $x_1=+\infty$. To see why the second object is a KK-fracton rather than an anti-fracton one can study the field $\phi_2$ in the uniton solution on $\mathbb{R}\times S^1$ with the $\Z_N$-symmetric background holonomy.  It can be shown that $\phi_2$ interpolates between a configuration with zero winding in the $x_2$ direction at $x_1=-\infty$
and a configuration with  $-1$ winding in the $x_2$ direction at $x_1=+\infty$:
\begin{align}
\phi_2(x_1,x_2)|_{\rm uniton} = \frac{-\pi x_2}{L}- \arctan\left[\tanh\left(\frac{\pi x_1}{L}\right) \tan\left(\frac{\pi x_2}{L}\right) \right] ,
\end{align}
where for simplicity we fixed the moduli $\lambda_1=\lambda_2=1$.
 Hence we see the precise sense in which the fractons are the constituents of  unitons.

\subsection{$SU(N)$ Fractons}
We now discuss the generalization of the $SU(2)$ fractons  to the $SU(N)$ PCM.   This can be done 
by embedding the  $SU(2)$ fractons into the $SU(N)$ model, by using Eq.~\eqref{eq:GeneralSU2Embedding}.  
This construction permits us to classify all the fracton  events with action 
\begin{align}
S^{(k)} =  k \times S_F,  \qquad  k \in \Z_{+}\,,
\label{level}
\end{align}
where 
\begin{align}
S_F = \frac{8\pi }{g^2 N} = \frac{S_{\rm uniton}}{N} 
\end{align}
is the action of the minimal fracton, and $k$ may be seen as ``level" of the associated action.

{\bf Naive approach:} Consider the root system of the Lie algebra  $\mathfrak{su} (N)$.  Denote the simple root system as 
\begin{align}
\Delta_1= \{ \alpha_1, \alpha_2, \ldots, \alpha_{N-1}\}  .
\label{r-s}
\end{align}
We can use $ \Delta_1 $ to  build the complete root system $A_{N-1}$. 
 For each root ${\bm \alpha}$ (simple or not), there is an associated   $\mathfrak{su} (2)$ sub-algebra. All fracton embeddings associated with simple  roots are minimal action,  level-1 events, with action:    
\begin{align}
 \mathcal{F}_{\bm \alpha_i}^{(1)}  \equiv  \mathcal{F}_i  : \qquad    S_{i,1} =   
  \frac{8 \pi \mu_{i+1,i}}{g^2}  = \frac{4  \xi }{g^2}=     
    \frac{8 \pi}{ g^2N} = \frac{S_{\rm uniton}}{N}   \qquad  i \in [1, N-1]\label{eq:F1event}
\end{align} 
where we set $\mu_i$ to their center-symmetric values.   There are $N-1$ simple roots, and hence there are $N-1$ minimal action fracton events.

 The positive roots which can be written as a sum of just two adjacent simple roots correspond to level-2 fractons. There are $N-2$ such positive roots. 
 The actions of these events are (in the center-symmetric background) 
 \begin{align}
 \mathcal{F}_{\bm \alpha_i + \bm \alpha_{i+1} }^{(2)}   : \qquad   S_{i,2} = \frac{8 \pi \mu_{i+2,i}}{g^2}  
 =  2 \times  \frac{8 \pi}{g^2N} = 2 \times  \frac{S_{\rm uniton}}{N} \qquad  i \in [1, N-2]\,.
\end{align} 

Similarly positive roots which can be written as a sum of just three (adjacent) simple roots correspond to level-3 fractons and so on.  The general construction is that the root space $A_{N-1}$ can be  split into simple roots and  roots that can be written as  $k$-linear combination  of simple roots: 
  \begin{align}
 \begin{array}{llll}
&k=1:  & \qquad (N-1)  {\rm \,\,positive \; roots} : & \qquad  \Delta_1^{+} \equiv  \{\alpha_1, \alpha_2, \ldots, \alpha_{N-1} \} , \cr 
&k=2: &  \qquad (N-2)  {\rm \,\,positive \; roots} : & \qquad   \Delta_2^{+} \equiv \{ \alpha_1+ \alpha_2,   \alpha_2+ \alpha_3,    \ldots, \alpha_{N-2}+ \alpha_{N-1} \}  \cr  
& k=3: &  \qquad (N-3)  {\rm  \,\,positive \; roots}  :  &\qquad    \Delta_3^{+} \equiv \{ \alpha_1+ \alpha_2 + \alpha_3,      \ldots, \alpha_{N-3}+ \alpha_{N-2} +  \alpha_{N-1} \}   \cr
& \cdots & \qquad \cdots & \qquad\cdots  \cr 
& k= N-2:  &  \qquad  2  {\rm  \,\,positive \; roots} :  & \qquad    \Delta_{N-2}^{+} \equiv \{ \alpha_1+ \alpha_2 +  \ldots + \alpha_{N-2}, 
   \alpha_2+ \alpha_3 + \ldots + \alpha_{N-1}  \} \cr \cr
   & k= N-1:  &   \qquad 1 {\rm  \,\,positive \;  root} :  & \qquad   \Delta_{N-1}^{+} \equiv  \{ \alpha_1+ \alpha_2 +  \ldots + \alpha_{N-2} + \alpha_{N-1} \} 
   \end{array}
\end{align} 
We can similarly define the negative roots, $ \Delta_j^{-} = -\Delta_j$.  
Obviously, the root space  $A_{N-1}$ can be decomposed as  
 \begin{align}
A_{N-1} \;= \;\bigoplus_{k=1}^{N-1} \left(  \Delta_j^{+} +  \Delta_j^{-}  \right)\,.
  \end{align}
This root space $A_{N-1}$ contains $N^2 -N$ roots, half positive and half negative, and together with the Cartan subalgebra, whose dimension is $N-1$, they form the complete set of generators of $SU(N)$.

As is well-known, there is an $\mathfrak {su}(2) $ sub-algebra of $\mathfrak{ su}(N)$ generated by 
$E_{\bm \alpha}, E_{ - \bm \alpha}, {\bm  \alpha} \cdot {\bm  H } $ where $E_{ \pm \bm \alpha}$ are  raising and lowering operators, and  $ {\bm  H } $ are Cartan sub-algebra matrices.  We can embed an    
 $\mathfrak {su}(2) $  fracton into  $\mathfrak {su}(N) $  by using any root   ${\bm  \alpha} $. Clearly, we have fracton embeddings associated with  ${\bm \alpha}$,
 \begin{align}
 \mathcal{F}_{\bm \alpha}^{(k)} :    \qquad   {\bm \alpha} \in  \Delta_k^{+},  \qquad   S_{i,k} =   
   \frac{8 \pi (\mu_{i+k} - \mu_i) }{g^2}  = 
    \frac{8 \pi k }{ g^2N} = k\times S_F = \frac{ k S_{\rm uniton}}{N}\,.   \qquad  
\end{align} 
According to this rationale, there are $N-1$ fractons at level-1, $N-2$ fractons at level-2, so and so forth,  and  only  one fracton at level $N-1$.  Clearly, the densities of events  of level-1, level-2, etc are also hierarchical, and obeys 
\begin{align}
e^{-S_F} \gg e^{-2 S_F} \gg \ldots \gg e^{-(N-1)S_F}   
\end{align}

{\bf More carefully,  accounting for the compactness of holonomy:}  The previous discussion does not take into account the fact that  the background holonomy is compact.  For example, one might naively think that events in  $\Delta_{N-1}^{+}$ are very rare tunneling paths, exponentially suppressed with respect to the   minimal fractons,  because their actions are $ S_{1, N-1} = (N-1) \times S_{i,1}$.  Indeed, this is largely correct.
 However, if choose with a $k=N-1$ configuration with winding number  $n=-1$ for $\phi_2$, as was done in the $SU(2)$ PCM in Sec.~\ref{sec:KKFractons}, and then apply the Kaluza-Klein-reduction to the action afterwards, we obtain
 \begin{eqnarray}
S&&= \frac{1}{g^2}\int_{\R }  \Big[ ( \partial_t  \theta )^2 + \left(\frac{2 \pi}{L} (-1+\mu_{N} - \mu_1)\right)^2 \sin^2 \theta  \Big] \,.
  \label{eq:actionQM2}
\end{eqnarray} 
This happens because  $\mu_i- \mu_j$ behaves as a fractional  momentum in the compact direction, and combining 
 $(\mu_{N} - \mu_1)$ with $-1$ unit of winding number results in a {\it twisted KK-reduction}. 
  In the full 2D QFT,  this tunneling path is exactly on the same footing with the elementary fractons 
 corresponding to $\Delta_{1}^{+}$,   and its action coincides with them (\ref{eq:F1event}) in the center-symmetric background of  Eq.~\eqref{eq:MinSpatial}: 
   \begin{align}
 S_{\rm twisted} =  \frac{8 \pi   | -1+ (\mu_{N} - \mu_1) |    }{g^2}  =    \frac{8 \pi}{g^2N}  = S_{i,1} = \frac{S_{\rm uniton}}{N}\,.
\end{align} 
We refer to this instanton-type event as the KK fracton, denoted $\F_N$.  It is associated with the affine root of the $\mathfrak{su}(N)$ Lie algebra, 
defined by 
\begin{align}
\alpha_N= - \sum_{i=1}^{N-1} \alpha_i\,,
\end{align}
which is itself a negative root.
The $\F_N$ fracton  is the counter-part of the so called KK-monopole-instanton (also called affine or twisted instanton) in gauge theories on $\R^3 \times S^1$ 
~\cite{Kraan:1998pm,Lee:1997vp}.

Similarly, by inserting $-1$ units of winding number into the two events living in $k=N-2$, we can turn them into tunneling events with action $\frac{2}{N} S_{\rm uniton}$. This pattern continues for all levels:
  \begin{align}
  \label{affine-full}
& \begin{array}{llll}
&k=1:  & \qquad  & \qquad  
\Delta_1^{\rm aff} \equiv  \Delta_1^{+} +  \Delta_{N-1}^{-}  \equiv  \{\alpha_i  \}\cr \cr
&k=2: &  \qquad & \qquad  \Delta_2^{\rm aff} \equiv    \Delta_2^{+}  +    \Delta_{N-2}^{-}  
\equiv \{ \alpha_i+ \alpha_{i+1}   \}  \cr \cr
& k=3: &  \qquad   &\qquad  \Delta_3^{\rm aff} \equiv     \Delta_3^{+}  
 +    \Delta_{N-3}^{-}  
\equiv \{ \alpha_i+ \alpha_{i+1} + \alpha_{i+2}    \}  \cr
& \ldots  \cr 
& k= N-2:  &  \qquad  & \qquad    \Delta_{N-2}^{\rm aff} \equiv   \Delta_{N-2}^{+}  +  \Delta_{2}^{-}
\equiv \{ \alpha_1+ \alpha_2 +    \ldots  +  \alpha_{i-1}  +   \alpha_{i+2}   \ldots  + \alpha_{N}  \} \cr \cr 
   & k= N-1:  &  & \qquad  \Delta_{N-1}^{\rm aff} \equiv    \Delta_{N-1}^{+}  +  \Delta_{1}^{-}
   \equiv  \{ \alpha_1+ \alpha_2 +  \ldots  +    \alpha_{i-1} +  \alpha_{i+1}    \ldots  + \alpha_N \}
   \end{array} 
 \end{align}
where  $ i \in [1,N] $. In particular, we learn that at level $k$, there are  {\it always} $N$ NP-saddles   associated with the roots  $\Delta_{k}^{\rm aff}$~\footnote{The advantage of $\Delta_k^{\rm aff} $ with respect to  $\Delta_k^{+}$ is that  it makes the fact that the 
KK-fractons are on the same footing with the regular fractons manifest, along with the $\Z_N$ cyclicity. 
Its main disadvantage is that one loses the  notion of positivity associated with level-$k$.  Recall that all roots living in   $\Delta_k^{+}$ were positive. Since  
\begin{align}
\Delta_k^{\rm aff}  = \Delta_k^{+} +  \Delta_{N-k}^{-} \; , 
\end{align}
 and  all roots  (involving the affine root in some way)  living in  $\Delta_{N-k}^{-}$ are negative,  
 $\Delta_k^{\rm aff}  $ is not comprised of positive roots only. 
 Unfortunately, one cannot have both properties at once. The concept of positive roots will play some role when we discuss correlated fracton events, and hence, we work with both representation.}.  This is actually a consequence of the cyclic $\Z_N$ symmetry of the background holonomy. 
All $N$-saddles in $\Delta_k^{\rm aff} $ have actions 
$
S^{(k)} =   
    \frac{8 \pi  }{ g^2N}  \times k = \frac{  S_{\rm uniton}}{N}   \times k$.

The existence of the KK fracton events is due to the compactness of the $SU(N)$ target space and of the background  holonomy $\Omega$ in Eq.~\eqref{twist2}. Equivalently,   although the long distance effective field theory on small $\R\times S^1_L$ may appear one-dimensional, the underlying microscopic theory is two-dimensional, and this is encoded in the structure of the effective field theory. 
 As we saw above both in the perturbative and non-perturbative contexts, the details of dimensional reduction at small-$L$ are quite subtle in the context of theories with adiabatic small-$L$ limits.

 The fractons associated with $ \Delta_1^{\rm aff} $
 are the leading (minimal action) topological configurations in the PCM. In the bosonic model, this will be of crucial importance in the determination of the mass gap. Furthermore, certain   correlated fracton-anti-fracton  events  (neutral bions)  will be crucial in resolving the renormalon ambiguity,  similar to the $\mathbb C \mathbb P^{N-1}$ model and YM theory\cite{Argyres:2012ka,Argyres:2012vv,Dunne:2012ae,Dunne:2012zk}.

  \subsection{Fracton amplitudes}
 
 The fracton amplitude is the same as the 1-d instanton amplitude associated with the quantum mechanical system in Eq.~\eqref{eq:BOHamiltonian}.    Within the Born-Oppenheimer approximation in the low energy effective theory we find there is effectively only one bosonic zero mode, the position modulus of the fracton.  The amplitude of the fracton event using dimensionless units introduced in Eq.~\eqref{canonical}, is given by
\begin{align}
 \mathcal{F}_i  
 = J_{\tau_0}  \; e^{-S_i} \;  \left[\frac{ \det' M }{ \det M_0}\right]^{- {1 \over2}}
 \end{align}
  where 
 $ M = \frac{ \delta^2 S}{\delta \theta_c(t_1) \delta \theta_c(t_1) } =  \left[ - \left( \frac{d}{dt_1}\right)^2 - V^{''} ( \theta_c(t_1) )  \right]\delta(t_1-t_2)$
  is the  quadratic fluctuation operator 
in the background of  the fracton,    ${\det' M }$ indicates that the zero mode is dropped in evaluation of the determinant, 
 and  $\det M_0$ is  for normalization. The determinant can be evaluated in 
 multiple different ways \cite{ZinnJustin:2002ru, Marino:2012zq}, for example, via the Gelfand-Yaglom method   \cite{Dunne:2007rt}.
 The Jacobian associated with the zero mode is  
 \begin{align}
 J_{\tau_0} =  \sqrt \frac{  S_i}{ 2 \pi}\,.
 \end{align}
 The result of this calculation is given in Section~41.2 of \cite{ZinnJustin:2002ru}, and is equal to 
 $ \I = \frac{4}{\sqrt{ \pi g_{\rm ZJ}}} e^{-8/g_{\rm ZJ}} $. Converting to  our notation, see Eq.~\eqref{canonical}, 
  $g_{\rm ZJ}= \frac{2 g^2 }{\xi}$, we  can 
 write the minimal  fracton amplitude $(i=1,\ldots, N)$ in the center-symmetric background as 
  \begin{align}
 \mathcal{F}_i 
 =  \sqrt \frac{ 2 S_i}{\pi} e^{-S_i} = \sqrt \frac{8 \xi }{\pi g^2 }  e^{-  \frac{4 \xi }{g^2} } =  \sqrt \frac{16  }{ g^2N }  e^{-  \frac{8 \pi }{g^2 N} } \,.
\label{amplitude}
 \end{align}
 There are  perturbative fluctuations around the fracton, and the  amplitude incorporating those fluctuations is given by  
 \begin{align}
[ \mathcal{F}] \times   \Phi_{\F} (g^2)   = \sqrt \frac{16  }{ g^2N }  e^{-  \frac{8 \pi }{g^2 N} }   \; \Phi_{\F} (g^2),  \qquad  \; \Phi_{\F} (g^2)  \equiv \sum_{n=0}^{\infty} a_n^{[\F]} g^{2n}\,.
\label{amplitude2}
 \end{align} 
  It should be noted that there is no non-perturbative ambiguity in the NP part of the  amplitude,  
  given by  $\sqrt \frac{16  }{ g^2N }  e^{-  \frac{8 \pi }{g^2 N} } $.  
  On the other hand we expect 
 $\Phi_{\F} (g^2) $ to be   a non-Borel summable asymptotic series,  similarly to perturbative expansion around the P-saddle   Eq.~\eqref{pt-low}.

\section{Resurgence triangle and 
emergent topological structure}
\label{sec:Resurgence}

Now that we have found the fractons,  the minimal-action non-perturbative field configurations  at weak coupling, we are in a position to see how  the renormalon ambiguities in the Borel resummation of the perturbative series,  such as Eq.~
\eqref{eq:Borel-res} are cancelled.  We will show that they cancel against corresponding ambiguities in the contributions from  the non-perturbative sector, leaving well-defined results order by order in an expansion in $e^{-S_F} = e^{-8\pi/(g^2N)}$.   
The phenomenon we are describing is an example of a semiclassical expansion which is not Borel summable, but \emph{is}  Borel-Ecalle summable. This is a generalization of  Eq.~\eqref{a-c-2} for ordinary integrals.  

We first emphasize again that there is no non-perturbative ambiguity in the fracton amplitude  $[\mathcal{F}_i]$ itself\footnote{Of course, the perturbative fluctuations  $ \Phi_{\F} (g^2) $ around it are still non-Borel summable and ambiguous, but that is a different story.}.
  Although there is no topological charge in the microscopic theory, there is an emergent topological structure in the low energy effective theory, and the fractons are 1d-instantons from that point of view, hence they cannot mix with perturbation theory.  In theories with a topological $\Theta$ angle it is always true that  events with non-zero instanton number cannot cure the ambiguity of perturbation theory
\cite{Dunne:2012ae} due to the $\Theta$ dependence of instanton amplitudes. In our case, despite the absence of a microscopic topological argument, at small-$L$ we find that fractons cannot mix with perturbation theory either. 
 
 Second, it is now worthwhile to re-inspect perturbation theory around the perturbative saddle, and rewrite Eq.~\eqref{pt-low} and Eq.~\eqref{eq:LargeOrderPT} as expansions in
\begin{align}
 \frac{1}{2S_F} = \frac{g^2}{8 \xi} = \frac{g^2N}{16 \pi} .
 \end{align}
Then
 \begin{align}
\label{p t-low-2}
\mathcal{E}_i(S_F)  &=   \textstyle{   \frac{1}{2}-  \frac{1}{2} \left(\frac{1}{2S_F} \right) -  \frac{1}{2}  \left(\frac{1}{2S_F} \right)^2 -  \frac{3}{2} \left(\frac{1}{2S_F} \right)^3 - \frac{53}{8}  \left(\frac{1}{2S_F} \right)^4  - \frac{297}{8} \left(\frac{1}{2S_F} \right)^5    - {3961 \over 16}  \left(\frac{1}{2S_F} \right)^6  -  \ldots }  \nonumber\\
 & \textstyle{    -\frac{2}{\pi} \left( \frac{1}{2S_F}  \right)^{n} n!  (  1+ O\left(\frac{1}{n}\right) )}.
 \end{align}
This emphasizes that the perturbative expansion parameter is the inverse of {\it two} times the fracton action.  As already mentioned in Footnote~\ref{fn6}, this is one of the major differences between ordinary integrals and path integrals.  
 The leading ambiguity of perturbation theory is given by the  right/left Borel resummation
Eq.~\eqref{eq:Borel-res}. Expressing these left/right resummations in terms of the action, 
 \begin{align}
{\cal S}_{0^\pm} {\mathcal{E}_i}   &= 
 \Re {\cal S}_{0} {\mathcal{E}}  \mp {2 \pi i} [\F_i ][\bar \F_i]=
 \Re {\cal S}_{0} {\mathcal{E}}  \mp  i  (4S_F) e^{-2S_F} =
 \Re {\cal S}_{0} {\mathcal{E}}  \mp  i \frac{32 \pi}{g^2N} e^{-\frac{16 \pi }{g^2N}} 
\label{eq:Borel-res-2}
\end{align}
makes it clear that the ambiguities can only be canceled  by contributions  which live in the fracton-anti-fracton sector. We show that this is the case in Section~\ref{sec:Cancellations}. 
The first step in seeing how this works is to understand the interactions of fractons and anti-fractons, which will be our task in Section~\ref{sec:FractonInteractions}. 
 
 \begin{figure}[tbp]
\centering
\includegraphics[width=0.8\textwidth]{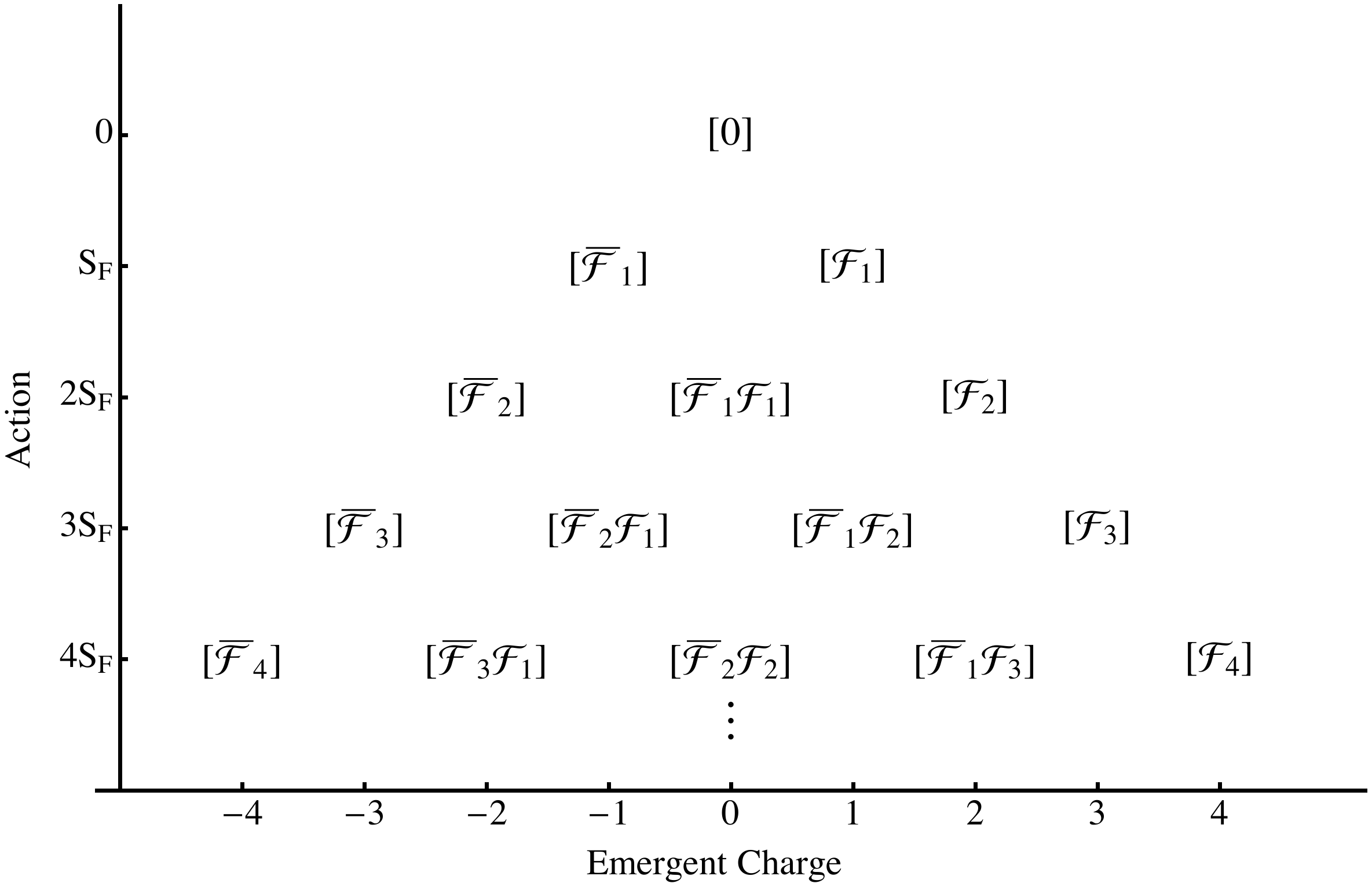}
\caption{Schematic version of the resurgence triangle classification of saddle points induced by resurgence theory for the PCM on small $\mathbb{R}\times S^1$. }
\label{fig:ResurgenceTriangle}
\end{figure}

Our main idea  about emergent topology and resurgence is encoded in the resurgence triangle 
\cite{Dunne:2012ae} classification of saddle points, which is illustrated for the PCM on $\mathbb{R}\times S^1$ in Fig.~\ref{fig:ResurgenceTriangle} in a schematic form, which does not take into account the ``ramification" phenomenon we will describe later in this Section.  The entries in the resurgence triangle are $[\overline{\F}_{\bar n}\F_n ]$, which denote \emph{correlated} events involving $n$-fracton and $\bar n$ anti-fractons.  The vertical axis is the action of the events, while the horizontal axis tracks the ``charge" $q$ of the events, defined as $q= n-\bar{n}$.  Then the perturbative vacuum is the  ``primary"  or ``level zero"  for the $q=0$ sector, $[\F_1]$ is the ``level zero" event for the $q=1$ sector,  $[\F_2]$ is the level zero event for the $q=2$ sector, and so on.   The cancellation of ambiguities happens between saddles in a given column, and different columns cannot mix in the cancellation of their respective ambiguities.    Note that unlike theories in which different columns are classified according to different homotopy classes of the microscopic theory, in the present example where there is no microscopic topological classification, the classification of columns takes place according to the topology in the low energy effective theory, which is \emph{emergent}.

It should be emphasized that the resurgence triangle and the variety of NP-saddles for the PCM  are just as rich as $\mathbb {CP}^{N-1}$ model \cite{Dunne:2012ae} or deformed Yang-Mills theory or QCD(adj) \cite{Argyres:2012ka}.  In other words, the fact that the homotopy group $\pi_2$ is trivial in the microscopic PCM is only a superficial difference from Yang-Mills theory.   At the deeper level of insight allowed by resurgence theory, 
it appears that the set of NP saddles in the PCM is in fact in \emph{one-to-one} correspondence to the set of NP saddles in Yang-Mills theory, because in both cases the set of saddles are determined by the properties of the underlying $\mathfrak{su}(N)$ Lie algebra!

\subsection{Fracton interactions and $[\F_i {\F}_i], \; [\bar \F_i  \bar {\F}_i], \;  [\F_i  \bar {\F}_i]$ amplitudes  }
\label{sec:FractonInteractions}
To understand how the weak coupling  realization of the  renormalon ambiguity explained around 
Eq.~\eqref{eq:Borel-res} and Eq.~\eqref{eq:Borel-res-2}
 is cured, we must understand the structure of the semi-classical expansion at second and higher order.   As we have described above,   at first order in the semi-classical expansion, one has the minimal-action fractons,  but their contributions  $ (\sim e^{-S_F})$ are {\bf 1)} exponentially  {\it larger} than the renormalon ambiguity  $ (\sim e^{-2S_F})$ and {\bf 2)} are unambiguous.  Hence the minimal-action fractons contributing at first order in the semi-classical expansion cannot contribute to resolving the issues of perturbation theory, which can only be cured at second order as shown by Eq.~\eqref{eq:Borel-res-2}.

 At second order in the semi-classical expansion,  we encounter {\it correlated}   fracton events, 
 such as $[\F_i {\F}_i], \; [\bar \F_i  \bar {\F}_i], \;  [\F_i  \bar {\F}_i]$, as well as  {\it uncorrelated} single fracton events associated with roots $ \alpha  \in  \Delta_2^{\rm aff}$ in Eq.~\eqref{affine-full}.  The uncorrelated events correspond to field configurations which are exact solutions of the equations of motion of the theory.  The correlated events have finite actions which are parametrically close to those of the uncorrelated events at this order in the $\lambda \ll 1$ limit, and the associated field configurations are quasi-solutions to the equations of motion rather than exact solutions, meaning that  the equations of motion are satisfied with parametrically good accuracy when $\lambda \ll 1$.  The correlated events are \emph{quasi}-saddle-points, and make critically important contributions to the path integral in the semiclassical limit.

  According to the emergent topological structure associated with the resurgence triangle of Fig.~\ref{fig:ResurgenceTriangle},  $[\F_i {\F}_i], \; [\bar \F_i  \bar {\F}_i]$ are the leading saddles  in their classes, and cannot cure the ambiguity of perturbation theory either. 
It is the   fracton-anti-fracton  configurations 
$[\F_{i}\bar{\F}_{i}], i = 1,2, \ldots, N$ (which we refer to as ``neutral bions" due to the universality of such configurations across a wide range of asymptotically-free theories) which can mix with perturbation theory.  They are the first sub-leading saddles in the column  associated with the perturbative saddle. 

Below, we demonstrate that the non-perturbative amplitudes (without incorporating the perturbative fluctuations) associated with the correlated events are  given by: 
\begin{align}
\label{ff-int}
[\F_j {\F}_j] =[\bar \F_j \bar {\F}_j] =  &  (- \log (4S_F) - \gamma)  [\F_j][{\F}_j]   \cr
 =&  (- \log (4S_F) - \gamma)  \frac{2S_F}{\pi} e^{-2S_F} \cr
=&  \left( - \log \left[\frac{32  \pi }{g^2 N} \right] -\gamma \right)   \frac{16  }{ g^2N }  e^{-  \frac{16 \pi }{g^2 N} } \,, \\  
\cr
[\F_j\bar{\F}_j]_{\pm} = [\bar \F_j {\F}_j]_{\pm} =  & (- \log (4S_F) - \gamma)  [\F_j][\bar{\F}_j]  \pm i \pi  [\F_j][\bar{\F}_j]   \cr
= & (- \log (4S_F) - \gamma)  \frac{2S_F}{\pi} e^{-2S_F}  \pm i \pi \times \frac{2S_F}{\pi} e^{-2S_F} \cr  
=&  \left( - \log \left[\frac{32  \pi }{g^2 N} \right] -\gamma \right)   \frac{16  }{ g^2N }  e^{-  \frac{16 \pi }{g^2 N} } \pm  i \frac{16 \pi }{ g^2N }  e^{-  \frac{16 \pi }{g^2 N} }\,.
\label{ffbar-int}
\end{align} 
Each of these representations is useful for slightly complementary reasons. 
Note that the correlated  $[\F_j {\F}_j]$ event amplitude is \emph{not} the same quantity as  $[\F_j][{\F}_j]$, and these are genuinely different (isolated, non-degenerate) saddles.  In fact, in the semi-classical evaluation of partition function, we must sum over both types of saddles independently.  For example, up to second order in NP trans-series expansion, 
we can write a grand canonical ensemble of all defects entering up to second order in the resurgence triangle:
   \begin{eqnarray}
e^{- E \beta} &\sim& e^{- \frac{\omega}{2} (1+ O(g)) \beta}  
\prod_{\T}  \left(\sum_{n_\T }  
 \frac{(\beta \T)^{n_{\T}}} {n_{\T}!}  \right)
  \cr 
  & =& e^{- \frac{\omega}{2} (1+ O(g))  \beta}   \left(\sum_{n_\F }  \frac{(\beta \F)^{n_{\F}}} {n_{\F}!}  \right) \left(\sum_{n_{ \bar \F} }   \frac{(\beta \bar \F)^{n_{ \bar \F}}} {n_{\bar \F}!}  \right)    \left(\sum_{n_{[\F\F]} }   \frac{(\beta [\F \F])^{n_{[\F \F]}}} {n_{[\F \F]}!}  \right)   
 \ldots
  \cr 
&= & e^{- \left( \frac{\omega}{2} (1+ O(g))  - \F -\bar \F -  [\F\F] - [\bar \F \bar \F]- [\F\bar \F] + \ldots  \right)  \beta} 
\label{sum2}
\end{eqnarray}
A pictorial description of this generalized grand canonical ensemble can be found in \cite{Dunne:2014bca}. It is also implicit in the Zinn-Justin's exact quantization formula which incorporates all orders in non-perturbative effects \cite{ZinnJustin:2002ru} via a resurgent trans-series.

\vspace{3mm}
\noindent{\bf  Computation of $[\F_j\bar{\F}_j]$ and $[\F_j {\F}_j]$ and  amplitude }
\vspace{3mm}

 \noindent Amplitudes for correlated two-events in the $ \mathbb{ CP}^{N-1}$ model on $\mathbb{R}\times S^1$ with $\Z_N$-twisted boundary conditions were  calculated in \cite{Dunne:2012ae}, and as we have remarked there is a striking degree of similarity between the NP saddles of the $\mathbb{CP}^{N-1}$ model and the NP saddles  of the PCM associated with the affine simple roots of $\mathfrak{su}(N)$.   Moreover, it turns out that at small-$L$ it is precisely (the interactions of) these $N$ minimum-action saddle points of the PCM which are responsible for the cancellation of renormalon ambiguities.   Hence our discussion below will amount to a review of the argument of \cite{Dunne:2012ae}, with a few additional clarifying comments. 
We suppress the index $i$ in what follows to lessen the clutter, since we are looking for the correlated amplitudes for $[\F_\alpha \bar{\F}_\alpha]$ and $[\F_\alpha {\F}_\alpha]$  for any $\alpha \in A_{N-1}$. 
Indeed, $\alpha$ may be chosen to be $\alpha_i \in \Delta_1^{\rm aff}$ in the classification of  Eq.~\eqref{affine-full}, without loss of generality.

Consider an   $SU(2)$ subgroup embedded into $SU(N)$  PCM, and let    $[\F  {\F} ]$  and   $[\F \bar{\F} ]$ configurations denote correlated events therein.  Then consider the ansatz for e.g. 
 $[\F \bar{\F} ]$:
\begin{align}
\theta_{\F \bar{\F} }(t_0; \tau) = \theta_{\F }(t_0+ \tau/2) + \theta_{\bar{\F} }(t_0 - \tau/2)\,.
\label{a-1}
\end{align}
Here,   $t_0$ is the center of action coordinate of the pair, while $\tau$ is the separation of the constituent fractons.  While $\theta_{\F}$ and $\theta_{\bar{\F}}$ are individually solutions to the equations of motion, we do \emph{not} expect   $\theta_{\F \bar{\F} }(t_0; \tau)$ to be a solution, because the equations of motion are nonlinear.   However, when $\tau \gg \xi$, we expect $\theta_{\F \bar{\F} }(t_0; \tau)$ to become a quasi-solution to the equations of motion, in the sense that the equations of motion are satisfied up to terms exponentially small in $\xi \tau$.  In the far-separated regime, we expect the action of $\theta_{\F \bar{\F} }(t_0; \tau)$ to approach $2 S_F$.  So we expect $\theta_{\F \bar{\F} }(t_0; \tau)$ to be a quasi-saddle-point of the theory.

Fluctuations around exact saddle points are either zero modes, which cost zero action,  or perturbative modes, which have an action cost of order $\xi$.  Fluctuations around quasi-saddle-points are more subtle.  In addition to zero modes and perturbative fluctuation modes, there are also quasi-zero modes, which have an action cost which is exponentially small in $\xi$.   For $\theta_{\F \bar{\F} }(t_0; \tau)$ it turns out that $t_0$ is an exact zero mode, while $\tau$ is a quasi-zero mode (with an `energy' parametrically separated from the perturbative fluctuations).
The contribution of $[\F \bar{\F}]$ events to the path integral is given by
\begin{align}
 [\F \bar{\F}] 
  &= [\F ] [\bar {\F}]   \int d\Omega \, d \tau \,  e^{- V(\tau) }\,,
\end{align} 
where $[\F ] [\bar {\F}] $  are the product of uncorrelated amplitudes. The explicit integration over  $\tau$,  the separation between the constituent fractons, appears because it is a quasi zero mode, due to the presence of the  ``interaction potential" between the fractons.  The integration over
 $d\Omega$ is an instruction to integrate over the 1d ``solid angle",  which simply counts for the two different  (distinguishable) orderings of the events namely, $[\F \bar{\F} ]$ and $[\bar \F  {\F} ]$. 
 
 One may evaluate the interactions between  two fracton events separated by a distance $\tau$ by directly computing the action on the ansatz in Eq.~\eqref{a-1}. 
We  find
 \begin{align} 
 S_{\rm 2-event} = 2S_{ \rm 1-event}  + V(\tau) \,,
 \end{align} 
 where  $V(\tau) $ characterize the interaction and are given by: 
  \begin{align}
  \label{full-int}
 V(\tau) = 
  + \frac{4 L \xi}{g^2} \frac{1}{\sinh^2 ( \frac{\xi \tau}{2})}   \left(1- \frac{\xi \tau}{\sinh(\xi \tau)} \right)  
 \xrightarrow[\tau \gg \xi^{-1} ]{}
   +\frac{16 L \xi}{g^2}   e^{-\xi \tau}  
  \qquad  {\rm for }    \;\;\; [\F  \F] \,,\cr \cr
  V(\tau) =  -\frac{4 L \xi}{g^2} \frac{1}{\cosh^2 ( \frac{\xi \tau}{2})}   \left(1+ \frac{\xi \tau}{\sinh(\xi \tau)} \right)  
\xrightarrow[\tau \gg \xi^{-1} ]{}
-\frac{16 L \xi}{g^2}   e^{-\xi \tau} 
\qquad  {\rm for}  \;\;\; [\F \bar \F]  .
 \end{align}
In the latter expressions,   we assumed the dilute fracton regime, where  the separation $\tau$ between fractons is much larger than fracton size $r_{\F} \sim \xi^{-1}$.   Note that $V(\tau)$ is exponentially small in $\xi \tau$, as advertised.
In Euclidean space, where we map the proliferation of fractons into a dilute classical gas,
these results can be interpreted as \emph{repulsive} interactions between  widely separated fracton-fracton events, and \emph{attractive} interactions between widely separated 
 fracton  anti-fracton events. 

According to the structure of the resurgence triangle of Fig.~\ref{fig:ResurgenceTriangle},  the  integral over the quasi-zero modes  for   $[\F {\F}]$ should not yield an imaginary ambiguous part,  while the one for $[\F \bar{\F}] $ should have an ambiguity.  The quasi-zero mode integrals are of the form 
 \begin{align}
\label{ff}   
I_1 (g^2) & =  \int d (\xi \tau) \,  \left[ e^{-  \frac{16 L \xi}{g^2}  e^{- \xi \tau}   }    -1 \right]
\qquad \text{for $[\F \F]$}\,,\\  
\label{ffbar}   
I_2 (g^2) &=  \int d (\xi \tau) \,   \left[ e^{+  \frac{16 L \xi}{g^2}  e^{- \xi \tau}   }       -1\right]
\qquad \text{for $[\F \bar \F]$}\,,
\end{align}
where $(-1)$ subtracts off the uncorrelated fracton-fracton events \cite{Bogomolny:1980ur, 
Dunne:2012ae,Dunne:2012zk}, and arises from the semi-classical expansion of the partition function automatically. In fact,  not subtracting this factor  would amount to double-counting the uncorrelated fracton events. 

For the $I_1(g^2)$ integral,  using an integration by parts,  
we obtain 
 \begin{align}
\label{ff2}   
I_1 (g^2) = \frac{16 L \xi}{g^2}    \int d (\xi \tau)   \tau   e^{-   \left( \frac{16 L \xi}{g^2}  e^{- \xi \tau}   + \xi \tau \right) }    =   \left(- \log \left[\frac{g^2}{16 L \xi}\right] -\gamma\right)\,.
\end{align} 
This has its main support at  the length scale  $\tau_*$:
\begin{align} 
\tau_*= \frac{1}{\xi} \log \frac{16 L \xi}{g^2}   \gg \frac{1}{\xi} \,.
\end{align}
The integrand dies off for $\tau \lesssim \tau_{*} $ because of the repulsion, and it dies off 
 for $\tau \gtrsim \tau_{*} $ because of the subtraction of uncorrelated events. Thus, 
we identify $ \tau_{*}  $ as the characteristic size of the correlated 2-events.   Since $\tau_{*}\xi \gg 1$, the correlated event is a \emph{quasi-solution} to the equations of motion.  Moreover, note that  $ \tau_{*}  $  is parametrically larger than the fracton size, but parametrically smaller than the inter-fracton separation, which is in turn
much smaller than the typical separation between 2-events.  So  we have a hierarchy of scales:
\begin{align}
\label{hierarchy}
\begin{matrix}
r_\F & \ll  & r_{\rm [\F\F]} & \ll&  d_{\F} & \ll & d_{[\F \F]} \\ 
  \downarrow   &&\downarrow&&\downarrow && \downarrow  \\
\frac{1}{\xi}  & \ll &  \frac{1}{\xi}  \textstyle{  \log( 4S_F )} & \ll& \frac{1}{\xi} e^{+S_F} &\ll&  \frac{1} {\xi} e^{2S_F} 
\end{matrix}
\end{align}
This hierarchy of scales  means that the use of  semi-classical  methods for 1-events \emph{and} 2-events  is simultaneously  justified. 

 Now consider the $I_2(g^2) $ integral. The integral is dominated at separations where we cannot meaningfully talk about  isolated fracton-anti-fracton events. This is neither a bug nor an  accident.  Rather, it is an important feature!  
  Recall that the  $\arg(g^2) >0$ line is a Stokes line, 
 along which the Borel resummation of the perturbative series is ambiguous.  The left or right Borel resummations are not ambiguous, but there is a Stokes jump associated with the crossing of the Stokes line.   The  $[\F \bar \F]$ configuration can mix with perturbation theory, and if perturbation theory is ambiguous with an ambiguity at order $e^{-2S_F}$, it is conceivable that configurations that can mix with perturbation theory may  also have ambiguities of the same order  (but not larger).
 
  In fact, rather than trying to compute the $[\F \bar \F]$ directly on a Stokes line, 
  we should instead calculate the right and left $[\F \bar \F]_{\pm}$ amplitudes. The simplest way to do this it to take $g^2 \rightarrow -g^2$,  and then observe that $I_2(-g^2) = I_1(g^2)$, 
 an integral already performed. The  analytic continuation back to $+g^2$ from $-g^2$  through the complex $g^2$ plane
is two-fold ambiguous, with the result depending on whether we approach the positive real axis from above or from below.  This method of evaluating the $[\F \bar \F]$ amplitude is called the Bogomolny-Zinn-Justin (BZJ) prescription\cite{Bogomolny:1980ur,ZinnJustin:1981dx}\footnote{The BZJ prescription is very general, and is also relevant for theories with massless fermions, for some important early works on this application see \cite{Balitsky:1985in,Yung:1987zp}.}. Following the BZJ method, we find 
\begin{align}
I_{2, \pm} (g^2) =   \left( - \log \left[\frac{16  \xi} {g^2} \right] -\gamma \pm i \pi\right)\,.
\end{align}
So the correlated $[\F \bar{\F}]_{\pm}$ events have an imaginary ambiguity.

\vspace{3mm}
\noindent{\bf  Remark on analytic continuation}
\vspace{3mm}

 \noindent The reader may feel concerned  by the following aspect of the previous derivation. It naively looks like we are taking $g^2 \rightarrow -g^2$ for $\F \bar \F$,  evaluating the QZM integration there, and then we take $-g^2 \rightarrow +g^2$ again either clock-wise or anti-clock-wise, producing a two-fold ambiguous result.    But, naively,  we are not 
performing  any analytic continuation while calculating amplitudes for $\F \F$ events.  Does this mean that we are treating the theory inconsistently, by treating one sector differently from  the other?

As a matter of fact,  we can (and should)  move off the $\arg(g^2)=0 $ 
line, the Stokes line,  both for  $\F \bar \F$ as well as $\F \F$.  The point is that we can do so by just taking $g^2 \rightarrow g^2 e^{\pm i \epsilon}$, where $0<\epsilon<\pi$, evaluate the integral , and then come back to the 
$\arg(g^2)=0 $  line by taking   $\epsilon \rightarrow 0$.  
Then, we find  that   
\begin{align}
&   [\F \bar \F]_{+} - [\F \bar \F]_{-}  =  2 i \pi \times \frac{2S_F}{\pi} e^{-2S_F},  \qquad  \cr
 & [\F \F]_{+} - [\F  \F]_{-}  =0\,,
\end{align}
so that the $[\F \bar \F]$ amplitude has a Stokes jump while the $[\F \F]$ does not!
 
In fact, this is how things must work out if the resurgence triangle of Fig.~\ref{fig:ResurgenceTriangle} is the consistent characterization of  the semi-classical regime of the  principle chiral model  or other 
QFTs  studied so far \cite{Cherman:2013yfa,Dunne:2012ae,Argyres:2012ka}. 
  As we will demonstrate in the next section, the ambiguity associated with $[\F \bar \F]_{\pm}$ events cancels the ambiguity associated with the non-Borel-summability of the perturbative vacuum $[0]$ which was calculated in Eqs.~\eqref{eq:Borel-res},\eqref{eq:Borel-res-2}. According to the emergent topological structure,  $[\F  \F]$ is the lowest action configuration  (the ``level zero"  or ``primary")
in the sector with ``charge" $+2$. 
If 
$[\F  \F]$ were to have an ambiguity in its NP-part,  that would imply that there must exist another configuration with lower action, and charge $+2$, which is impossible by construction.

\vspace{3mm}
\noindent{\bf  SU(N)}
\vspace{3mm}

 \noindent
We now discuss the generalization of this analysis to $SU(N)$, where the minimal action fractons are labeled by the  roots 
$\alpha \in  \Delta_1^{\rm aff} $ of the  $\mathfrak{su}(N)$ algebra.   Based on some general arguments we expect that the interaction  potential between the fractons are given by  
\begin{align}
 V^{(ij)}(\tau) = \frac{ 8L \xi}{g^2}   (\alpha_i \cdot \alpha_j)  e^{-\xi  \tau } \,,
\end{align}
where   $\alpha_i \cdot \alpha_j = 2 \delta_{i,j} - \delta_{i, j +1} -  \delta_{i, j -1} $, which reduces to
Eq.~\eqref{full-int} for $i=j$. 


The classification of 1- and 2-events  in the bosonic $SU(N)$ PCM on $\R \times S^1$ is quite analogous to the classification of tunneling events in deformed Yang-Mills on  $\R^3 \times S^1$  and $\mathbb{CP}^{N-1} $ on 
 $\R \times S^1$.   There are tunneling 1-events in field space associated with the change of the field by   $ \alpha \in A_{N-1} $,  where $\alpha$ is any element of root space.  These are uncorrelated single events which can be embedded as exact solution into $SU(N)$ PCM.   At level-2 and above, there are a number of subtle issues which we address below.  
 
\begin{itemize}
\item {{\bf Neutral bions:}  These are correlated  $[{\F_i \bar \F_i}]_{\pm}= [\B_{ii}]_{\pm} $ events,  a tunneling occurring in an SU(2) subgroup associated with root $\alpha_i$ followed by an anti-tunneling associated with  $-\alpha_i$.  In the parametrization of Eq.~\eqref{algebra-par}, 
we have 
\begin{align}
W_0 \rightarrow W_0 + \pi H_{\alpha_i}  - \pi H_{\alpha_i} \,.
\end{align}
 These  exist  for all positive entries of the extended Cartan matrix $\hat A_{ii}$.  For $\hat A_{ii}>0$, the interaction between the constituents is attractive.  Since neutral bions have the quantum numbers of the perturbative vacuum, they can (and do)  play a role in the cancellation of the ambiguities of perturbation theory. Furthermore, since ${\alpha_i}  - {\alpha_i} =0$, there is no single uncorrelated 
 event associated with neutral bion. By its very nature, it is a correlated event. 
 It also generates a non-perturbative contribution for the background holonomy  potential
\eqref{twist2} on top of the one-loop  perturbative potential \eqref{spotential}, and may play a role in the 
deconfinement phase transition (rapid crossover, for finite $N$) in PCM. }
\item { {\bf Charged bions:} These are correlated  $[\F_i \bar \F_{j}] =\B_{ij} $ events that exist for all
  negative  entries of the extended Cartan matrix, $\hat A_{ij} $, thus $j= i \pm 1$, which can be described as 
  \begin{align}
&W_0 \rightarrow W_0 + \pi H_{\alpha_i}  - \pi H_{\alpha_j}  \nonumber \\
&\underbrace{\longrightarrow}_{{\rm e.g.}, j=i+1} W_0 + \pi  
 \left ( \begin{array}{cccccc }
0 & &&&  &\cr
  & \ddots  &&& &  \cr
  &&  
   \left ( \begin{array}{lll}
1_{i, i} &  0&  0\cr
0 &-2_{i+1, i+1}&0  \cr
0 & 0&  1_{i+2, i+2}
    \end{array} \right) 
   &&  & \cr 
      &&&& \ddots & \cr
      &&&& &0
  \end{array} \right) \,.
\end{align}
 Since $\hat A_{i, i \pm1} < 0$, the interaction between the constituents is repulsive, and there is no ambiguity associated with these two-events. Furthermore, since ${\alpha_i}  - {\alpha_{i\pm1}} $ 
 is not a simple root, there 
  there is no single uncorrelated   event associated with charged bion. Like the neutral bion, 
  by its very nature, the charged bions are fundamentally correlated events. 
  In theories with fermions, they play the leading role in the generation of the mass gap. 
 }
 \item {{\bf Higher action elementary fractons:} 
 For all  roots of the Lie algebra of $\mathfrak{su}(N)$, an exact solution can be embedded into an $SU(2)$ subgroup of $SU(N)$. In particular this includes the roots  $\alpha_i +  \alpha_{i+1} \in \Delta_2^{\rm aff} $  associated with the  tunneling event
   \begin{align}
W_0 \rightarrow W_0 + \pi H_{\alpha_i + \alpha_{i+1}}   \longrightarrow  W_0 + \pi  
 \left ( \begin{array}{cccccc }
0 & &&&  &\cr
  & \ddots  &&& &  \cr
  &&  
   \left ( \begin{array}{lll}
1_{i, i} &  0&  0\cr
0 & 0 &0  \cr
0 & 0&  -1_{i+2, i+2}
    \end{array} \right) 
   &&  & \cr 
      &&&& \ddots & \cr
      &&&& &0
  \end{array} \right) \,.
\end{align}
Since this tunneling event is associated with a root, unlike the neutral and charged bions events, it should be considered as an elementary event, even though it carries two units of action $2S_F$. There is no ambiguity  in the NP-part of the elementary events, while P-fluctuations around them will always have ambiguities.  }

 \item {{\bf Higher action composite fracton-fracton pairs:} Consider  two roots $\alpha$ and $\beta$ 
 whose sum $\alpha+ \beta$  is not a root itself,  for example, $2 \alpha_i$. In all such cases, the tunneling event must be seen as composite.  This is because one cannot embed a simple  $SU(2)$ fracton associated with the sum  $\alpha+ \beta$ into $SU(N)$. 
 For {\it all} such correlated events, the interaction between the constituents is {\it always} repulsive, and so there is no ambiguity associated with the NP-part of such events:
   \begin{align}
W_0 \rightarrow W_0 + \pi H_{\alpha} +   \pi H_{\beta}\,.   
\end{align} 
For example,  for  $SU(N), N \geq 3$,  $\F_{\alpha_1 + \alpha_2}$ is a single event  with action $2S_F$ 
while $[\F_{\alpha_1} \F_ { \alpha_1}]$ is a correlated event with action $2S_F$.  This is because $\alpha_1 + \alpha_2$  is a root, while $2 \alpha_1 $ is not a root. }
\end{itemize}

With these remarks in mind, we observe that the rows of the resurgence triangle of Fig.~\ref{fig:ResurgenceTriangle} get ``ramified" according to a Lie algebraic structure.  For example, the first three rows 
of the resurgence triangle in Fig.~\ref{fig:ResurgenceTriangle}
should really have been written as :
  \begin{eqnarray}
&[0] & \nonumber \\ \cr
 [\bar{\F}_{\alpha_i}]  
&& 
[\F_{\alpha_i} ] \\ \cr
 \{  [ \bar  \F_{\alpha_i}^2], \; [\bar \F_{\alpha_i + \alpha_{i+1}}]   \} \qquad \qquad & 
\{   [\F_{\alpha_i} \bar \F_{\alpha_i}]_{\pm}, \; [\F_{\alpha_i} \bar \F_{\alpha_{i \pm 1}}]  \}  
  &\qquad \qquad\{  [\F_{\alpha_i}^2], \; \{ [\F_{\alpha_i + \alpha_{i+1}}]  \} 
\nonumber   
\label{triangle3}
\end{eqnarray}
This carries rather refined information about the structure of the semi-classical expansion. 
There are essentially \emph{six} type of events at order 
$e^{-2S_F}$ in the semiclassical expansion. This is the same order as the leading ambiguity in perturbation theory.  Only \emph{one} class out of the six, the neutral bion, participates in fixing the ambiguity of perturbation theory around the perturbative vacuum.   The other classes of events have other roles in the rich inner life of the theory.

Despite the fact that all of these events appear at the same order in the semi-classical expansion, 
  ($e^{-2S_F}$), their amplitudes \footnote{As one can see, most of these amplitudes contain logarithms.  Similar logarithms also appear in the case of resonances in resurgent trans-series\cite{Aniceto:2011nu,Schiappa:2013opa,Aniceto:2013fka}. } differ is crucial ways:
 \begin{eqnarray}
 \label{level-2amp}
 [\F_{\alpha_i}^2] &=&    (- \log (4S_F) - \gamma)  \frac{2S_F}{\pi} e^{-2S_F}\,,   \cr
  [\F_{\alpha_i + \alpha_{i+1}}]  &=&  \sqrt {\frac{ 4 S_F}{\pi}}e^{-2S_F}\,, \cr
   [\F_{\alpha_i} \bar  \F_{\alpha_i}]_{\pm} &=&    (- \log (4S_F) - \gamma \pm i \pi)  \frac{2S_F}{\pi} e^{-2S_F}\,,  \cr    [\F_{\alpha_i} \bar \F_{\alpha_{i \pm 1}}] &=&  (- \log (2S_F) - \gamma)  \frac{2S_F}{\pi} e^{-2S_F}\,.
       \end{eqnarray}
Even the two events  which  carry ``charge" 2 under the emergent topological structure, $\{  [\F_{\alpha_i}^2], \; [\F_{\alpha_i + \alpha_{i+1}}]   \}$ differ. The reason for this, as explained above, is that $2\alpha_i$ is not a root, and consequently the corresponding event is a correlated one.  On the other hand, ${\alpha_i + \alpha_{i+1}}$ is a root, and is associated with a single tunneling within the corresponding $SU(2)$ subgroup of $SU(N)$. 

The expressions given in Eq.~\eqref{level-2amp} do not include perturbative fluctuations around the NP saddles. 
Incorporating the perturbative fluctuations around the 2-events, we get
\begin{align} 
 [\F \bar {\F}]_{\pm}   \times \Phi_{[\F \bar  {\F}]} \equiv
 \left( - \log \left[\frac{32  \pi }{g^2 N} \right] -\gamma  \pm i \pi \right)   \frac{16  }{ g^2N }  e^{-  \frac{16 \pi }{g^2 N} } 
  \times  \sum_{n=0}^{\infty} a_n^{[\F  \bar {\F}]}  g^{2n} \, ,
\end{align}
for the   $[\F \bar {\F}]_{\pm}$ event, and similar expression for other events.

\subsection{Reality of resurgent trans-series for real $\lambda$ and BE-summability}
\label{sec:Cancellations}
The principal chiral model is a matrix field theory with a real action and a stable ground state.  Hence its partition function must be real and unambiguous, similar to our toy  example Eq.~\eqref{part}.    In the preceding sections, however, we have calculated the leading perturbative and non-perturbative contributions to the the partition function, and have found that:
\begin{enumerate}
\item Perturbation theory is non-Borel resummable on the $\arg(g^2)= 0$ Stokes line, meaning that the Borel sum of the perturbative series has a two-fold ambiguous imaginary part. 
\item  At second order in semi-classical expansion, we have calculated the neutral bion amplitude 
(via the BZJ prescription) and showed that  it also has  a two-fold ambiguous imaginary part.
\end{enumerate}

Each of these ambiguities would by themselves be disastrous, indicating that the theory is  not well-defined.   The framework of resurgence suggests the resolution.   Perturbation theory by itself is indeed not well-defined.  The semi-classical expansion by itself is not well-defined either. 
 However, neither is a direct physical observable, only their sum is.  In fact, the ambiguities  at order $e^{-2S_F}$ cancel exactly to yield a result which is ambiguity free up to order $e^{-4S_F}$. Resurgence is the statement that these cancellations repeat order by order in the  resurgent trans-series expansion for every physical observable.

 Using Eq.~\eqref{eq:Borel-res-2} and  Eq.~\eqref{Stokes-auto},  the right/left Borel resummation of the 
perturbative series  can now be written in terms of fracton amplitudes as: 
 \begin{align}
 {\cal S}_{0^\pm} {\mathcal{E}}   &= 
 \Re {\cal S}_{0} {\mathcal{E}}  \pm  \frac{\rm s}{2}  [\F] [\bar \F],  \qquad {\rm  s} = - 4 \pi i   \,,
\label{disc4} 
\end{align}
where ${\rm  s} = - 4 \pi i$ is the purely imaginary  Stokes constant (analytic invariant) of the problem.  
Similarly, using Eq.~\eqref{ff-int}, the right/left neutral bion amplitudes are given by
\begin{align}
[\F \bar{\F} ]_{\pm} +  [\bar \F  {\F}]_{\pm}  = &  
2 (- \log (4S_F) - \gamma)  [\F ][\bar{\F}]  \mp  \frac{\rm s}{2}  [\F][\bar{\F}]   .
\label{ffbar-int-2}
\end{align} 
 Note that the \emph{same} Stokes constant appears in the imaginary ambiguous part of the neutral bion amplitude.  

The ambiguities in both quantities are a manifestation of the fact that we are performing an expansion on a Stokes line. Consequently, the perturbative series is non-Borel summable, and exhibits a Stokes jump, 
which is {\it mirrored} by the jump  in the  neutral bion amplitude, leading to the cancellation of ambiguities: 
 \begin{align}
 {\rm Im} \left[  {\cal S}_{0^\pm} {\cal E}+ \(  [{\cal F} \overline {\cal F}]_{0^\pm}  +  [\overline {\cal F}  {\cal F}]_{0^\pm} \)    \right] =0.  \;\;  
\label{a-c-3}
\end{align}
This is the  counterpart of the cancellation of ambiguities we saw in  $d=0$ example Eq.~\eqref{a-c-2}, and it is an explicit realization of the median resummation \cite{Ecalle:1981,Aniceto:2013fka} and BE-summability. 

The sum is ambiguity free up to higher order effects, and the non-canceling terms are of the form 
 \begin{align}
{\cal S}_{0^\pm} {\mathcal{E}}    +  [\F \bar{\F}]_{\pm} +  [\bar \F  {\F}]_{\pm}  = 
 \Re {\cal S}_{0} {\mathcal{E}}  +  2  \left( - \log \left[\frac{32  \pi }{g^2 N} \right] -\gamma \right)   \frac{16  }{ g^2N }  e^{-  \frac{16 \pi }{g^2 N} } \,.
 \label{eq:BE}
\end{align}
Physically, this quantity is the average of the ground state and first excited state. The difference of the first excited state and the ground state is the mass gap, and will be discussed in the next section. 

 Despite the fact that we have only shown Eq.~\eqref{a-c-3} to be true at order $O(e^{-2S_F})$, resurgence actually implies that  
  \begin{align}
 {\rm Im} \left(  {\cal S}_{0^\pm} \Phi_0 + 2   [{\cal F} \overline {\cal F}]_{0^\pm} \times   {\cal S}_{0^\pm} 
 \Phi_{[{\cal F} \overline {\cal F}]}   \right) =0  \;\; { \rm up \; to}  \; O(e^{-4S_F}) \,.
\label{a-c-4}
\end{align}
 This expression can be inverted, using   
  Eq.~\eqref{dispersion},  to obtain the non-alternating part of the late terms in the perturbative expansion around the perturbative vacuum:
  \begin{align}
a_n^{[0]} \sim - \frac{2 }{\pi } \frac{\Gamma(n+1)}{(2S_F)^n} \left[ a_0^{ [{\cal F} \overline {\cal F}] }  +  a_1^{    [{\cal F} \overline {\cal F}] }  \frac{(2S_F)}{n} + a_2^{[{\cal F} \overline {\cal F}]} \frac{ (2S_F)^2 }{n (n-1)}  +  \ldots  \right]   + \ldots 
\end{align}
which indeed agrees with Eq.~\eqref{eq:LargeOrderPT}. Note that this expression is very much in the same spirit as our zero dimensional example Eq.~\eqref{late-1}.

 Finally, although we do not attempt to derive it here (it is beyond the scope of our present work), 
 we believe that all of the formal series appearing in our problem form an infinite dimensional algebra, the  resurgence algebra,  closed under the action of the singularity (``alien") derivative. For  example,  we expect to have the relations 
\begin{align}  
&\Delta_{2S_F}  \Phi_{0}  = {\rm s_1}  \Phi_{[{\cal F} \overline {\cal F}] }\,, \\
&\Delta_{4S_F}  \Phi_{0}  = {\rm s_2}  \Phi_{[{\cal F}^2 \overline {\cal F}^2] }\,, \\
&\Delta_{-2S_F}   \Phi_{[{\cal F} \overline {\cal F}] }   = \tilde {\rm s}_1  \Phi_{0}\,, \\
&\notag \cdots
\end{align} 
 where ${\rm s}_i, \tilde {\rm s}_1 $ are Stokes constants and $\Delta$ are Ecalle's alien derivatives.
The first  one of these relations means that the action of the alien derivative  on  $\Phi_{0}$   at $ 2S_F$    yields the formal perturbative series around the $ [{\cal F} \overline {\cal F}] $ saddle point.   The second relation means that the action of the alien derivative on $\Phi_0$ at $4S_F$ yields the formal perturbative series describing fluctuations around the $[\mathcal{F}^2 \overline{\mathcal{F}}^2  ] $ saddle.  The third relation means that 
the action of the alien derivative  on  $ \Phi_{[{\cal F} \overline {\cal F}] } $  at $ -2S_F$  should   yield the formal series around the perturbative saddle. It would be very interesting to derive the entire resurgence algebra for QM or QFT.

\subsection{Lefschetz thimbles and  geometrization  of ambiguity cancellation}

In this section, we briefly sketch our conjecture of the geometric reasons for the ambiguity cancellation, in connection with semi-classics and Lefschetz thimbles. 

Our results strongly suggest that the geometric explanation of the cancellations  imaginary parts is very similar in the $d=0$ and $d=1,2$ examples. 
It is tied up with the steepest descent  (or semi-classical) expansion. The lesson of resurgence theory is that whenever we consider a semi-classical  expansion, we should in fact always work with a \emph{complexified} version of the path integral.  That this is the case becomes clear when one appreciates the nature of our approach in the preceding sections, which involved analytically continuing $g$, and hence to make sense of the ambiguities and use resurgence theory, we had to work with a complexified version of the path integral.  

More specifically, in our example we first have to generalize 
\begin{align}
U(x) \in SU(N) \longrightarrow Z(x) \in SL(N, \mathbb C)\,,
\end{align}
where $SL(N, \mathbb C)$ is the complex special linear group. 
In a lattice version of the model, 
the infinite-dimensional path integral is regularized into  a finite dimensional integral\footnote{Note that a deformation of the domain of integration to a complexified extension of the path integral, and subsequently integrating over Lefschetz thimbles where the phase is stationary, might be of extreme importance in theories affected by the sign problem \cite{Cristoforetti:2012su,Aarts:2013fpa}.}, and the connection of resurgence theory to the Lefshetz thimble decompositions of integration cycles (which was explained in detail in \cite{Witten:2010cx,Witten:2010zr}) becomes particularly clear. Let the two dimensional lattice have $L_1 L_2$ sites. Then the original partition function is an integral over 
$ [SU(N)]^{L_1 L_2}  $, an $(N^2-1)L_1L_2$ dimensional space.   When we complexify,  the dimension of space is doubled,     
\begin{align}
{\rm dim}\left( [SL(N, \mathbb C)]^{L_1L_2}  \right)=  
2(N^2-1)L_1L_2 .
\end{align}
But since our goal is to find an analytic continuation of the original integral, we must pick special integration cycles within the complexified field space.  Hence the integration runs over a sub-manifold $\Sigma$  which is again   $(N^2-1)L_1L_2$ dimensional, same as original purely real cycle. 

For each saddle in the discretized theory, there exists a unique  Lefschetz thimble attached to it, ${\cal J}_{[\rm saddle]}(\theta)$, whose structure depends on  the phase $\arg(g^2)= \theta$. In fact, in the finite dimensional case,  ${\cal J}_{[\rm saddle]}(\theta)$ lives in  $[SL(N, \mathbb C)]^{L_1L_2}$ 
and  its dimension is half of the complex space.  The analytic continuation of the  original integration cycle may be expressed as a linear combination of  Lefschetz thimbles: 
\begin{align}
\Sigma(\theta) = \sum_{i \in \rm saddles} n_i {\cal J}_{[i]}(\theta)\,,
\end{align} 
and $n_i$ are piece-wise constant (in Stokes wedges) but jump at the Stokes lines.

For example, consider the  Lefschetz thimble attached to perturbative vacuum, ${\cal J}_{[0]}(\theta)$ at  $\theta=0^+$ and $\theta=0^{-}$.   The integration cycle must have a dramatic change  (upon crossing a Stokes line) which does not alter the real part of the integration,  but leads  to a jump in the imaginary part.  Fig.~\ref{fig:jump2}  provides a cartoon of this phenomenon for an ordinary integral. In fact, Fig.~\ref{fig:jump2} is related to our present problem via a dimensional reduction, in the one-site  limit of the lattice model $L_1=L_2=1$ with 
a twisted reduction in the $L_2$ direction. Fig.~\ref{fig:jump2} shows that 
the real part of the cycle must remain  unaltered  upon a Stokes jump at $\theta=0$, while the ``tail" in imaginary direction is reversed. We believe that a generalization of these phenomena to the continuum limit, where the integrals become infinite dimensional, is operative in the path integrals of quantum mechanics and field theory\footnote{See also \cite{Ferrante:2013hg}.}, although the explicit construction of the Lefshetz thimbles in the infinite dimensional case may be quite subtle\cite{Witten:2010zr,Harlow:2011ny} and may require  a generalization of the techniques explained in \cite{Witten:2010cx,Witten:2010zr}.   

We expect infinitely many thimbles associated with the infinite number of saddle points 
of the path integral in the semi-classical domain. In fact, we can make the thimble version of our resurgence triangle (in the  unramified notation to avoid clutter):
 \begin{eqnarray}
&{\cal J}_{[0]} & \nonumber \\ \cr
{\cal J}_{ [\bar{\F}_1] } 
&& 
{\cal J}_{ [\F_1] } \nonumber\\ \cr
{\cal J}_{[\bar \F_2]}
\qquad \qquad & 
{\cal J}_{ [\bar \F_1\F_1]  }&
\qquad \qquad
 {\cal J}_{ [\F_2]}
\\   \cr
{\cal J}_{ [\bar \F_3]}
\qquad \qquad
{\cal J}_{ [\bar \F_2 \F_1]}
&& 
{\cal J}_{ [\bar \F_1 \F_2]}
\qquad \qquad
{\cal J}_{ [\F_3] }  \nonumber\\ \cr
 &\vdots & \qquad \qquad\qquad \qquad\qquad \nonumber
\label{triangle4}
\end{eqnarray}

The combination of the solvable $d=0$ dimensionally reduced model and the non-Borel summability of the perturbative series Eq.~\eqref{eq:Borel-res-2} tells us that the integral over just the 
 P-thimble 
${\cal J}_{[0]}(\theta)$ will have a pathological  $\theta=0$ limit, resulting in a two-fold ambiguous result, depending on  the direction of the approach to  $\theta=0$ in the complex $g^2$ plane. 
 We have already seen that this pathology can be  fixed  by integration over other NP-thimbles.  
 The integration over the fracton-anti-fracton thimble  
 ${\cal J}_{[\F \bar  \F]}(0^{\pm})$ contributes at the same order as the ambiguity of the P-thimble on the $\theta=0$ direction.  In fact, we expect 
  \begin{align}
 \int_{{\cal J}_{[0]} (0^-) +  {\cal J}_{[\F \bar  \F]} (0^-) }  DZ(x_1,x_2) e^{-S[Z(x_1,x_2)]} 
   \end{align}
   to be ambiguity free at order $e^{-2S_F}$, but to have some  
ambiguities at order $e^{-4S_F}$, which are cancelled thanks to the fact that the full integration contour includes thimbles which pass through the appropriate higher-action NP saddle points, and so on. 
It would be very interesting to understand the structure of all the thimbles and Stokes phenomena in this problem.  We believe that this would provide a geometric understanding of the intricate relations between P and NP data which leads to the  cancellation of ambiguities in resurgent trans-series, as described in e.g. \cite{Aniceto:2013fka}. 



\section{Mass gap flow and Borel flow}
\label{sec:Borelflow}
In one of the first works on renormalons, 't Hooft speculated that they may be connected to the mass gap and confinement in gauge theories\cite{tHooft:1977am}.  Using resurgence, we find a refinement 
and a  confirmation of this idea in a semi-classical regime continuously connected to the strongly coupled regime of gauge theories and non-linear sigma models. 

 In every semi-classically calculable example studied so far, it turns out that the mass gap is due to {\it half} a renormalon in the semi-classical domain\cite{Argyres:2012ka,Argyres:2012vv,Dunne:2012ae,Dunne:2012zk,Cherman:2013yfa}. 
  In deformed Yang-Mills on small $\R^3 \times S^1$,  
 the mass gap is due to monopole-instantons ${\cal M}$ \cite{Unsal:2008ch}, while $[{\cal M}  \bar {\cal M}]_{\pm}$ yields the 
 leading semi-classical realization of the renormalons\cite{Argyres:2012ka,Argyres:2012vv}. 
  In ${\cal N}$=1 SYM and QCD(adj)  on small $\R^3 \times S^1$,   the mass gap is due to magnetic bions ${\cal B}$ \cite{Unsal:2007jx}, while the 
 leading semi-classical realization of the renormalon is  the neutral bion $[{\cal B}  \bar {\cal B}]_{\pm}$ 
 \cite{Argyres:2012ka,Argyres:2012vv}, {\it etc}.

In the SU(2) PCM,  the evaluation of the mass gap in the small-$L$  regime follows very closely the calculation of the mass gap in the  $\mathbb {CP}^1$ model.    In Section~\ref{sec:Cancellations},  we already argued that the weak coupling realization of the renormalon is again the neutral bion,  $[{\cal F}  \bar {\cal F}]_{\pm}$. Below, we show that  
 the mass gap  at  leading order  in semi-classical expansion in the 
small $\R \times S^1$ regime is due to 
fractons ${\cal F}$.  
This is also true   for  the $SU(N)$ model in the  
$\frac{N L \Lambda}{2 \pi} \ll 1$ small circle limit adiabatically connected to $\R^2$.

\begin{figure}[tbp]
\centering
\includegraphics[width=0.40\textwidth]{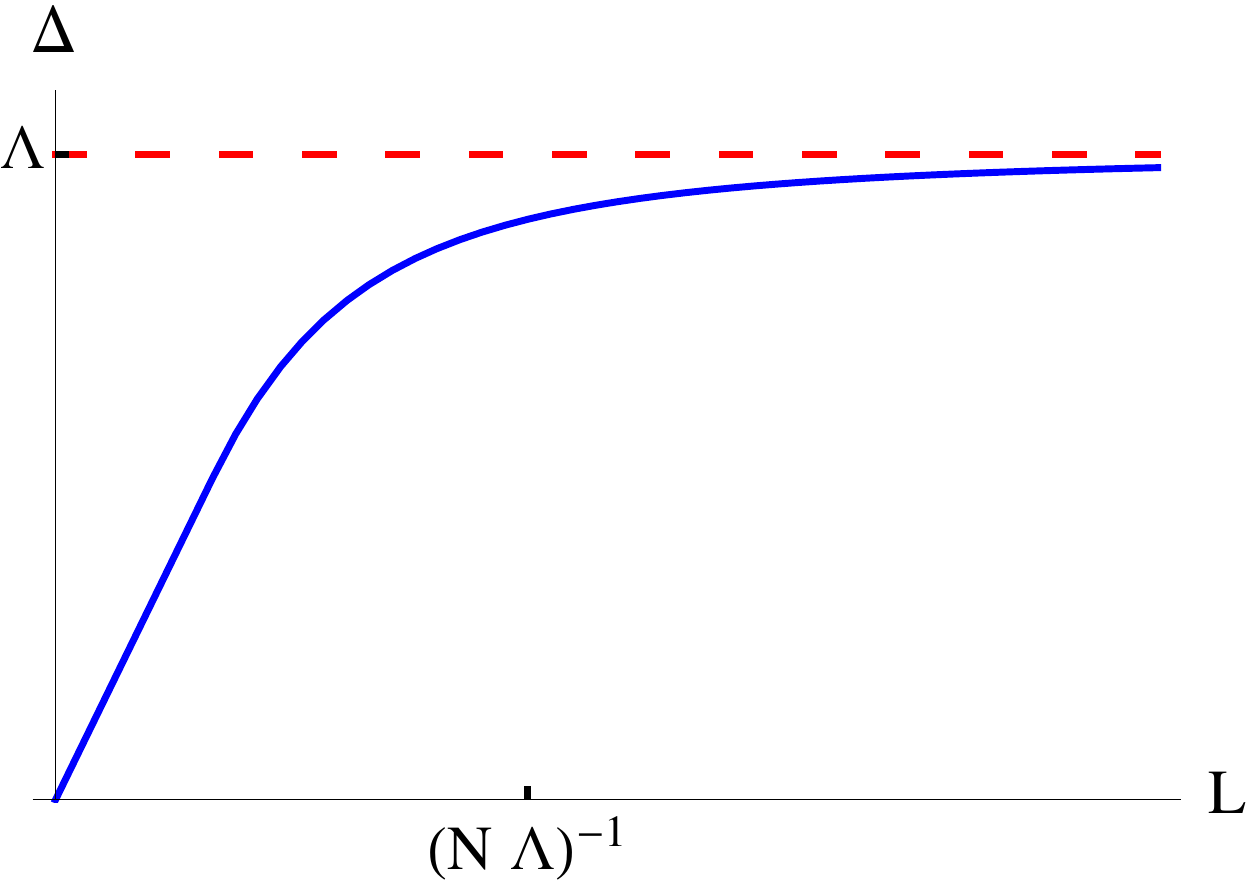}
\caption{{\bf Mass gap flow:}  The mass gap on  the  small $\R^1 \times S^1$ regime, corresponding to  $ \frac{ LN \Lambda}{2 \pi} \ll1$   is semi-classically calculable.  At leading order, it is a   one fracton effect.  On $\R^2$ or large  $\R^1 \times S^1$,    a  reliable analytical method which can address the mass gap question is at present unknown.  
Our small $\R^1 \times S^1$  theory is adiabatically connected to the theory on $\R^2$. 
   }
\label{fig:massgapflow}
\end{figure}

 The mass gap is defined as the energy required to excite the system from the ground state  $E^{(0)}$ to the first excited state $E^{(1)}$. 
   As discussed in Section~\ref{sec:pert-th}, in the small-$L$ regime, we can work with 
   the Hamiltonian Eq.~\eqref{Ham-master} which describes the dynamics of a small-$L$ EFT with zero KK-momentum,  or of a small-$L$ EFT with $-1$ units of winding number on the $S^1$. In the Born-Oppenheimer approximation,  we can further focus on 
     Eq.~\eqref{eq:BOHamiltonian}, since the states  which carry non-zero $P_{\phi_1}$ and $P_{\phi_2}$ momentum      acquire a gap of order $g^2/L$, while (as we show below) the low lying states of Eq.~\eqref{Ham-master} are 
     split by a non-perturbatively small amount, justifying the use of the  Born-Oppenheimer approximation.
     
        In the $SU(2)$  PCM the ground state is  two-fold degenerate to all orders in perturbation theory.  This degeneracy is lifted by non-perturbative  fracton effects.   
 The two lowest lying eigenstates described by Eq.~\eqref{Ham-master} have the quantum numbers   $| \pm, n_1=0, n_2=0 \rangle $, where  $\pm$ are the eigenvalues of the parity  operator $P$ which acts on the polar coordinate $\theta$ in the Hopf parametrization as $\theta \to \pi - \theta$.
 At leading non-perturbative order in the semiclassical expansion the $SU(2)$ PCM mass gap is then given by 
\begin{align}
\Delta = (E_{-}-E_{+}) =  \xi  (2 \F_1  + 2 \F_2)= \frac{4 \pi}{L} \sqrt \frac{16  }{ g^2N }  e^{-  \frac{4 \pi }{g^2 } }    \sim \Lambda ( \Lambda L)\,.
\end{align}
For $SU(N)$, parametrically\footnote{At leading order in the $SU(N)$ PCM, the mass gap gets contributions from the $N$ minimal-action fractons which describe tunneling between the ground state and the $N$ directions in field space parametrized by of the $N$ affine simple roots.  To compute an exact expression for the mass gap in the $SU(N)$ case, one must diagonalize the resulting ``tunneling matrix" and compute its smallest eigenvalue.  The expression we show in Eq.~\eqref{gap-2} is the parametric form of the result which this calculation would give.}, the mass gap is of order
 \begin{align}
\Delta = (E_{-}-E_{+})  \sim  \frac{2 \pi }{LN}  \sqrt \frac{16  }{ g^2N }  e^{-\frac{8 \pi }{g^2 N} }   \sim 
  \frac{1 }{LN} e^{-S_{\rm uniton}/N}
  \sim 
\Lambda ( \Lambda L N)\,,
\label{gap-2}
 \end{align} 
in the weak coupling semi-classical regime $\frac{N L \Lambda}{2 \pi} \ll 1$.\footnote{
The appearance of the non-perturbative factor in the mass gap is the major difference with respect to thermal compactfication. In thermal theory at small $S^1_\beta \times \R$, 
usual KK-reduction works, and the gap is  given $\Delta_{\rm gap}^{\rm thermal} \sim  \frac{g^2}{L}$. The thermal low energy theory does not remember its two dimensional origin, in contrast to the adiabatic small-$L$ limit we have constructed.}  Note that the mass gap at small-$L$ is $N$-independent in the large $N$ limit, just as it is on $\mathbb{R}^2$, because when $N L \Lambda \ll 1$ and $N \gg 1$, $L$ scales as $L\sim 1/N$.  This lends further support for the claim that our small-$L$ limit is adiabatic.
  At the boundary of the region of validity of the semi-classical regime ${N L \Lambda} \sim 1$ where we can no longer   rely on semiclassics, we observe that the mass gap acquires a strong scale value $m_g \sim \Lambda$.   In the strong coupling regime, ${N L \Lambda} \gg1 $, we expect  
  the mass gap to be independent of the size of the circle.  In fact, the onset of $\mathbb{R}^2$ behavior at the compactification scale $(\Lambda/N)^{-1}$ rather than $\Lambda^{-1}$ is a hallmark of large $N$ volume independence, which can be shown to apply to the theory we are working with.  Therefore, provided we are given 
  a value $m_0$ for the gap at some ${N L_0 \Lambda} \gg1 $,   as $N$ is varied $m_0$  can only change by order $O(1/N^2)$ corrections in the $SU(N)$ model.  In other words, we expect the mass gap to plateau and remain fixed in this regime as shown in Fig.~\ref{fig:massgapflow}, which shows the expected form of the mass gap as a function of $L$.

The connection with renormalons should now be clear.  The field configurations $\F_i$ that give rise to the mass gap at order $e^{-S_F}$ then produce the leading renormalon singularities at the next order of the semi-classical expansion $e^{-2S_F}$.  So the mass gap is tied to ``half" of a renormalon.  This is a  concrete realization of 't Hooft's idea\cite{tHooft:1977am}.

\begin{figure}[tbp]
\centering
\includegraphics[width=0.65\textwidth]{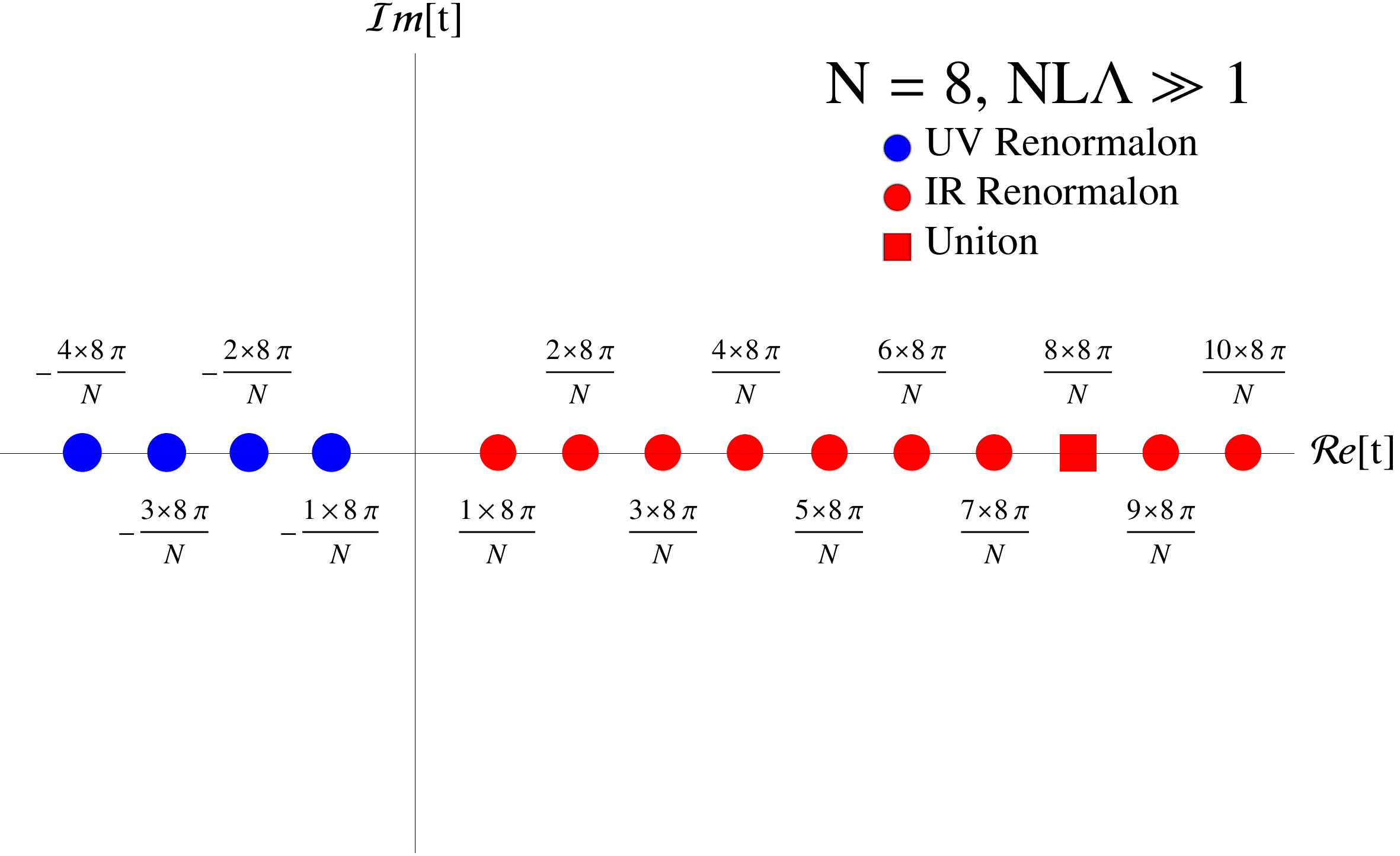}\\
\vspace{1cm}
\includegraphics[width=0.65\textwidth]{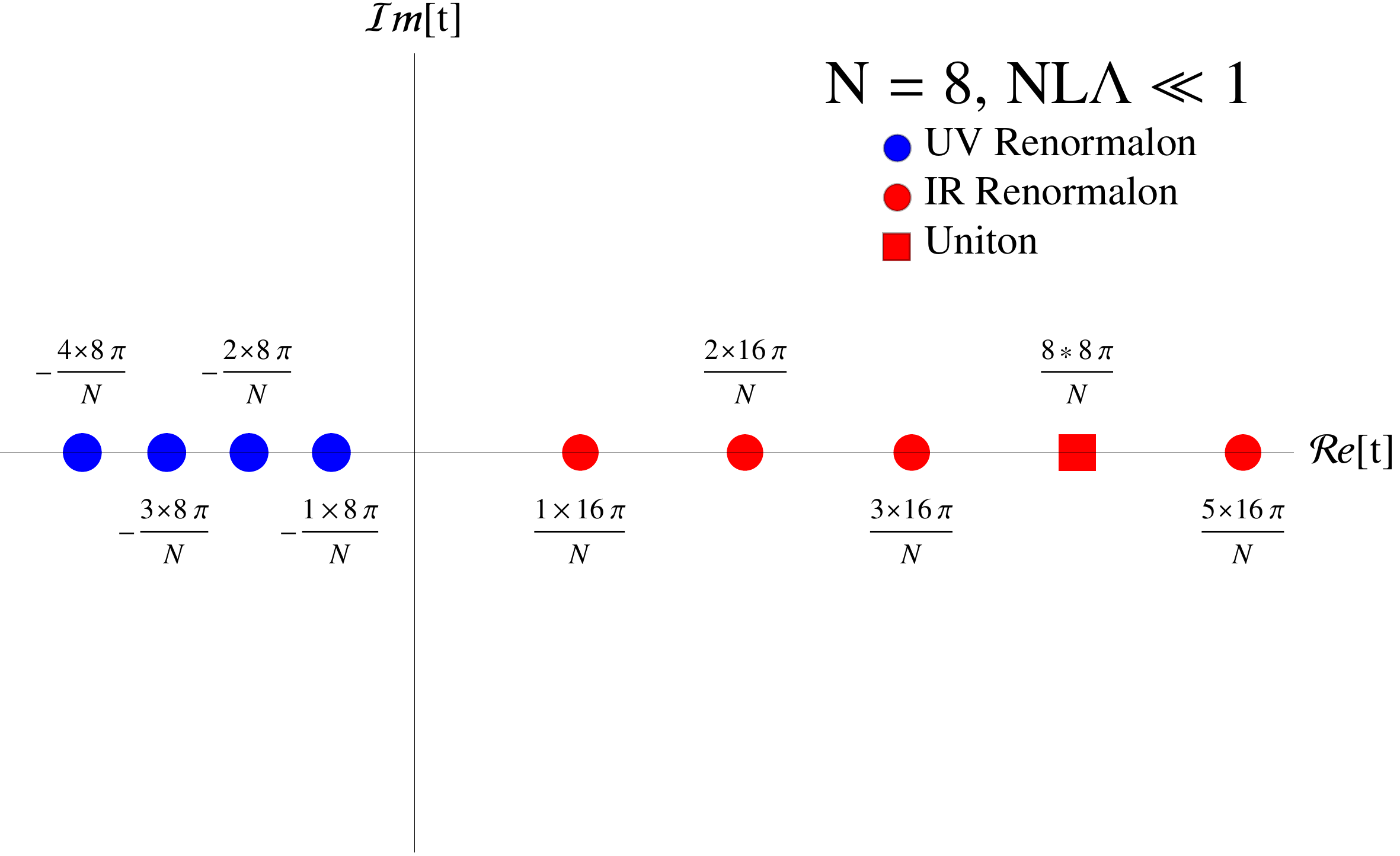}
\caption{{\bf Borel flow:} {(Top)}   The expected structure of the Borel plane on $\R^2$. 
{ (Bottom)} The Borel plane structure on small $\R^1 \times S^1$.  The small-$L$ effective field theory (in which small-$L$ physics is integrated out) does not capture the UV renormalon structure by construction, but does capture the IR-renormalon singularities of the small-$L$ theory.
   }
\label{fig:Borelflow}
\end{figure}

\vspace{3mm}
\noindent{\bf  Borel flow}
\vspace{3mm}

 \noindent
The idea of Borel flow is a more abstract version of the mass gap flow.
Borel flow is tied up with all non-perturbative observables in the problem. 
 The IR-singularities in the Borel plane on the small $S^1_L \times \R$ regime and on $\R^2$ 
  are located at 
 \begin{align}
 t^{ S^1_L \times \R}_m &= \frac{16 \pi}{N} \; \;  m, \qquad   m= 1,2, \ldots  \cr 
  t^{  \R^2}_m  &= \frac{8 \pi}{N} \; \;  m,  \qquad  m= 1,2, \ldots  
 \end{align}
while the location of the UV-renormalon singularities remain  unchanged no matter the value of $L$.  See Fig.~\ref{fig:Borelflow}.
  The most dominant singularities $(m=1)$   lead to  ambiguities of order 
 \begin{align}
S^1_L \times \R  :& \qquad   \pm i e^{-\frac{16 \pi}{g^2N(1/LN)} }\sim   \pm i  (\Lambda L N)^4 \,,\cr 
  \R^2 : & \qquad  \pm i e^{-\frac{8 \pi}{g^2N(Q)} }\sim   \pm i  ( \Lambda^2 Q^{-2}) \,,
  \label{small-large}
 \end{align}
 where at small $S^1_L \times \R$, the 't Hooft coupling is evaluated  at distance $LN$, while on $\R^2$, it is determined at a high (Euclidean)  momentum scale $Q$ (entering through an
   OPE with an external momentum insertion $Q$.)   The crucial point is that  $Q \gg \Lambda$   and $\frac{1}{LN}  \gg \Lambda$ so that   the coupling is weak at the scale of $Q$ and $\frac{1}{LN}$, and in both cases, this gives a control parameter over the  small NP-induced term.   On   $S^1_L \times \R$ these ambiguities are cancelled respectively by the ambiguity in neutral bion events shown in Eq.~\eqref{ffbar-int}, while on
  $\R^2$
they are cancelled by the ambiguity in the condensate Eq.~\eqref{condensate}. 
 
 In the semiclassical regime,  the mass gap is generated by half a renormalon, i.e., a fracton.  The dimensionless mass gap is   $m_g LN$, and takes the form 
  \begin{align}
S^1_L \times \R  :& \qquad   m_g LN \sim   (\Lambda^2 L^2N^2) \longrightarrow  
m_g \sim   \Lambda (\Lambda L N )\,,
\end{align} 
demonstrating explicitly the relation between the mass gap and the leading renormalon. Compare this with the first line of Eq.~\eqref{small-large}.  If we accept the behavior we have seen in the semi-classical regime  as a rough guide to the behavior we should expect in the strongly coupled domain, we would deduce that the dimensionless mass gap  (now measured in units of some external large momentum $Q$) behaves as 
 \begin{align}
 \R^2  :& \qquad    m_g Q^{-1} \sim   (\Lambda Q^{-1})  \longrightarrow  
m_g \sim   \Lambda \,,
\end{align}
which is a  sensible result  on $\R^2$.  
Thus, we are tempted to sharpen 't Hooft speculation.  In asymptotically free non-linear sigma models (including for instance $\mathbb{ CP}^{N-1}$, $O(N)$, Grassmannian, and principal chiral model-type matrix field theories)    there may exist a quantitative relation between the mass gap of the theory and the location of the first renormalon singularity for all values of $L$.  

As illustrated in Fig.~\ref{fig:Borelflow} the IR-singularities in the Borel plane are twice as  dense on $\R^2$ with respect to   $\R \times S^1_L$.    The crucial point for adiabatic continuity is the fact that in both regimes the singularities are spaced by units of $\sim \frac{1}{N}$.
Our framework strongly suggests that as we dial the radius from small  to large, the singularities must exhibit a  flow towards the origin rendering them twice as dense.  The same phenomenon should also take place in deformed Yang-Mills, in which the dilution factor between the weak coupling regime and strong coupling regime is $\frac{11}{3}$ \cite{Argyres:2012vv}.

Clearly, the flow of the singularities in the Borel plane as the radius is dialed, i.e, the Borel flow, and the flow of the mass gap as the theory is dialed from a weak coupling to strong coupling are intimately related. They are very likely manifestations of the same underlying dynamics.  We believe that developing a thorough understanding of these flow equations would constitute a major step towards the solution of the mass gap problem in a large variety of asymptotically free theories. 

\section{Discussion and prospects}
\label{sec:Conclusions}

 We have employed resurgence and adiabatic continuity to give a classification of the 
 P- and NP- saddles in the principal chiral model. Due to insights from various techniques, such as lattice Monte Carlo calculations and 
 integrability, these theories were believed to have highly non-trivial similarities to Yang-Mills theory.  But the theory had  no known NP-saddles, except for the uniton saddle discussed mostly in the mathematics literature \cite{uhlenbeck1989harmonic},  whose role in the quantum version of the PCM never became clear. So, from a semi-classical point of view, the only relevant saddle seemed to be the perturbative vacuum.  This appeared to be a dramatic difference from Yang-Mills theory.

In this work we have constructed an infinite class of NP-saddles such as the fractons, as well as the zoo of correlated events, 
which turn out to play a crucial role in the dynamics of the PCM. The fractons lead to the generation of the mass gap of the theory, while neutral bions (correlated fracton-anti-fracton events) lead to the semi-classical realization of the  IR
 renormalons  in the weak coupling calculable regime $\frac{NL \Lambda}{2\pi} \ll 1$. 
 Our analysis is inspired by the recent treatment of the 2D in $\mathbb{CP}^{N-1}$ model \cite{Dunne:2012ae,Dunne:2012zk}, and has  many parallels with the analysis of gauge theories on $\mathbb{R}^3 \times S^1$ initiated in \cite{Unsal:2007jx,Unsal:2008ch,Shifman:2008ja}, and recently revisited in the context of resurgence in \cite{Argyres:2012vv,Argyres:2012ka}.

In the present work, in the small-$L$ regime, we were able to demonstrate the existence of fractons, whose action is  $S_F= S_{\rm uniton}/N$,  by three  independent methods: 
\begin{itemize}
\item{Large order analysis of the small-$L$   perturbative series describing fluctuations around the perturbative saddle point implies, via resurgence, that the non-perturbative completion of the problem involves saddles with action  $2 S_F$.  Moreover, the notion of emergent topology encoded in the resurgence triangle leads to the conclusions that there must also be other saddles, with action 
$S_F$, which are precisely the fractons. }
\item{The effective field theory obtained via adiabatic continuity, which is just a particular quantum mechanical theory, allows a simple study of the NP-saddles. The NP saddles with the smallest action have actions $S_F$, and they come in $N$ different types.}
\item{When the $\Z_N$ symmetric background holonomy is turned on we have seen that the uniton splits into $N$ lumps, each of which carries an action $S_F$, as nicely shown by the plots in Fig.~\ref{fig:SU3SU4Unitons}.}
\end{itemize}

This last phenomenon is morally similar to the splitting of calorons (periodic instantons) into $N$-monopole instantons in gauge theories \cite{Kraan:1998pm,Lee:1997vp}, but it is again worth emphasizing that the PCM does not have any instantons.

\subsection{Future directions}
There are a large number of interesting possible extensions of the study we have performed here.  A few of them are:
\begin{enumerate}
\item{\bf WZW term and sign problem:} Addition of a WZW-term to the action modifies the IR dynamics on $\R^2$ and introduces a sign problem if one were to try to attack the system using Monte Carlo simulations. It would be interesting to study this system on the calculable small-$L$ regime. 

\item{ \bf Borel and mass gap flows:} An  understanding of  the non-perturbative dynamics on $\R^2$ can perhaps  be achieved by summing the resurgent trans-series at small-$L$, and extrapolating the sum to large-$L$.  Less ambitiously, one may develop an understanding of an infinitesimal version of Borel flow from a more detailed analysis of the construction of the action of the small-$L$ effective field theory, by exploring the effects of the higher-derivative terms in the action induced by integrating out high-momentum modes.

\item {\bf PCM with  fermions:}  
The PCM model, like other non-linear sigma models, is asymptotically free for any number of fermion flavors (unlike QCD-like theories on $\R^4$.) It would be interesting to perform a detailed investigation of the impact of fermions on the dynamics, and to determine the boundary between the confining and IR-conformal regimes.  

\item  {\bf Large-N reduced model:}  We can reduce the theory with $N_f \geq 1$ Majorana fermions to a one-site lattice theory, by imposing 't Hooft twisted boundary conditions. Let $\Phi$ represent the bosonic/fermionic degrees of freedom $U, \,\psi_i$ with $\bf{e_1},\bf{e_2}$ the two lattice vectors and then reduce the matrix model to one-site by imposing  \begin{align} 
\Phi ( {\bf x +  e_1}) = \Omega_1 \Phi ( {\bf x }) \Omega_1^{\dagger}\,,  \qquad  \Phi ( {\bf x + e_2}) = \Omega_2 \Phi ( {\bf x }) \Omega_2^{\dagger}\,, \qquad  \Omega_1 \Omega_2= e^{i \frac{2\pi}{N}} \Omega_2 \Omega_1\,,
\end{align}
where the last condition is ensures compatibility of the fields at $\mathbf{x} \sim {\bf x +  e_1 + e_2}$.  It would be useful to examine the dynamics of the associated reduced large-N model. 

\item  {\bf Renormalons in the large-$N$ reduced model:} In the single-site reduction which is enabled by large-$N$ volume independence,  the space-time volume of the theory on  $\R^2$ is mapped to the matrix size  $(N=\infty)$ of the 1-site matrix model. Within planar perturbation theory there is an exact mapping between summation over spacetime momenta, and summation over the adjoint $SU(N)$ indices.  This implies that there must exist a matrix field interpretation for both IR and UV renormalons. It would be interesting to understand this in detail. 
\end{enumerate}

\acknowledgments
We  are very grateful to  Ines Aniceto,  Gokce Basar,  Gerald Dunne, David Gross, Daniel Harlow, Vladimir Kazakov, Peter Koroteev, Sungjay Lee, Jonathan  Maltz, Nick Manton, Soo-Jong Rey, Ricardo Schiappa, Misha Shifman, Edward Witten, Kenny Wong and Larry Yaffe for stimulating and enjoyable discussions.  We are also particularly grateful to Gerald Dunne for collaboration on related work.  M.\"U. thanks KITP, Kobayashi-Maskawa Institute at Nagoya University and Seoul National University for hospitality.  He also thanks 
the participants of the ``New Methods in Nonperturbative Quantum Field Theory" program for fruitful discussions. 
 D. D. is grateful for the support of European Research Council Advanced Grant No. 247252, Properties and Applications of the Gauge/Gravity Correspondence.
 

\appendix
\section{Resurgence Terminology}
\label{sec:ResurgenceAppendix}
In this Appendix we briefly discuss some of the mathematics behind asymptotic series and resurgence methods.  For further details, see e.g.~\cite{Ecalle:1981,Delabaere1999,Marino:2012zq,2007arXiv0706.0137S}.

As we already mentioned at the beginning of our paper, most perturbative series appearing in physics are not convergent. When we compute a generic physical quantity by means of perturbative expansion, generic observables take the form
\begin{equation}
f(g) = \sum_{n=0}^\infty a_n \,g^n\,,\label{series}
\end{equation}
with $g$ is the coupling constant. 
When the constant term $a_0$ in (\ref{series}) vanishes the asymptotic series is called a \textit{small} power series.
As noted long time ago by Dyson and Lipatov\cite{Dyson:1952tj,Lipatov:1976ny}, a generic feature of quantum field theory is the factorial growth $n!$ of the coefficients $a_n$ (i.e. combinatorics of Feynman diagrams or phase space integraltion of UV renormalons), which effectively makes the series Eq.~\eqref{series} divergent for all non-zero $g$.
Indeed, it turns out that series usually found in physics are only asymptotic series, meaning that the difference between the function $f(g)$ and the partial sum tends to zero as 
\begin{equation}
\lim_{g\rightarrow 0} g^{-N}\vert f(g)-\sum_{n=0}^N a_n\,g^n \vert=0\,,
\end{equation}
this for all $N$. Clearly this does not imply convergence, and a naive finite-order partial sum may differ enormously from the actual function $f$.

\begin{figure}[tbp]
\centering
\includegraphics[scale=0.3]{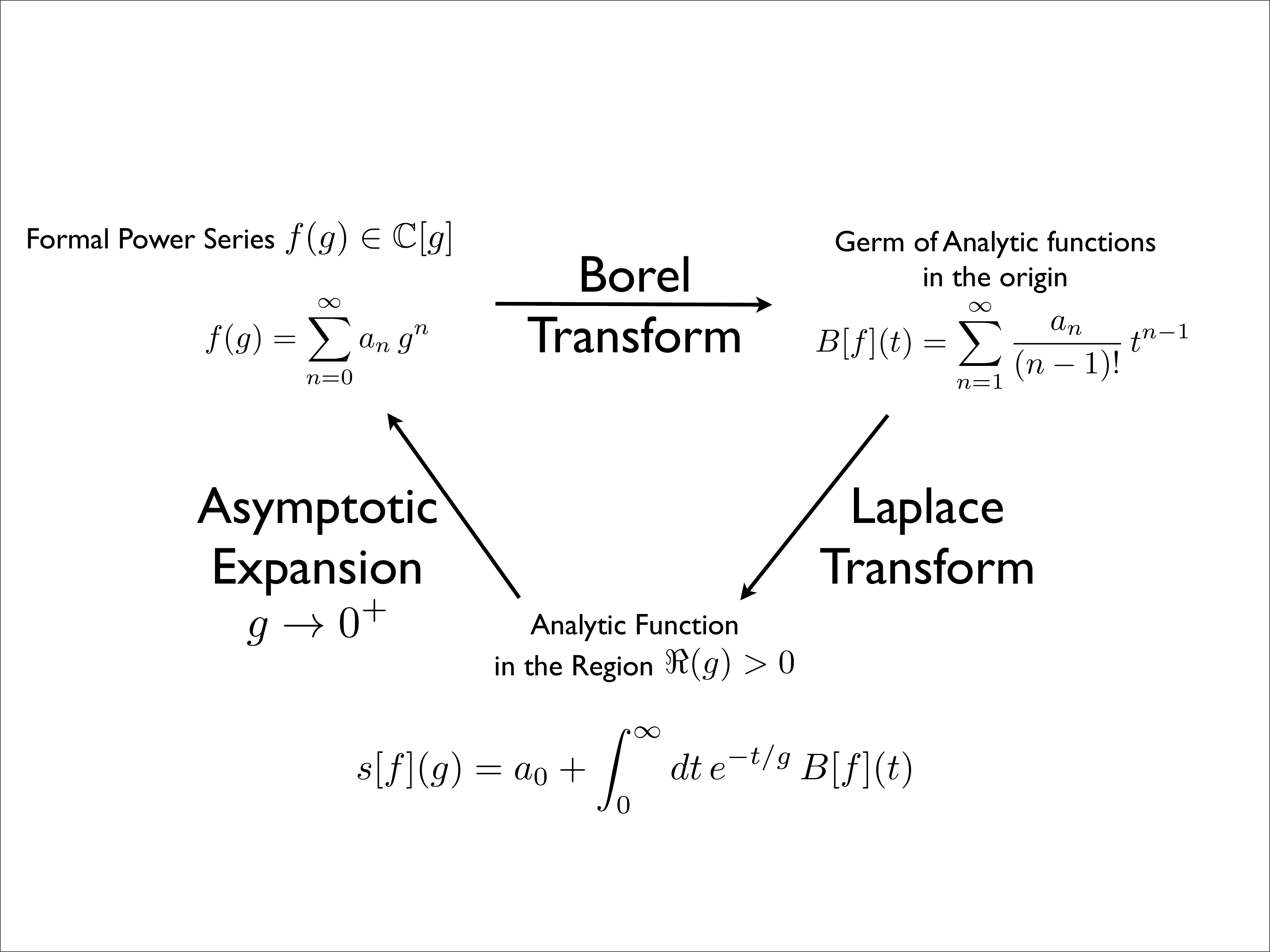}
\caption{Schematic representation of the Borel resummation procedure. }
\label{fig:BorelDiagram}
\end{figure}

We now define an important unitary subalgebra of the algebra of formal power series with coefficients in $\mathbb{C}$, $\mathbb{C}[g]$.  Expansions of the form of Eq.~\eqref{series} are examples of  \textit{Gevrey order 1} formal power series with coefficients $a_n g^n$, which are defined by the property that $\vert a_n \vert /n!$ is growing at most as a geometric series.
Some Gevrey-$1$ series can be assigned a meaningful sum by the method of Borel summation.  To define a Borel sum of a Gevrey-$1$ series, we first insert the factor "$1$" into the series using the well known formula
\begin{equation}
\frac{1}{n!} \int_0^\infty dt\,t^n\,e^{-t} = 1\,.
\end{equation}
This defines the Borel transform of the series.  If one then commutes the integral with the sum, one gets the Borel sum $\mathcal{S}[f](g)$ of the original series
\begin{equation}
\mathcal{S}[f](g) =a_0+ \int_0^\infty dt\, e^{-t/g} B[f](t)\,.\label{BorelSum} 
\end{equation}
Here the Borel operator $B$ takes the formal power series $f(g)$ and gives
\begin{equation}
B[f](t) =\sum_{n=1}^\infty \frac{a_n}{(n-1)!} t^{n-1}\,, \label{BorelTransform}
\end{equation}
called the Borel transform of $f$.
By a change of variables in the integral one can see that the expansion for $g\sim0$ of Eq.~\eqref{BorelTransform} leads to our initial expansion Eq.~\eqref{series}.
The Borel sum, represented by the operator $\mathcal{S}$, is then simply the Laplace transform of the analytic continuation of the Borel transform $B[f]$. This leads to a well defined expression for $\mathcal{S}[f](g)$ as an analytic function in the half-plane $\Re (g) >0$, as shown schematically in Fig.\ref{fig:BorelDiagram}.

As an example of the use of this machinery, consider the following formal series
\begin{equation}
E(g) = \sum_{n=0}^\infty (-1)^n \,n!\, g^{n+1}\,.\label{EulerSeries}
\end{equation}
This is a formal solution to Euler's equation\footnote{Note that usually Euler's equation is written in terms of $x=1/g$.}
\begin{equation}
g^2 E'(g) + E(g) = g\,.\label{Euler}
\end{equation}
The formal series diverges for all $g\neq 0$. However, the coefficients $a_n=(-1)^{n-1}(n-1)!$ alternate in sign, leading to a well defined and unique Borel sum.
The Borel transform can be obtained from the definition Eq.~\eqref{BorelTransform}
\begin{equation}
B[E](t) =\frac{1}{1+t}\,.
\end{equation}
Next one can compute the Laplace transform of $B[E](t)$ to obtain an analytic function in the half-plane $\Re (g)>0$, which solves Euler's equation (\ref{Euler}).

Consider now a simple modification of the formal series Eq.~\eqref{EulerSeries}:
\begin{equation}
F(g) = \sum_{n=0}^\infty n!\, g^{n+1}\,.\label{FSeries} 
\end{equation}
This is a solution of the ordinary differential equation
\begin{equation}
g^2 F'(g) - F(g) = -g\,.
\end{equation}
The coefficients $a_n= (n-1)!$ are now non-alternating,
and the Borel transform of $F(g)$ is
\begin{equation}
B[F](t)=\frac{1}{1-t}\,.
\end{equation}
 Due to the non-alternating nature of the coefficients $a_n$ we obtain a two-fold ambiguity in the Laplace transform, since the integration contour $t\in[0,\infty)$ is a Stokes line.  That is, it contains a pole for $t =1$.
If we allow the contour of integration to move into the complex $t$ plane, also called Borel plane, we can avoid the singularity by either  passing above it $\mathrm{arg}(t) >0 $ or below it $\mathrm{arg}(t)<0$.  Hence we can define two Laplace transform of $B[F]$, $\mathcal{S}_+[F]$ and $\mathcal{S}_-[F]$, obtained by integrating on the two different contours.
The functions $\mathcal{S}_+[F]$ and $\mathcal{S}_-[F]$ give two possible analytic continuation of the original formal series Eq.~\eqref{FSeries}.  One can show that their difference is related to the residue around $t=1$, and is given by
\begin{equation}
\mathcal{S}_+[F](g)-\mathcal{S}_-[F](g) = 2\pi\,i\,e^{-1/g}\,.
\end{equation}
This exponential term is the hallmark of non-perturbative physics, and the example just presented is just a simplified version of what normally happens in generic asymptotically-free quantum field theories.  As a result, the Borel sum is ambiguous since the integration line is then a Stokes line containing  poles (or more generically branch cuts) of the Borel transform.

For a generic asymptotic series $f$ of Gevrey type 1 we will require its Borel transform to have only a "few" singularities in the Borel plane. More precisely  we require the germ of analytic functions $B[f]$ to be \textit{endlessly continuable} on $\mathbb{C}$, meaning that for all $L>0$ there exists only a finite set $\Omega_L(B[f]) \subset \mathbb{C}$, called the set of $L$-accessible singularities, such that $B[f]$ has an analytic continuation along every path whose length is less than $L$, while avoiding the set $\Omega_L(B[f])$.
This definition is slightly stronger than the original definition given by Ecalle. Both our previous example obtained from the $E$ and $F$ series satisfy this requirement.  A conjecture consistent with all results available so far is that in fact the perturbative series arising from physical QFTs also satisfy this requirement, which is necessary for the technology of resurgence theory to be useful.

We will also say that $B[f]$ has only \textit{simple singularities} if for all paths $\gamma$ ending at a singular point $t_\star$, the analytic continuation $B[f]_\gamma$ of the germ $B[f]$, along the path $\gamma$, in a neighborhood of $t_\star$ takes the form
\begin{equation}
B[f]_\gamma(t) = \frac{a_\gamma}{2\pi \,i\, (t-t_\star)}+b_\gamma(t-t_\star) \frac{\log(t-t_\star)}{2\pi\,i}+h_\gamma(t-t_\star)\,,\label{simpleSing}
\end{equation}
where $a_\gamma\in\mathbb{C}$, while $b_\gamma$ and $h_\gamma$ are some analytic germs around the origin.
The germs $b_\gamma$ and $h_\gamma$ are themselves endlessly continuable functions with simple singularities.
With the concepts just introduced, we will define the formal power series $F$ to be a \textit{simple resurgent function} if it is of Gevrey order 1 and its Borel transform is an endlessly continuable function with only simple singularities.
The set of simple resurgent function is actually a subalgebra of $\mathbb{C}[g]$ denoted by $^+ \mathcal{R}(1)$ as proven by Ecalle\cite{Ecalle:1981}.

It is useful at this point to introduce the concept of \textit{directional Borel summation}.
Given a simple resurgent function $f$, we can compute its Borel transform $B[f]$ Eq.~\eqref{BorelTransform}.  Then from the germ $B[f]$ we can compute the directional Laplace transform
\begin{equation}
\mathcal{S}_\theta[f](g) =a_0+ \int_{0}^{e^{i\theta}\infty }dt\, e^{-t/g} B[f](t)\,,
\end{equation}
where the contour of integration is the line, lying in the complex Borel plane, starting from the origin, $t=0$, and going to infinity in the direction $\mathrm{arg}(t) = \theta$.
This integral is convergent in the half-plane defined by $P_\theta = \{ g \in \mathbb{C} \,s.t.\,\mathrm{arg}( t / g) > 0 \}$, by Cauchy's theorem $\mathcal{S}_\theta[f]$ and $\mathcal{S}_\phi[f]$ will coincide on $P_\theta \cap P_\phi$, so they are analytic continuation of one another.
Furthermore they all have the same asymptotic expansion given by Eq.~\eqref{series}.
A direction $\theta$ which contains singularities for $B[f](t)$ is called a Stokes line.  Thus we define lateral Borel sums by considering $\mathcal{S}_{\theta_+}[f]$ and  $\mathcal{S}_{\theta_-}[f]$, where we avoid the singularity by slightly deforming the contour of integration $\mathrm{arg}(t) = \theta+\epsilon$ or $\mathrm{arg}(t) =\theta - \epsilon$.

If we take a singular direction $\theta$, and we assume for simplicity that along this direction the Borel transform $B[f]$ of the simple resurgent function $f$ has only one singularity at $t_\star$ given by the form Eq.~\eqref{simpleSing}, then we can compute
\begin{equation}
\mathcal{S}_{\theta_+}[f](g)-\mathcal{S}_{\theta_-}[f](g) = \int_{C_{\theta}} dt\,e^{-t/g}\,B[f](t)\,,
\end{equation}
where the contour $C_{\theta}$ comes from infinity in the direction $\mathrm{arg}(t)=\theta$, turns around the singular point $t_\star$ clockwise, and then goes back to infinity once again in the direction $\theta$, as displayed in Fig. \ref{fig:Hankel}.
Given the form of the singularity Eq.~\eqref{simpleSing}, one can show that
\begin{equation}
(\mathcal{S}_{\theta_+}-\mathcal{S}_{\theta_-})[f](g)= -e^{-t_\star/g} \,a_\theta-e^{-t_\star/g}\, \mathcal{S}_\theta \circ B^{-1}[b_\theta] (g)\,,
\end{equation}
where $B^{-1}$ is the inverse of the Borel transform.
As we can see from this last equation, the ambiguity in the Borel sum $\mathcal{S}_\theta[f]$ is related to the presence of infinitesimal terms $e^{-t_\star/g}(1+O(g))$. These terms cannot be captured by our perturbative series Eq.~\eqref{series}.  On a Stokes line it is essential to take into account non-perturbative contributions in order to be able to assign a well-defined meaning to the sum of the perturbative series.  Hence we have to replace our asymptotic series Eq.~\eqref{series} with a \emph{trans-series} of the schematic form
\begin{equation}
f_{TS}(g)=\sum_{n=0}^\infty a_n \,g^n+\sum_{c} e^{-t_c/g} \sum_{n=0}^\infty a_{c,n} \label{trans-series}\,.
\end{equation}
This is precisely what we would expect from an observable computed using a saddle-point method, see Eq.~\eqref{eq:NaiveTransSeries}.
The $ e^{-t_c/g}$ factors are non-analytic for $g\to 0$, so they have to be treated as objects external to the algebra of simple resurgent functions, and induce a grading on it.

As shown above, when the direction $\theta$ is a singular one, the Borel summation jumps as we cross this \textit{Stokes line},
and the full discontinuity across this direction plays a crucial role in linking perturbative and non-perturbative terms.
The \textit{Stokes automorphism} $\underline{\mathfrak{S}}_\theta$ is defined by
\begin{align}
&\mathcal{S}_{\theta^+} = \mathcal{S}_{\theta^-} \circ {\underline{\mathfrak{S}}}_\theta = \mathcal{S}_{\theta^-}\circ \left(\mbox{Id}-\mbox{Disc}_\theta\right)\label{eq:Stokes}\,,\\
&\mathcal{S}_{\theta^+}-\mathcal{S}_{\theta^-}=-\mathcal{S}_{\theta^-}\circ \mbox{Disc}_\theta\,,
\end{align}
where $\mbox{Disc}_\theta$ encodes the full discontinuity across $\theta$.

When the Stokes automorphism in a particular direction $\theta$ acts as the identity operator, it means that the Borel transform of $f(g)$ has no singularities along the $\theta$ direction and is given by a Borel-summable power series. Across a Stokes line $\mathfrak{S}_\theta$ is non-trivial and it encodes the jump between the two lateral resummations. Passing from a standard asymptotic series Eq.~\eqref{series} to a trans-series Eq.~\eqref{trans-series} with the inclusion of non-analytic (non-perturbative) terms of the form $e^{-t/g}$ is crucial.  While these terms are exponentially suppressed for $g\sim 0$ compared to terms of the form $a_n g^n$, when sitting on a Stokes line these terms are critical for making observables well-defined, and they must be taken into account.

By a contour deformation it is possible to show that the difference between the $\theta^+$ and $\theta^-$ deformation is nothing but a sum over Hankel's contours, and the discontinuity of $\mathcal{S}$ across $\theta$ is arising as an infinite sum of contribution coming from each one of the singular points.
The logarithm of the Stokes automorphism defines the \textit{alien derivative}
\begin{equation}
\underline {\frak S}_\theta = \exp \left(\sum_{t_\star \in\Gamma_\theta} e^{-t_\star / g}\Delta_{t_\star }\right)\,,\label{eq:expAlien}
\end{equation} 
where we denoted with $\Gamma_\theta$ the set of singular points of the Borel transform along the $\theta$ direction.
It is possible to show that this operator is a real derivation acting on the space of simple resurgent functions\cite{2007arXiv0706.0137S}.
When the Borel transform of $f$ has only one simple singularity Eq.~\eqref{simpleSing} at $ t_\star$ in the direction 
$\theta=\arg(t_\star)$, the alien derivative takes the simpler form
\begin{align}
&\Delta_{t} f(g) = 0 \,,\qquad\qquad \qquad\qquad\qquad t\neq t_\star\,,\\
&\Delta_{t_\star} f(g) = a_\gamma+B^{-1} [b_\gamma](g) \,\,,
\end{align}
where the path $\gamma$ is the line emanating from the origin in the direction $\theta = \arg(t_\star)$.
A more general definition is possible when there are multiple singular points along the chosen Stokes line but we will not need it for the present work \cite{2007arXiv0706.0137S}.

In the particular case in which $t_\star$ is the only simple singularity for the asymptotic series Eq.~\eqref{series} along $\theta$, we can rewrite the Stokes automorphism using Eq.~\eqref{eq:expAlien}
\begin{equation}
\underline {\frak S}_\theta f(g) = \left( 1+ e^{-t_\star/g}\,\Delta_{t_\star} +\frac{e^{-2\,t_\star/g}}{2}\,\Delta_{t_\star}^2+...  \right)f(g)\,.
\end{equation}
When $B^{-1} [b_\gamma](g)$ has no singularities in this direction, as in the $0$ dimensional example in the main text (\ref{hyper-disc}), the above equation simplifies even further giving
\begin{equation}
\underline {\frak S}_\theta f(g) =f(g) + e^{-t_\star/g}\left( a_\gamma+B^{-1} [b_\gamma](g) \right)\,,
\end{equation}
 which is just a manifestation of Stokes phenomena written in the language of alien derivatives.

\bibliography{PCM}

\end{document}